\documentclass[12pt]{article}
\usepackage{amssymb,amsmath}
\usepackage{mathrsfs}
\newcommand{\eq}[1]{(\ref{#1})}
\newcommand{\be}{\begin{equation}}
\newcommand{\ee}{\end{equation}}
\newcommand{\Sc}{Schr\"{o}dinger}
\newcommand{\sa}{self-adjoint}
\newcommand{\se}{Schr\"odinger's equation}
\newcommand{\BM}{Bohmian mechanics}
\newcommand{\qf}{quantum formalism}

\newcommand{\qt}{quantum theory}
\newcommand{\wf}{wave function}
\newcommand{\ewf}{effective wave function}
\newcommand{\cwf}{conditional wave function}
\newcommand{\qe}{quantum equilibrium}
\newcommand{\PV}{projection-valued-measure}

\newcommand{\nrao}{naive realism about operators}
\newtheorem{remark}{\small \bf Remark}[section]
\newcommand{\br}{ \begin{remark}\rm}
\newcommand{\er}{ \end{remark}}

\renewcommand{\dagger}{\ast}
\newcommand{\mybold}[1]{\mbox{\boldmath $#1$}}

\newcommand{\avec}[2]{\mbox{${#1}_{1},\ldots,{#1}_{#2}$}}
\newcommand{\rvect}[2]{(#1_1,\ldots,#1_#2)}
\newcommand{\pder}[2]{\frac{\partial #1}{\partial #2}}
\newcommand{\oder}[2]{\frac{ d #1}{ d #2}}
\renewcommand{\Im}{\mbox{\textup{\textrm{Im}}}\,}
\renewcommand{\div}{\mbox{\textup{\textrm{div}}}\,}
\newcommand{\tr}{\mbox{${\rm tr}\,$}}
\newcommand{\id}{I}
\renewcommand{\a}{\alpha}

\newcommand{\la}{\lambda_{\a}}
\newcommand{\lam}{\lambda}

\newcommand{\suma}{\sum_{\a }}

\newcommand{\ot}{\otimes}

\newcommand{\biga}{\bigoplus_{\a}}
\newcommand{\psia}{\psi_{\a}}
\newcommand{\Phia}{\Phi_{\a}}

\newcommand{\Ha}{{\H}_{\a}}
\renewcommand{\H}{\mbox{$\mathcal{H}$}}
\newcommand{\Pa}{ P_{ {\mathcal{H}_{\a} } } }

\newcommand{\As}{{R}}
\newcommand{\Aa}{R_{\a}}
\newcommand{\Aad}{R^{\dagger}_{\a}}
\newcommand{\Al}{R_{\lambda}}
\newcommand{\Ald}{R^{\dagger}_{\lambda}}

\newcommand{\CC}{\mathbb{C}}
\newcommand{\R}{\mathbb{R}}
\newcommand{\E}{\mbox{$\mathscr{E}$}}
\newcommand{\EE}{\mathscr{E}}
\newcommand{\Ex}{\mbox{$\mathcal{E}$}}

\newcommand{\M}{\mbox{$\mathcal{M}$}}
\renewcommand{\P}{\mbox{$\mathbb{P}$}}
\newcommand{\prob}{\mbox{Prob}_{\psi}}
\newcommand{\Prob}{\mbox{Prob}}
\newcommand{\pro}{\mbox{Prob}}
\newcommand{\norm}{\|}
\newcommand{\oqt}{orthodox quantum theory}
\newcommand{\rv}{Z}
\begin{document}
\title{Quantum Equilibrium and the Role of Operators as Observables in
   Quantum Theory\footnote{Dedicated to Elliott Lieb on the occasion of
     his 70th birthday. Elliott will be (we fear unpleasantly) surprised
     to learn that he bears a greater responsibility for this paper
     than he could possibly imagine. We would of course like to think
     that our work addresses in some way the concern suggested by the
     title of his recent talks, {\it The Quantum-Mechanical World View:
       A Remarkably Successful but Still Incomplete Theory}, but we
     recognize that our understanding of incompleteness is much more
     naive than Elliott's. He did, however, encourage us in his capacity
     as an editor of the Reviews of Modern Physics to submit a paper on
     the role of operators in quantum theory.  That was 12 year ago.
     Elliott is no longer an editor there and the paper that developed
     is not quite a review.} }
\author{ Detlef  D\"{u}rr\\
   Mathematisches Institut der Universit\"{a}t M\"{u}nchen\\
   Theresienstra{\ss}e 39, 80333 M\"{u}nchen, Germany\\
   E-mail: duerr@mathematik.uni-muenchen.de \and
   Sheldon  Goldstein\\
   Departments of Mathematics, Physics, and Philosophy, Rutgers
   University\\
   110 Frelinghuysen Road, Piscataway, NJ 08854-8019, USA\\
   E-mail: oldstein@math.rutgers.edu \and
   Nino Zangh\`{\i}\\
   Dipartimento di Fisica dell'Universit\`a di Genova\\Istituto
   Nazionale di Fisica Nucleare
   --- Sezione di Genova\\
   via Dodecaneso 33, 16146 Genova, Italy\\
   E-mail: zanghi@ge.infn.it} \date{} \maketitle
\begin{abstract}
   \BM\ is the most naively obvious embedding imaginable of \Sc's
   equation into a completely coherent physical theory.  It describes a
   world in which particles move in a highly non-Newtonian sort of way,
   one which may at first appear to have little to do with the spectrum
   of predictions of quantum mechanics.  It turns out, however, that as
   a consequence of the defining dynamical equations of \BM, when a
   system has \wf\ $\psi$ its configuration is typically random, with
   probability density $\rho$ given by $|\psi|^2$, the \qe\
   distribution.  It also turns out that the entire \qf, operators as
   observables and all the rest, naturally emerges in Bohmian mechanics
   {}from the analysis of ``measurements.''  This analysis reveals the
   status of operators as observables in the description of quantum
   phenomena, and facilitates a clear view of the range of
   applicability of the usual quantum mechanical formulas.
\end{abstract}

\maketitle

\tableofcontents

\section{Introduction}
\setcounter{equation}{0}

It is often argued that the quantum mechanical association of
observables with self-adjoint operators is a straightforward
generalization of the notion of classical observable, and that quantum
theory should be no more conceptually problematic than classical
physics {\it once this is appreciated}.  The classical physical
observables---for a system of particles, their positions
$q=(\mathbf{q}_k)$, their momenta $p=(\mathbf{p}_k)$, and the
functions thereof, i.e., functions on phase space---form a commutative
algebra.  It is generally taken to be the essence of quantization, the
procedure which converts a classical theory to a quantum one, that
\(q\), \(p\), and hence all functions \(f(q,p)\) thereof are replaced
by appropriate operators, on a Hilbert space (of possible wave
functions) associated with the system under consideration.  Thus
quantization leads to a noncommutative operator algebra of
``observables,'' the standard examples of which are provided by
matrices and linear operators.  Thus it seems perfectly natural that
classical observables are functions on phase space and quantum
observables are self-adjoint operators.

However, there is much less here than meets the eye.  What should be
meant by ``measuring" a quantum observable, a self-adjoint operator?
We think it is clear that this must be specified---without such
specification it can have no meaning whatsoever. Thus we should be
careful here and use safer terminology by saying that in quantum
theory observables are {\it associated} with self-adjoint operators,
since it is difficult to see what could be meant by more than an
association, by an {\it identification} of observables, regarded as
somehow having independent meaning relating to observation or
measurement (if not to intrinsic ``properties"), with such a
mathematical abstraction as a self-adjoint operator.

We are insisting on ``association" rather than identification in
quantum theory, but not in classical theory, because there we begin
with a rather clear notion of observable (or property) which is
well-captured by the notion of a function on the phase space, the
state space of {\it complete descriptions}.  If the state of the
system were observed, the value of the observable would of course be
given by this function of the state $(q,p)$, but the observable might
be observed by itself, yielding only a partial specification of the
state.  In other words, measuring a classical observable means
determining to which level surface of the corresponding function the
state of the system, the phase point---which is at any time {\it
   definite} though probably unknown---belongs.  In the quantum realm
the analogous notion could be that of function on Hilbert space, not
self-adjoint operator.  But we don't measure the wave function, so
that functions on Hilbert space are not physically measurable, and
thus do not define ``observables.''  \medskip

The problematical character of the way in which measurement is treated
in orthodox quantum theory has been stressed by John Bell:

\begin{quotation}\setlength{\baselineskip}{12pt}\noindent
   The concept of `measurement' becomes so fuzzy on reflection that it
   is quite surprising to have it appearing in physical theory {\it at
     the most fundamental level.\/} Less surprising perhaps is that
   mathematicians, who need only simple axioms about otherwise
   undefined objects, have been able to write extensive works on
   quantum measurement theory---which experimental physicists do not
   find it necessary to read. \dots Does not any {\it analysis\/} of
   measurement require concepts more {\it fundamental\/} than
   measurement?  And should not the fundamental theory be about these
   more fundamental concepts?~\cite{Bel81}
\end{quotation}

\begin{quotation}\setlength{\baselineskip}{12pt}\noindent
   \dots in physics the only observations we must consider are position
   observations, if only the positions of instrument pointers.  It is a
   great merit of the de Broglie-Bohm picture to force us to consider
   this fact.  If you make axioms, rather than definitions and
   theorems, about the `measurement' of anything else, then you commit
   redundancy and risk inconsistency.~\cite{Bel82}
\end{quotation}

The Broglie-Bohm theory, Bohmian mechanics, is a physical theory for
which the concept of `measurement' does not appear at the most
fundamental level---in the very formulation of the theory. It is a
theory about concepts more fundamental than `measurement,' in terms of
which an analysis of measurement can be performed. In a previous work
\cite{DGZ92a} we have shown how probabilities for positions of
particles given by $|\psi|^2$ emerge naturally {}from an analysis of
``equilibrium'' for the deterministic dynamical system defined by
Bohmian mechanics, in much the same way that the Maxwellian velocity
distribution emerges {}from an analysis of classical thermodynamic
equilibrium.  Our analysis entails that Born's statistical rule
$\rho=|\psi|^{2}|$ should be regarded as a local manifestation of a
global equilibrium state of the universe, what we call \emph{quantum
   equilibrium}, a concept analogous to, but quite distinct {}from,
thermodynamic equilibrium: a universe in quantum equilibrium evolves
so as to yield an appearance of randomness, with empirical
distributions in agreement with all the predictions of the quantum
formalism.

While in our earlier work we have proven, {}from the first principles
of \BM{}, the ``quantum equilibrium hypothesis'' that \emph{when a
   system has \wf\ $\psi$, the distribution $\rho$ of its configuration
   satisfies $\;\rho = |\psi|^2$}, our goal here is to show that it
follows {}from this hypothesis, not merely that \BM{} makes the same
predictions as does orthodox quantum theory for the results of any
experiment, but that \emph{the quantum formalism of operators as
   observables emerges naturally and simply as the very expression of
   the empirical import of \BM{}}.

More precisely, we shall show here that \sa{} operators arise in
association with specific experiments: insofar as the statistics for
the values which result {}from the experiment are concerned, the
notion of \sa{} operator compactly expresses and represents the
relevant data.  It is the association ``$\E\mapsto A$'' between an
experiment \E{} and an operator $A$---an association that we shall
establish in Section 2 and upon which we shall elaborate in the other
sections---that is the central notion of this paper. According to this
association the notion of operator-as-observable in no way implies
that anything is measured in the experiment, and certainly not the
operator itself.  We shall nonetheless speak of such experiments as
measurements, since this terminology is unfortunately standard. When
we wish to emphasize that we really mean measurement---the
ascertaining of the value of a quantity---we shall often speak of
genuine measurement.

Much of our analysis of the emergence and role of operators as
observables in \BM{}, including the von Neumann-type picture of
measurements at which we shall arrive, applies as well to orthodox
\qt{}.  Indeed, the best way to understand the status of the quantum
formalism---and to better appreciate the minimality of Bohmian
mechanics---is Bohr's way: What are called quantum observables obtain
meaning \emph{only} through their association with specific
\emph{experiments}.  We believe that Bohr's point has not been taken
to heart by most physicists, even those who regard themselves as
advocates of the Copenhagen interpretation.

Indeed, it would appear that the argument provided by our analysis
against taking operators too seriously as observables has even greater
force {}from an orthodox perspective: Given the initial \wf{}, at
least in \BM{} the outcome of the particular experiment is determined
by the initial configuration of system and apparatus, while for
orthodox quantum theory there is nothing in the initial state which
completely determines the outcome.  Indeed, we find it rather
surprising that most proponents of standard quantum measurement
theory, that is the von Neumann analysis of measurement \cite{vNe55},
beginning with von Neumann, nonetheless seem to retain an uncritical
identification of operators with properties.  Of course, this is
presumably because more urgent matters---the measurement problem and
the suggestion of inconsistency and incoherence that it entails---soon
force themselves upon one's attention.  Moreover such difficulties
perhaps make it difficult to maintain much confidence about just what
{\it should\/} be concluded {}from the ``measurement'' analysis, while
in \BM, for which no such difficulties arise, what should be concluded
is rather obvious.

Moreover, a great many significant real-world experiments are simply
not at all associated with operators in the usual way.  Because of
these and other difficulties, it has been proposed that we should go
beyond operators-as-observables, to {\it generalized observables\/},
described by mathematical objects (positive-operator-valued measures,
POVMs) even more abstract than operators (see, e.g., the books of
Davies \cite{Dav76}, Holevo \cite{Hol82} and Kraus \cite{Kra83}).  It
may seem that we would regard this development as a step in the wrong
direction, since it supplies us with a new, much larger class of
abstract mathematical entities about which to be naive realists.  We
shall, however, show that these generalized observables for \BM\ form
an extremely natural class of objects to associate with experiments,
and that the emergence and role these observables is merely an
expression of \qe\ together with the linearity of \Sc's evolution.  It
is therefore rather dubious that the occurrence of generalized
observables---the simplest case of which are \sa{} operators---can be
regarded as suggesting any deep truths about reality or about
epistemology.

As a byproduct of our analysis of measurement we shall obtain a
criterion of measurability and use it to examine the genuine
measurability of some of the properties of a physical system. In this
regard, it should be stressed that measurability is theory-dependent:
different theories, though empirically equivalent, may differ on what
should be regarded as genuinely measurable \emph{within} each theory.
This important---though very often ignored---point was made long ago
by Einstein and has been repeatedly stressed by Bell.  It is best
summarized by Einstein's remark~\cite{Hei58}: \emph{``It is the theory
   which decides what we can observe.''}

We note in passing that measurability and reality are different
issues.  Indeed, for \BM{} most of what is ``measurable'' (in a sense
that we will explain) is not real and most of what is real is not
genuinely measurable. (The main exception, the position of a particle,
which is both real and genuinely measurable, is, however, constrained
by absolute uncertainty \cite{DGZ92a}).

In focusing here on the role of operators as observables, we don't
wish to suggest that there are no other important roles played by
operators in quantum theory.  In particular, in addition to the
familiar role played by operators as generators of symmetries and
time-evolutions, we would like to mention the rather different role
played by the {\em field operators} of quantum field theory: to link
abstract Hilbert-space to space-time and structures therein,
facilitating the formulation of theories describing the behavior of an
indefinite number of particles \cite{crea1,crea2}.

Finally, we should mention what should be the most interesting sense
of measurement for a physicist, namely the determination of the
coupling constants and other parameters that define our physical
theories. This has little to do with operators as observables in
quantum theory and shall not be addressed here.

\subsection*{Notations and Conventions}
$Q= \rvect{\mathbf{Q}}{N}$ denotes the actual configuration of a
system of $N$ particle with positions $\mathbf{Q}_k$;
$q=\rvect{\mathbf{q}}{N}$ is its generic configuration. Whenever we
deal with a system-apparatus composite, $x$ ($X$) will denote the
generic (actual) configuration of the system and $y$ ($Y$) that of the
apparatus. Sometimes we shall refer to the system as the $x$-system
and the apparatus as the $y$-system.  Since the apparatus should be
understood as including all systems relevant to the behavior of the
system in which we are interested, this notation and terminology is
quite compatible with that of Section \ref{sec:CEWF}, in which $y$
refers to the environment of the $x$-system.

For a system in the state $\Psi$, $\rho_{\Psi}$ will denote the
quantum equilibrium measure, $\rho_{\Psi}(dq)= |\Psi(q)|^2dq$.  If
$Z=F(Q)$ then $\rho_{\Psi}^Z$ denotes the measure induced by $F$, i.e.
$\rho_{\Psi}^Z= \rho_{\Psi}\circ F^{-1}$.

\section{Bohmian Experiments}\label{sec:BE}
\setcounter{equation}{0}
\label{sec:BM}

According to \BM{}, the complete description or state of an
$N$-particle system is provided by its \wf{} $\Psi(q,t)$, where
$q=\rvect{\mathbf{q}}{N} \in \R ^{3N}$, \emph{and} its configuration
$Q= \rvect{\mathbf{Q}}{N}\in \R ^{3N}$, where the $\mathbf{Q}_k$ are
the positions of the particles.  The \wf{}, which evolves according to
\se{},
\begin{equation}
i\hbar\pder{\Psi}{t} = H\Psi \,,
\label{eq:eqsc}
\end{equation}
choreographs the motion of the particles: these evolve according to
the equation
\begin{equation}
\oder{\mathbf{Q}_{k}}{t} = \frac{\hbar}{m_{k}} {\rm Im}\frac{
\Psi^*\boldsymbol{\nabla}_{k}\Psi}{\Psi^*\Psi}\,
\rvect{\mathbf{Q}}{N}
\label{eq:velo}
\end{equation}
where $\mybold{\nabla}_{\!k}=\partial/\partial \mathbf{q}_{\!k}.$ In
equation (\ref{eq:eqsc}), $H$ is the usual nonrelativistic \Sc\
Hamiltonian; for spinless particles it is of the form
\begin{equation}
H=-{\sum}_{k=1}^{N}
\frac{{\hbar}^{2}}{2m_{k}}\boldsymbol{\nabla}^{2}_{k} + V,
\label{sh}
\end{equation}
containing as parameters the masses $m_1\dots, m_N$ of the particles
as well as the potential energy function $V$ of the system.  For an
$N$-particle system of nonrelativistic particles, equations
(\ref{eq:eqsc}) and (\ref{eq:velo}) form a complete specification of
the theory (magnetic fields\footnote{When a magnetic field is present,
   the gradients $\boldsymbol{\nabla}_{k}$ in the equations
   (\ref{eq:eqsc} and (\ref{eq:velo}) must be understood as the
   covariant derivatives involving the vector potential
   $\boldsymbol{A}$.  } and spin,\footnote{See Section \ref{secSGE}.}
as well as Fermi and Bose-Einstein statistics,\footnote{For
   indistinguishable particles, a careful analysis \cite{DGZ94} of the
   natural configuration space, which is no longer $\R^{3N}$, leads to
   the consideration of \wf s on $\R^{3N}$ that are either symmetric or
   antisymmetric under permutations.} can easily be dealt with and in
fact arise in a natural manner \cite{Bel66,Boh52,Nel85,Gol87,DGZ94}).
There is no need, and indeed no room, for any further \emph{axioms},
describing either the behavior of other observables or the effects of
measurement.

\subsection{Equivariance and Quantum Equilibrium}\label{sec.e}

It is important to bear in mind that regardless of which observable
one chooses to measure, the result of the measurement can be assumed
to be given configurationally, say by some pointer orientation or by a
pattern of ink marks on a piece of paper.  Then the fact that \BM{}
makes the same predictions as does orthodox quantum theory for the
results of any experiment---for example, a measurement of momentum or
of a spin component---\emph{provided we assume a random distribution
   for the configuration of the system and apparatus at the beginning
   of the experiment given by $|\Psi(q)|^2$}---is a more or less
immediate consequence of (\ref{eq:velo}).  This is because of the
quantum continuity equation
\begin{displaymath}
\pder{|\Psi|^2}{t} + \div J^{\Psi} =0,
\end{displaymath}
which is a simple consequence of \se{}.  Here $ J^{\Psi}=
\rvect{\mathbf{J}^\Psi}{N} $ with
\begin{displaymath}
\mathbf{J}^\Psi_k= \frac{\hbar}{m_k} \Im
(\Psi^*\mybold{\nabla}_{\!k}\Psi)
\end{displaymath}
the \emph{quantum probability current}.  This equation becomes the
classical continuity equation
\begin{equation}
\pder{\rho}{t} + \div \rho\, v =0
\label{eq:contieq}
\end{equation}
for the system of equations $dQ/dt=v$ defined by
(\ref{eq:velo})---governing the evolution of the probability density
$\rho$ under the motion defined by the guiding equation
(\ref{eq:velo}) for the particular choice $\rho=|\Psi|^2=\Psi^*\Psi$.
In other words, if the probability density for the configuration
satisfies $\rho(q,t_0)=|\Psi(q,t_0)|^2$ at some time $t_0$, then the
density to which this is carried by the motion (\ref{eq:velo}) at any
time $t$ is also given by $\rho(q,t)=|\Psi(q,t)|^2$.  This is an
extremely important property of any Bohmian system, as it expresses a
certain compatibility between the two equations of motion defining the
dynamics, which we call the
\emph{equivariance}\footnote{\label{fn:equivariance} Equivariance can
   be formulated in very general terms: consider the transformations $
   U: \Psi \to U \Psi$ and $f: Q \to f(Q) $, where $U$ is a unitary
   transformation on $L^{2}(dq)$ and $f$ is a transformation on
   configuration space that may depend on $\Psi$.  We say that the map
   $\Psi\mapsto\mu_{\Psi}$ from \wf s to measures on configuration
   space is equivariant with respect to $U$ and $f$ if $ \mu_{\,U\Psi}
   = \mu_{\Psi}\circ f^{-1} $.  The above argument based on the
   continuity equation (\ref{eq:contieq}) shows that $ \Psi\mapsto
   |\Psi|^{2} dq$ is equivariant with respect to $U\equiv U_{t} =
   e^{-i\,\frac{t}{\hbar}\,H}$, where $H$ is the \Sc{} Hamiltonian
   (\ref{sh}) and $f\equiv f_{t}$ is the solution map of
   (\ref{eq:velo}).  In this regard, it is important to observe that
   for a Hamiltonian $H$ which is not of \Sc{} type we shouldn't expect
   \eq{eq:velo} to be the appropriate velocity field, that is, a field
   which generates an evolution in configuration space having $
   |\Psi|^{2} $ as equivariant density.  For example, for $H=
   c\frac{\hbar}{i}\frac{\partial}{\partial q}$, where $c$ is a
   constant (for simplicity we are assuming configuration space to be
   one-dimensional), we have that $ |\Psi|^{2} $ is equivariant
   \emph{provided} the evolution of configurations is given by
   ${dQ}/{dt} = c$.  In other words, for $U_{t}= e^{ct
     \frac{\partial}{\partial q}}$ the map $ \Psi\mapsto |\Psi|^{2} dq$
   is equivariant if $f_{t}: Q\to Q +ct$.}  of $|\Psi|^2$.

The above assumption guaranteeing agreement between \BM{} and quantum
mechanics regarding the results of any experiment is what we call the
``quantum equilibrium hypothesis'':
\begin{equation}
\mbox{
\begin{minipage}{0.85\textwidth}
   \emph{When a system has \wf\ $\Psi$ its configuration $Q$ is random
     with probability distribution given by the measure
     $\rho_{\Psi}(dq)=|\Psi(q)|^2dq$.}
\end{minipage}}
\label{def:qe}
\end{equation}
When this condition is satisfied we shall say that the system is in
quantum equilibrium and we shall call $\rho_{\Psi}$ the quantum
equilibrium distribution.  While the meaning and justification of
(\ref{def:qe}) is a delicate matter, which we have discussed at length
elsewhere \cite{DGZ92a}, it is important to recognize that, merely as
a consequence of (\ref{eq:velo}) and (\ref{def:qe}), \BM{} is a
counterexample to all of the claims to the effect that a deterministic
theory cannot account for quantum randomness in the familiar
statistical mechanical way, as arising {}from averaging over
ignorance: \BM{} is clearly a deterministic theory, and, as we have
just explained, it does account for quantum randomness as arising
{}from averaging over ignorance given by $|\Psi(q)|^2$.

\subsection{Conditional and Effective Wave Functions}
\label{sec:CEWF}
Which systems should be governed by \BM ?  An $n$-particle subsystem
of an $N$-particle system ($n<N$) need not in general be governed by
\BM, since no \wf\ for the subsystem need exist.  This will be so even
with trivial interaction potential $V$, if the \wf\ of the system does
not properly factorize; for nontrivial $V$ the \Sc\ evolution would in
any case quickly destroy such a factorization.  Therefore in a
universe governed by \BM\ there is a priori only one wave function,
namely that of the universe, and there is a priori only one system
governed by \BM, namely the universe itself.

Consider then an $N$-particle non relativistic universe governed by
\BM{}, with (universal) \wf{} $\Psi$.  Focus on a subsystem with
configuration variables $x$, i.e., on a splitting $q=(x,y)$ where $y$
represents the configuration of the \emph{environment} of the
\emph{$x$-system}.  The actual particle configurations at time $t$ are
accordingly denoted by $X_t$ and $Y_t$, i.e., $Q_t=(X_t ,Y_t)$.  Note
that $\Psi_t=\Psi_t(x,y)$.  How can one assign a \wf{} to the
$x$-system?  One obvious possibility---\emph{afforded by the existence
   of the actual configuration}---is given by what we call the
\emph{conditional} wave function of the $x$-system
\begin{equation}
\psi_t(x) = \Psi_t (x,Y_t).
\label{eq:con}
\end{equation}

To get familiar with this notion consider a very simple one
dimensional universe made of two particles with Hamiltonian
($\hbar=1$)
\begin{displaymath}
H =H^{(x)}+H^{(y)} +H^{(xy)} = -\frac{1}{2}\big(
\frac{\partial^2}{\partial x^2} + \frac{\partial^2}{\partial
y^2}\big) \ + \frac{1}{2} (x-y)^2 .
\end{displaymath}
and initial \wf{}
\begin{displaymath}
\Psi_0 = \psi \ot \Phi_0 \quad\hbox{with}\quad
\psi(x)=\pi^{-\frac{1}{4}} e^{-\frac{x^2}{2}}\quad\hbox{and}\quad
\Phi_0 (y)=\pi^{-\frac{1}{4}} e^{-\frac{y^2}{2}}.
\end{displaymath}
Then (\ref{eq:eqsc}) and (\ref{eq:velo}) are easily solved:
\begin{displaymath}
\Psi_t (x,y) =\pi^{-\frac{1}{2}} (1+it)^{-\frac{1}{2}}
e^{-\frac{1}{4}\big[(x- y)^2+\frac{(x+y)^2}{1+2it}\big]},
\end{displaymath}
\begin{displaymath}
X_t = a(t)X + b(t)Y \quad\hbox{and}\quad Y_t = b(t)X +a(t) Y ,
\end{displaymath}
where $a(t)= \frac{1}{2}[ (1+t^2)^{\frac{1}{2}}+1] $, $b(t)=
\frac{1}{2}[ (1+t^2)^{\frac{1}{2}}-1] $, and $X$ and $Y$ are the
initial positions of the two particles.  Focus now on one of the two
particles (the $x$-system) and regard the other one as its environment
(the $y$-system).  The conditional \wf{} of the $x$-system
\begin{displaymath}
\psi_t(x) = \pi^{-\frac{1}{2}} (1+it)^{-\frac{1}{2}}
e^{-\frac{1}{4}\big[(x- Y_{t})^2+\frac{(x+Y_{t})^2}{1+2it}\big]},
\end{displaymath}
depends, through $Y_t$, on \emph{both} the initial condition $Y$ for
the environment \emph{and} the initial condition $X$ for the particle.
As these are random, so is the evolution of $\psi_t$, with probability
law determined by $|\Psi_0|^2$.  In particular, $\psi_t$ does not
satisfy \Sc's equation for any $H^{(x)}$.

We remark that even when the $x$-system is dynamically decoupled
{}from its environment, its conditional \wf\ will not in general
evolve according to \Sc's equation. Thus the conditional \wf\ lacks
the {\it dynamical} implications {}from which the \wf\ of a system
derives much of its physical significance. These are, however,
captured by the notion of \ewf:
\begin{equation}
\mbox{%
\begin{minipage}{0.85\textwidth}\openup 1.2\jot
   \setlength{\baselineskip}{12pt}\emph{Suppose that $\;\Psi(x,y) =
     \psi(x)\Phi(y) + \Psi^{\perp}(x,y)\, ,\;$ where $\Phi$ and
     $\Psi^{\perp}$ have macroscopically disjoint $y$-supports.  If $\;
     Y \in \hbox{\rm supp}\;\Phi \;$ we say that $\psi$ is the
     \emph{\ewf} of the $x$-system.}
\end{minipage}}
\label{def:ewf}
\end{equation}
Of course, $\psi$ is also the \cwf{} since nonvanishing scalar
multiples of \wf s are naturally
identified.\footnote{\label{foosti}Note that in \BM\ the \wf\ is
   naturally a projective object since \wf s differing by a
   multiplicative constant---possibly time-dependent---are associated
   with the same vector field, and thus generate the same dynamics.  }

\subsection{Decoherence}
\label{sec:DD}

One might wonder why systems possess an effective wave function at
all.  In fact, in general they don't!  For example the $x$-system will
not have an \ewf{} when, for instance, it belongs to a larger
microscopic system whose \ewf{} doesn't factorize in the appropriate
way.  However, the \emph{larger} the environment of the $x$-system,
the \emph{greater} is the potential for the existence of an \ewf{} for
this system, owing in effect to the abundance of ``measurement-like''
interaction with a larger environment.\footnote{To understand how this
   comes about one may suppose that initially the $y$-supports of
   $\Phi$ and $\Psi^{\perp}$ (cf.  the definition above of effective
   \wf{}) are just ``sufficiently'' (but not macroscopically) disjoint.
   Then, due to the interaction with the environment, the amount of
   $y$-disjointness will tend to increase dramatically as time goes on,
   with, as in a chain reaction, more and more degrees of freedom
   participating in this disjointness.  When the effect of this
   ``decoherence'' is taken into account, one finds that even a small
   amount of $y$-disjointness will often tend to become ``sufficient,''
   and quickly ``more than sufficient,'' and finally macroscopic.}

We remark that it is the relative stability of the macroscopic
disjointness employed in the definition of the \ewf, arising {}from
what are nowadays often called mechanisms of decoherence---the
destruction of the coherent spreading of the wave function, the
effectively irreversible flow of ``phase information'' into the
(macroscopic) environment---which accounts for the fact that the
\ewf{} of a system obeys \Sc's equation for the system alone whenever
this system is isolated.  One of the best descriptions of the
mechanisms of decoherence, though not the word itself, can be found in
Bohm's 1952 ``hidden variables'' paper \cite{Boh52}.

Decoherence plays a crucial role in the very formulation of the
various interpretations of \qt\ loosely called decoherence
theories(Griffiths \cite{Gri84}, Omn\`es \cite{Omn88}, Leggett
\cite{Leg80}, Zurek \cite{Zur82}, Joos and Zeh \cite{JZ85}, Gell-Mann
and Hartle \cite{GMH90}).  In this regard we wish to emphasize,
however, as did Bell in his article ``Against Measurement''
\cite{Bel90}, that decoherence in no way comes to grips with the
measurement problem itself, being arguably a {\it necessary}, but
certainly not a sufficient, condition for its complete resolution.  In
contrast, for Bohmian mechanics decoherence is purely
phenomenological---it plays no role whatsoever in the formulation (or
interpretation) of the theory itself\footnote{However, decoherence
   plays an important role in the emergence of Newtonian mechanics as
   the description of the macroscopic regime for Bohmian mechanics,
   supporting a picture of a macroscopic Bohmian particle, in the
   classical regime, guided by a macroscopically well-localized wave
   packet with a macroscopically sharp momentum moving along a
   classical trajectory.  It may, indeed, seem somewhat ironic that the
   gross features of our world should appear classical because of
   interaction with the environment and the resulting wave function
   entanglement, the characteristic quantum innovation.}---and the very
notion of \ewf\ accounts at once for the reduction of the wave packet
in quantum measurement.

According to orthodox quantum measurement theory \cite{vNe55, Boh51,
   Wig63, Wig83}, after a measurement, or preparation, has been
performed on a quantum system, the $x$-system, the \wf\ for the
composite formed by system and apparatus is of the form
\begin{equation}
\sum_\a{\psi_\a\otimes\Phi_\a}
\label{eq:msum}
\end{equation}
with the different $\Phi_\a$ supported by the macroscopically distinct
(sets of) configurations corresponding to the various possible
outcomes of the measurement, e.g., given by apparatus pointer
orientations. Of course, for \BM\ the terms of \eq{eq:msum} are not
all on the same footing: one of them, and only one, is selected, or
more precisely supported, by the outcome---corresponding, say, to
$\a_0$---which {\it actually\/} occurs. To emphasize this we may write
(\ref{eq:msum}) in the form
$$
\psi\otimes\Phi+\Psi^\perp
$$
where $\psi=\psi_{\a_0}$, $\Phi=\Phi_{\a_0}$, and
$\Psi^\perp=\sum_{\a\neq\a_0}{\psi_\a\otimes\Phi_\a}$.  By comparison
with (\ref{def:ewf}) it follows that after the measurement the
$x$-system has \ewf\ $\psi_{\a_0}$.  This is how {\it collapse} (or
{\it reduction}) of the \ewf\ to the one associated with the outcome
$\a_0$ arises in \BM.

While in orthodox quantum theory the ``collapse'' is merely
superimposed upon the unitary evolution---without a precise
specification of the circumstances under which it may legitimately be
invoked---we have now, in \BM{}, that the evolution of the \ewf{} is
actually given by a stochastic process, which consistently embodies
\emph{both} unitarity \emph{and} collapse as appropriate.  In
particular, the \ewf{} of a subsystem evolves according to \se{} when
this system is suitably isolated.  Otherwise it ``pops in and out'' of
existence in a random fashion, in a way determined by the continuous
(but still random) evolution of the conditional \wf{} $\psi_t$.
Moreover, it is the critical dependence on the state of the
environment and the initial conditions which is responsible for the
random behavior of the (conditional or effective) \wf{} of the system.

\subsection{Wave Function and State}
\label{sec:WFS}

As an important consequence of (\ref{eq:con}) we have, for the
conditional probability distribution of the configuration $X_{t}$ of a
system at time $t$, given the configuration $Y_{t}$ of its
environment, the \emph{fundamental conditional probability formula}
\cite{DGZ92a}:
\begin{equation}
\Prob_{\Psi_{0}} \bigl(X_t \in dx \bigm| Y_t\bigr)=|\psi_t(x)|^2\,dx,
\label{eq:fpf}
\end{equation}
where
\begin{displaymath}
\Prob_{\Psi_{0}}(dQ)={|\Psi_0(Q)|}^2\,dQ,
\end{displaymath}
with $Q=(X,Y)$ the configuration of the universe at the (initial) time
$t=0$.  Formula (\ref{eq:fpf}) is the cornerstone of our analysis
\cite{DGZ92a} on the origin of randomness in \BM{}.  Since the right
hand side of (\ref{eq:fpf}) involves only the effective \wf{}, it
follows that \emph{the \wf{} $\psi_t$ of a subsystem represents
   maximal information about its configuration $X_t$.} In other words,
given the fact that its \wf{} is $\psi_t$, it is in principle
impossible to know more about the configuration of a system than what
is expressed by the right hand side of (\ref{eq:fpf}), even when the
detailed configuration $Y_t$ of its environment is taken into account
\cite{DGZ92a}
\begin{equation}
\Prob_{\Psi_{0}}\bigl(X_t \in dx \bigm| Y_t\bigr)=
\Prob_{\Psi_{0}}\bigl(X_t \in dx \bigm|
\psi_t\bigr)=|\psi_t(x)|^2\,dx.
\label{eq:fpfp}
\end{equation}

The fact that the knowledge of the configuration of a system must be
mediated by its \wf{} may partially account for the possibility of
identifying the \emph{state} of a system---its complete
description---with its \wf{} without encountering any \emph{practical}
difficulties.  This is primarily because of the \wf{}'s statistical
role, but its dynamical role is also relevant here.  Thus it is
natural, even in \BM{}, to regard the \wf{} as the ``\emph{state}'' of
the system.  This attitude is supported by the asymmetric roles of
configuration and \wf{}: while the \emph{fact} that the \wf{} is
$\psi$ entails that the configuration is distributed according to
$|\psi|^2$, the \emph{ fact} that the configuration is $X$ has no
implications whatsoever for the \wf{}.\footnote{The ``fact'' (that the
   configuration is $X$) shouldn't be confused with the ``knowledge of
   the fact'': the latter does have such implications \cite{DGZ92a}!}
Indeed, such an asymmetry is grounded in the dynamical laws \emph{and}
in the initial conditions.  $\psi$ is always assumed to be fixed,
being usually under experimental control, while $X$ is always taken as
random, according to the \qe{} distribution.

When all is said and done, it is important to bear in mind that
regarding $\psi$ as the ``state'' is only of practical value, and
shouldn't obscure the more important fact that the most detailed
description---\emph{the complete state description}---is given (in
Bohmian mechanics) by the \wf{} \emph{and} the configuration.

\subsection{The Stern-Gerlach Experiment}
\label{secSGE}
Information about a system does not spontaneously pop into our heads
or into our (other) ``measuring'' instruments; rather, it is generated
by an \emph{experiment}: some physical interaction between the system
of interest and these instruments, which together (if there is more
than one) comprise the \emph{apparatus} for the experiment.  Moreover,
this interaction is defined by, and must be analyzed in terms of, the
physical theory governing the behavior of the composite formed by
system and apparatus.  If the apparatus is well designed, the
experiment should somehow convey significant information about the
system.  However, we cannot hope to understand the significance of
this ``information''---for example, the nature of what it is, if
anything, that has been measured---without some such theoretical
analysis.

As an illustration of such an analysis we shall discuss the
Stern-Gerlach experiment {}from the standpoint of \BM. But first we
must explain how {\it spin} is incorporated into \BM: If $\Psi$ is
spinor-valued, the bilinear forms appearing in the numerator and
denominator of (\ref{eq:velo}) should be understood as
spinor-inner-products; e.g., for a single spin $\frac{1}{2}$ particle
the two-component \wf{} $$
\Psi \equiv \left(\begin{array}{c} \Psi_+
     ({\bf x})\\ \Psi_- ({\bf x})
\end{array} \right) $$
generates the velocity
\begin{equation}
\label{vspin}
{\bf v}^{\Psi} = \frac{\hbar}{m}{\rm Im}
\frac{(\Psi,\boldsymbol{\nabla}\Psi)} {(\Psi, \Psi)}
\end{equation}
where $(\,\cdot\,,\,\cdot\,)$ denotes the scalar product in the spin
space $\CC^{2}$.  The \wf\ evolves via (\ref{eq:eqsc}), where now the
Hamiltonian $H$ contains the Pauli term, for a single particle
proportional to $\mathbf{B}\cdot\boldsymbol{ \sigma}$, that represents
the coupling between the ``spin'' and an external magnetic field
$\mathbf{B}$; here $\boldsymbol{\sigma} =(\sigma_x,\sigma_y,\sigma_z)$
are the Pauli spin matrices which can be taken to be
$$
\sigma_x \,=\, \left(\begin{array}{cc} 0 & 1 \\ 1 & 0\end{array}
\right) \quad \sigma_y\,=\, \left( \begin{array}{cc} 0 & -i \\ i& 0
\end{array} \right) \quad \sigma_z\,=\, \left( \begin{array}{cc} 1 &
0\\ 0& -1 \end{array} \right)
$$

Let's now focus on a Stern-Gerlach ``measurement of the operator
$\sigma_z$'': An inhomogeneous magnetic field $\mathbf{B}$ is
established in a neighborhood of the origin, by means of a suitable
arrangement of magnets.  This magnetic field is oriented in the
positive $z$-direction, and is increasing in this direction.  We also
assume that the arrangement is invariant under translations in the
$x$-direction, i.e., that the geometry does not depend upon
$x$-coordinate.  A particle with a fairly definite momentum is
directed towards the origin along the negative $y$-axis.  For
simplicity, we shall consider a neutral spin-$1/2$ particle whose
\wf{} $\Psi$ evolves according to the Hamiltonian
\begin{equation}
\label{sgh}
H = -\frac{\hbar^{2}}{2m} \boldsymbol{\nabla}^{2}  -
\mu\boldsymbol{\sigma}{\bf \cdot B}.
\end{equation}
where $\mu$ is a positive constant (if one wishes, one might think of
a fictitious electron not feeling the Lorentz force).

The inhomogeneous field generates a vertical deflection of $\Psi$ away
{}from the $y$-axis, which for \BM\ leads to a similar deflection of
the particle trajectory according to the velocity field defined by
\eq{vspin}: if its \wf\ $\Psi$ were initially an eigenstate of
$\sigma_z$ of eigenvalue $1$ or $-1$, i.e., if it were of the form
$$
\Psi^{(+)}=\psi^{(+)}\otimes\Phi_0 ({\bf x})\qquad\text{or}\qquad
\Psi^{(-)}=\psi^{(-)}\otimes\Phi_0 ({\bf x})
$$
where
\begin{equation}
\psi^{(+)} \equiv \left(\begin{array}{c} 1\\  0
\end{array} \right) \quad \mbox{and} \quad
\psi^{(-)} \equiv \left(\begin{array}{c} 0\\  1
\end{array} \right)
\label{eq:spinbasis}
\end{equation}
then the deflection would be in the positive (negative) $z$-direction
(by a rather definite angle).  This limiting behavior is readily seen
for $\Phi_0 = \Phi_0(z)\phi(x,y)$ and ${\bf B}=(0,0,B)$, so that the
$z$-motion is completely decoupled {}from the motion along the other
two directions, and by making the standard (albeit unphysical)
assumption \cite{Boe79}, \cite{Boh51}
   \begin{equation}
\label{consg}
\frac{\partial B}{\partial z} = const > 0\,.
\end{equation}
whence
$$
\mu\boldsymbol{\sigma}{\bf \cdot B}=( b+ az) \sigma_{z}
$$
where $a>0$ and $b$ are constants.  Then
$$
\Psi^{(+)}_{t} = \left(\begin{array}{c}
     \Phi^{(+)}_{t}(z)\phi_{t}(x,y)\\ 0
\end{array} \right) \quad \mbox{and} \quad
\Psi^{(-)}_{t} = \left(\begin{array}{c} 0\\
     \Phi^{(-)}_{t}(z)\phi_{t}(x,y)
\end{array} \right)
$$
where $\Phi^{(\pm)}_{t}$ are the solutions of
\begin{equation}
i\hbar\frac{\partial {\Phi_t}^{(\pm)}}{\partial t}=
-\frac{\hbar^2}{2m} \frac{\partial^2 {\Phi_t}^{(\pm)} }{\partial
z^2}\, \mp \, (b + a\,z){\Phi_t}^{(\pm)},
\label{eq:SGequ}
\end{equation}
for initial conditions ${\Phi_0}^{(\pm)}=\Phi_0(z)$.  Since $z$
generates translations of the $z$-component of the momentum, the
behavior described above follows easily. More explicitly, the limiting
behavior for $t\to\infty$ readily follows by a stationary phase
argument on the explicit solution\footnote{Eq. (\ref{eq:SGequ}) is
   readily solved:
   $$
   \Phi^{(\pm)}_{t}(z) = \int G^{(\pm)}(z,z';t) \Phi_{0}(z')\,
   dz'\,,
\label{eq:solpauli}
$$
where (by the standard rules for the Green's function of linear and
quadratic Hamiltonians)
$$
G^{(\pm)}(z,z';t) = \sqrt{\frac{m}{ 2 \pi i \hbar t}}\,
e^{\frac{i}{\hbar} \left( \frac{m}{2t}\left( z-z' -
       (\pm)\frac{at^{2}}{m}\right)^{2} + \frac{(\pm) at}{2} \left(
       z-z' - (\pm)\frac{at^{2}}{m}\right) -(\pm) (az'+b) t +
     \frac{at^{3}}{3m} \right)}
\label{eq:proppauli}
$$} of (\ref{eq:SGequ}). More simply, we may consider the initial
Gaussian state $$\Phi_{0}= \frac {e^{( - \,\frac {z^{2}}{4d^{2}}) }}{
   (2 d^{2}\pi)^\frac{1}{4} } $$
for which $|\Phi_{t}^{\pm}(z)|^2$, the
probability density of the particle being at a point of $z$-coordinate
$z$, is, by the linearity of the interaction in (\ref{eq:SGequ}), a
Gaussian with mean and mean square deviation given respectively by
\begin{equation}
\bar{z}(t) =(\pm) \frac {a\,t^{2}}{2 m}\quad\qquad d(t) =d \sqrt{1+
\frac{\hbar^{2} t^{2} }{2 m^{2}d^{4}} }\,.
\label{eq:mmsd}
\end{equation}

For a more general initial \wf,
\begin{equation}
\label{sgpsi}
\Psi = \psi  \otimes \Phi_0\qquad \psi=
\alpha
\psi^{(+)}\,+\,
\beta\psi^{(-)}\,
\end{equation}
passage through the magnetic field will, by linearity, split the \wf\
into an upward-deflected piece (proportional to $\psi^{(+)}$) and a
downward-deflected piece (proportional to $\psi^{(-)}$), with
corresponding deflections of the trajectories.  The outcome is
registered by detectors placed in the paths of these two possible
``beams.''  Thus of the four kinematically possible outcomes
(``pointer orientations'') the occurrence of no detection and of
simultaneous detection can be ignored as highly unlikely, and the two
relevant outcomes correspond to registration by either the upper or
the lower detector.  Accordingly, for a measurement of $\sigma_z$ the
experiment is equipped with a ``calibration'' (i.e., an assignment of
numerical values to the outcomes of the experiment) $\lambda_{+}=1$
for upper detection and $\lambda_{-}=-1$ for lower detection (while
for a measurement of the $z$-component of the spin angular momentum
itself the calibration is given by $\frac 12\hbar\lambda_{\pm}$).

Note that one can completely understand what's going on in this
Stern-Gerlach experiment without invoking any putative property of the
electron such as its actual $z$-component of spin that is supposed to
be revealed in the experiment.  For a general initial \wf\ there is no
such property. What is more, the transparency of the analysis of this
experiment makes it clear that there is nothing the least bit
remarkable (or for that matter ``nonclassical'') about the {\it
   nonexistence\/} of this property.  But the failure to pay attention
to the role of operators as observables, i.e., to precisely what we
should mean when we speak of measuring operator-observables, helps
create a false impression of quantum peculiarity.

\subsection{A Remark on the Reality of Spin in Bohmian Mechanics}

Bell has said that (for \BM) spin is not real.  Perhaps he should
better have said: {\it ``Even\/} spin is not real,'' not merely
because of all observables, it is spin which is generally regarded as
quantum mechanically most paradigmatic, but also because spin is
treated in orthodox \qt\ very much like position, as a ``degree of
freedom''---a discrete index which supplements the continuous degrees
of freedom corresponding to position---in the \wf.

Be that as it may, his basic meaning is, we believe, this: Unlike
position, spin is not {\it primitive\/}, i.e., no {\it actual\/}
discrete degrees of freedom, analogous to the {\it actual\/} positions
of the particles, are added to the state description in order to deal
with ``particles with spin.''  Roughly speaking, spin is {\it
   merely\/} in the \wf.  At the same time, as explained in Section
\ref{secSGE}, ``spin measurements'' are completely clear, and merely
reflect the way spinor \wf s are incorporated into a description of
the motion of configurations.

In this regard, it might be objected that while spin may not be
primitive, so that the result of our ``spin measurement'' will not
reflect any initial primitive property of the system, nonetheless this
result {\it is\/} determined by the initial configuration of the
system, i.e., by the position of our electron, together with its
initial \wf, and as such---as a function $X_{\sigma_z}(\mathbf{q},
\psi)$ of the state of the system---it is some property of the system
and in particular it is surely real.  We shall address this issue in
Sections \ref{sec:context} and \ref{sec:agcontext}.

\subsection{The Framework of Discrete Experiments}
\label{sec:FDE}
We shall now consider a generic experiment.  Whatever its
significance, the information conveyed by the experiment is registered
in the apparatus as an \emph{output}, represented, say, by the
orientation of a pointer.  Moreover, when we speak of a generic
experiment, we have in mind a fairly definite initial state of the
apparatus, the ready state $\Phi_0=\Phi_{0}(y)$, one for which the
apparatus should function as intended, and in particular one in which
the \emph{pointer} has some ``null'' orientation, as well as a
definite initial state of the system $\psi=\psi(x)$ on which the
experiment is performed.  Under these conditions it turns out
\cite{DGZ92a} that the initial $t=0$ \wf{} $\Psi_0=\Psi_{0}(q)$ of the
composite system formed by system and apparatus, with generic
configuration $q=(x,y)$, has a product form, i.e.,
$$
\Psi_0 = \psi \ot \Phi_0 .$$
Such a product form is an expression
of the \emph{independence} of system and apparatus immediately before
the experiment begins.\footnote{It might be argued that it is somewhat
   unrealistic to assume a sharp preparation of $\psi$, as well as the
   possibility of resetting the apparatus always in the same initial
   state $\Phi_0$.  We shall address this issue in Section 6}

For \BM\ we should expect in general, as a consequence of the \qe\
hypothesis, that the outcome of the experiment---the final pointer
orientation---will be random: Even if the system and apparatus
initially have definite, known \wf s, so that the outcome is
determined by the initial configuration of system and apparatus, this
configuration is random, since the composite system is in \qe, and the
distribution of the final configuration is given by
$|\Psi_{T}(x,y)|^2$, where $\Psi_{T}$ is the \wf\ of the
system-apparatus composite at the time $t=T$ when the experiment ends,
and $x$, respectively $y$, is the generic system, respectively
apparatus, configuration.

Suppose now that $\Psi_{T}$ has the form (\ref{eq:msum}), which
roughly corresponds to assuming that the experiment admits, i.e., that
the apparatus is so designed that there is, only a finite (or
countable) set of possible outcomes, given, say, by the different
possible macroscopically distinct pointer orientations of the
apparatus and corresponding to a partition of the apparatus
configuration space into macroscopically disjoint regions $G_{\a}$,
$\a=1,2, \ldots$.\footnote{Note that to assume there are only
   finitely, or countably, many outcomes is really no assumption at
   all, since the outcome should ultimately be converted to digital
   form, whatever its initial representation may be.} We arrive in this
way at the notion of \emph{discrete experiment}, for which the time
evolution arising {}from the interaction of the system and apparatus
{}from $t=0$ to $t=T$ is given by the unitary map
\begin{equation}
U : \;{\H}\ot\Phi_0 \to \bigoplus_{\a} {\H} \ot\Phia \, , \quad
\psi \ot \Phi_0 \mapsto \Psi_{T} =\suma \psia \ot \Phia
\label{eq:ormfin}
\end{equation}
where $\mathcal{H}$ is the system Hilbert space of square-integrable
wave functions with the usual inner product
$$
\langle\psi,\phi\rangle=\int{{\psi}^*( x)\,\phi( x)\,dx}.
$$
and the $\Phia$ are a \emph{fixed} set of (normalized) apparatus
states supported by the macroscopically distinct regions $G_{\a}$ of
apparatus configurations.

The experiment usually comes equipped with an assignment of numerical
values $\la$ (or a vector of such values) to the various outcomes
$\a$.  This assignment is defined by a ``calibration'' function $F$ on
the apparatus configuration space assuming on each region $G_{\a}$ the
constant value $\la$.  If for simplicity we assume that these values
are in one-to-one correspondence with the outcomes\footnote{We shall
   consider the more general case later on in Subsection
   \ref{sec:StrM}.} then
\begin{equation}
p_{\a} = \int_{F^{-1}(\la)} |\Psi_{T}(x,y)|^{2}
dx\,dy=\int_{G_{\a}} |\Psi_{T}(x,y)|^{2} dx\,dy
\label{eq:page}
\end{equation}
is the probability of finding $\la$, for initial system \wf{} $\psi$.
Since $\Phi_{\a'}(y)=0$ for $y\in G_{\a}$ unless $\a=\a'$, we obtain
\begin{equation}
p_{\a}=\int dx\int_{G_{\a}} |\sum_{\a'} \psi_{\a'} (x)
\Phi_{\a'} (y)|^{2}\,dy = \int | \psia (x) |^2 dx = \|\psia \|^2 .
\label{eq:pr}
\end{equation}
Note that when the result $\la$ is obtained, the effective wave
function of the system undergoes the transformation $\psi \to
\psi_\a.$

A simple example of a discrete experiment is provided by the map
\begin{equation}
U: \psi\ot\Phi_0 \mapsto \suma c_\a
\psi\ot\Phia,
\label{eq:extrva}
\end{equation}
where the $c_{\a}$ are complex numbers such that $\suma
|c_{\a}|^{2}=1$; then $ p_{\a}=|c_{\a}|^{2}$.  Note that the
experiment defined by \eq{eq:extrva} resembles a coin-flip more than a
measurement since the outcome $\a$ occurs with a probability
independent of $\psi$.

\subsection{Reproducibility and its Consequences}
\label{sec:RC}
Though for a generic discrete experiment there is no reason to expect
the sort of ``measurement-like'' behavior typical of familiar quantum
measurements, there are, however, special experiments whose outcomes
are somewhat less random than we might have thought possible.
According to \Sc{} \cite{Sch35}:

\begin{quotation}\setlength{\baselineskip}{12pt}\noindent
   The systematically arranged interaction of two systems (measuring
   object and measuring instrument) is called a measurement on the
   first system, if a directly-sensible variable feature of the second
   (pointer position) is always reproduced within certain error limits
   when the process is immediately repeated (on the same object, which
   in the mean time must not be exposed to additional influences).
\end{quotation}

To implement the notion of ``measurement-like'' experiment considered
by \Sc{}, we first make some preliminary observations concerning the
unitary map (\ref{eq:ormfin}).  Let $P_{[\Phia]}$ be the orthogonal
projection in the Hilbert space $\biga \mathcal{H}\ot\Phia$ onto the
subspace ${\H}\ot\Phia$ and let $\widetilde{\mathcal{H}_{\a}}$ be the
subspaces of $\H$ defined by
\begin{equation}
P_{[\Phia]}\left[ U({\H}\ot\Phi_0) \right]
=\widetilde{\mathcal{H}_{\a}}\ot\Phia\,.
\label{eq:htilde}
\end{equation}
(Since the vectors in $\widetilde{\mathcal{H}}_{\a}$ arise {}from
projecting $\Psi_{T}=\suma \psia \ot \Phia$ onto its $\a$-component,
$\widetilde{\mathcal{H}_{\a}}$ is the space of the ``collapsed''
\wf{}s associated with the occurrence of the outcome $\a$.)  Then
\begin{equation}
U({\H}\ot\Phi_0) \subseteq
\biga\widetilde{\mathcal{H}_{\a}}\ot\Phia.
\label{eq:rep2}
\end{equation}
Note, however, that it need not be the case that
$U({\H}\ot\Phi_0)=\biga\widetilde{\mathcal{H}_{\a}}\ot\Phia$, and that
the spaces $\widetilde{\mathcal{H}_{\a}}$ need be neither orthogonal
nor distinct; e.g., for (\ref{eq:extrva})
$\widetilde{\mathcal{H}_{\a}}=\H$ and $U({\H}\ot\Phi_0)=\H\ot\sum_\a
c_\a\Phia\neq\biga {\H}\ot\Phia$.\footnote{Note that if \H\ has finite
   dimension $n$, and the number of outcomes $\a$ is $m$, $\mbox{dim
   }[U({\H}\ot\Phi_0)]= n$, while $\mbox{dim }[\biga {\H}\ot\Phia] =
   n\cdot m$.}

A ``measurement-like'' experiment is one which is reproducible in the
sense that it will yield the same outcome as originally obtained if it
is immediately repeated.  (This means in particular that the apparatus
must be immediately reset to its ready state, or a fresh apparatus
must be employed, while the system is not tampered with so that its
initial state for the repeated experiment is its final state produced
by the first experiment.)  Thus the experiment is \emph{reproducible}
if
\begin{equation}
U(\widetilde{\mathcal{H}_{\a}}\ot\Phi_0) \subseteq
\widetilde{\mathcal{H}_{\a}}\ot\Phia \label{eq:repconold}
\end{equation}
or, equivalently, if there are spaces ${\Ha}'\subseteq
\widetilde{\mathcal{H}_{\a}}$ such that
\begin{equation}
U(\widetilde{\mathcal{H}_{\a}}\ot\Phi_0) = {\Ha}'\ot\Phia
\label{eq:repcon}\,.
\end{equation}

Note that it follows {}from the unitarity of $U$ and the orthogonality
of the subspaces $\widetilde{\Ha}\ot\Phia$ that the subspaces
$\widetilde{\mathcal{H}_{\a}}\ot\Phi_0$ and hence the
$\widetilde{\mathcal{H}_{\a}}$ are also orthogonal.  Therefore, by
taking the orthogonal sum over $\a$ of both sides of
(\ref{eq:repcon}), we obtain
\begin{equation}
\biga U(\widetilde{\mathcal{H}_{\a}}\ot\Phi_0)=  U \left( \biga
\widetilde{\mathcal{H}_{\a}}\ot\Phi_0\right) = \biga {\Ha}'\ot\Phia.
\label{eq:orto}
\end{equation}
If we now make the simplifying assumption that the subspaces
$\widetilde{\mathcal{H}_{\a}}$ are finite dimensional, we have {}from
unitarity that $ \widetilde{\mathcal{H}_{\a}}= {\Ha}'$, and thus, by
comparing \eq{eq:rep2} and (\ref{eq:orto}), that equality holds in
\eq{eq:rep2} and that
\begin{equation}
{\H}=\biga {\Ha}
\label{eq:sum}
\end{equation}
with
\begin{equation}
U({\H_\a} \ot \Phi_0 )=  {\Ha} \ot \Phia
\label{eq:rep4}
\end{equation}
for $$\Ha \equiv \widetilde{\mathcal{H}_{\a}}={\Ha}' \, .$$

Therefore if the \wf\ of the system is initially in $\H_\a$, outcome
$\a$ definitely occurs and the value $\lambda_\a$ is thus definitely
obtained (assuming again for simplicity one-to-one correspondence
between outcomes and results).  It then follows that for a general
initial system \wf
\begin{displaymath}
\psi =\suma \Pa \psi ,
\end{displaymath}
where $\Pa$ is the projection in $\H$ onto the subspace $\Ha$, that
the outcome $\a$, with result $\la$, is obtained with (the usual)
probability
\begin{equation}
p_\a = \| \Pa\psi\|^2= \langle\psi,\Pa \psi \rangle,
\label{eq:prr}
\end{equation}
which follows {}from (\ref{eq:rep4}), \eq{eq:pr}, and \eq{eq:ormfin}
since $U\big(\Pa\psi\ot\Phi_0\big) = \psia \ot\Phia$ and hence $ \|
\Pa\psi\| = \| \psia \|$ by unitarity.  In particular, when the $\la$
are real-valued, the expected value obtained is
\begin{equation}
\suma{p_\a \la}=\suma{\la{ \| \Pa \psi\|}^2} = \langle
\psi, A\psi \rangle \label{eq:meanz}
\end{equation}
where
\begin{equation}
A=\suma{\la \Pa}
\label{eq:A}
\end{equation}
is the \sa{} operator with eigenvalues $\la$ and spectral projections
$\Pa$.

\subsection{Operators as Observables}\label{subsec.oao}

What we wish to emphasize here is that, insofar as the statistics for
the values which result {}from the experiment are concerned,
\begin{equation}
\mbox{%
\begin{minipage}{0.85\textwidth}\openup 1.4\jot
   \setlength{\baselineskip}{12pt}\emph{the relevant data for the
     experiment are the collection $\{ \Ha\}$ of special orthogonal
     subspaces, together with the corresponding calibration $\{\la \},$
   }
\end{minipage}}
\label{def:exptoa}
\end{equation}
and \emph{this data is compactly expressed and represented by the
   self-adjoint operator $A$, on the system Hilbert space $\H$, given
   by \eq{eq:A}.} Thus, under the assumptions we have made, with a
reproducible experiment $\E$ we naturally associate an operator
$A=A_{\E}$, a single mathematical object, defined on the system alone,
in terms of which an efficient description \eq{eq:prr} of the
statistics of the possible results is achieved; we shall denote this
association by
\begin{equation}
\E\mapsto A\,.
\label{eq:fretoe}
\end{equation}
If we wish we may speak of ``operators as observables,'' and when an
experiment \E{} is associated with a \sa{} operator $A$, as described
above, we may say that \emph{the experiment \E{} is a ``measurement''
   of the observable represented by the \sa{} operator $A$.} If we do
so, however, it is important that we appreciate that in so speaking we
merely refer to what we have just derived: the role of operators in
the description of certain experiments.\footnote{Operators as
   observables also naturally convey information about the system's
   \wf\ after the experiment.  For example, for an ideal measurement,
   when the outcome is $\a$ the \wf\ of the system after the experiment
   is (proportional to) $P_{\H_\a}\psi$.  We shall elaborate upon this
   in the next section.}

So understood, the notion of operator-as-observable in no way implies
that anything is genuinely measured in the experiment, and certainly
not the operator itself!  In a general experiment no system property
is being measured, even if the experiment happens to be
measurement-like.  (Position measurements in \BM{} are of course an
important exception.)  What in general is going on in obtaining
outcome $\alpha$ is completely straightforward and in no way suggests,
or assigns any substantive meaning to, statements to the effect that,
prior to the experiment, observable $A$ somehow had a value
$\lambda_\a$---whether this be in some determinate sense or in the
sense of Heisenberg's ``potentiality'' or some other ill-defined fuzzy
sense---which is revealed, or crystallized, by the experiment.  Even
speaking of the observable $A$ as having value $\lambda_\a$ when the
system's \wf\ is in $\H_\a$, i.e., when this \wf\ is an eigenstate of
$A$ of eigenvalue $\lambda_\a$---insofar as it suggests that something
peculiarly quantum is going on when the \wf\ is not an eigenstate
whereas in fact there is nothing the least bit peculiar about the
situation---perhaps does more harm than good.

It might be objected that we are claiming to arrive at the \qf\ under
somewhat unrealistic assumptions, such as, for example,
reproducibility or finite dimensionality.  We agree.  But this
objection misses the point of the exercise.  The \qf\ itself is an
idealization; when applicable at all, it is only as an approximation.
Beyond illuminating the role of operators as ingredients in this
formalism, our point was to indicate how naturally it emerges.  In
this regard we must emphasize that the following question arises for
quantum orthodoxy, but does not arise for \BM: For precisely which
theory is the \qf\ an idealization?  \bigskip

We shall discuss how to go beyond the idealization involved in the
quantum formalism in Section 4---after having analyzed it thoroughly
in Section 3.  First we wish to show that many more experiments than
those satisfying our assumptions can indeed be associated with
operators in exactly the manner we have described.

\subsection{The General Framework of Bohmian Experiments}
\label{sec:E}\label{sec:GFE}

According to (\ref{eq:page}) the statistics of the results of a
discrete experiment are governed by the probability measure
$\rho_{\Psi_T}\circ F^{-1}$, where $\rho_{\Psi_T}(dq)
=|\Psi_{T}(q)|^{2}dq$ is the quantum equilibrium measure.  Note that
discreteness of the value space of $F$ plays no role in the
characterization of this measure.  This suggests that we may consider
a more general notion of experiment, not based on the assumption of a
countable set of outcomes, but only on the \emph{unitarity} of the
operator $U$, which transforms the initial state $\psi\otimes\Phi_{0}$
into the final state $\Psi_{T}$, and on a generic \emph{calibration
   function} $F$ {}from the configuration space of the composite system
to some value space, e.g., $\R$, or ${\R}^m$, giving the result of the
experiment as a function $ F(Q_T)$ of the final configuration $Q_T$ of
system and apparatus.  We arrive in this way at the notion of
\emph{general experiment}
\begin{equation}
\E{}\equiv\{\Phi_{0}, U, F\},
\label{eq:generalexperiment}
\end{equation}
where the unitary $U$ embodies the interaction of system and apparatus
and the function $F$ could be completely general.  Of course, for
application to the results of real-world experiments $F$ might
represent the ``orientation of the apparatus pointer'' or some
coarse-graining thereof.

Performing \E{} on a system with initial \wf{} $\psi$ leads to the
result ${Z}= F(Q_T)$ and since $Q_{T}$ is randomly distributed
according to the quantum equilibrium measure $\rho_{\Psi_T}$, the
probability distribution of $Z$ is given by the induced measure
\begin{equation}
\rho^{ Z}_{\psi}= \rho_{\Psi_T}\circ
F^{-1}\,.
\label{eq:indumas}
\end{equation}
(We have made explicit only the dependence of the measure on $\psi$,
since the initial apparatus state $\Phi_{0}$ is of course fixed,
defined by the experiment \E{}.)  Note that this more general notion
of experiment eliminates the slight vagueness arising {}from the
imprecise notion of macroscopic upon which the notion of discrete
experiment is based.  Note also that the structure
\eq{eq:generalexperiment} conveys information about the wave function
\eq{eq:con} of the system after a certain result $F(Q_T)$ is obtained.

Note, however, that this somewhat formal notion of experiment may not
contain enough information to determine the detailed Bohmian dynamics,
which would require specification of the Hamiltonian of the
system-apparatus composite, that might not be captured by $U$. In
particular, the final configuration $Q_T$ may not be determined, for
given initial \wf{}, as a function of the initial configuration of
system and apparatus. \E{} does, however, determine what is relevant
for our purposes about the random variable $Q_T$, namely its
distribution, and hence that of $Z=F(Q_T)$.

Let us now focus on the right had side of the equation (\ref{eq:prr}),
which establishes the association of operators with experiments:
$\langle\psi,\Pa \psi \rangle$ is the probability that ``the operator
$A $ has value $\la$'', and according to standard quantum mechanics
the statistics of the results of measuring a general \sa{} operator
$A$, not necessarily with pure point spectrum, in the (normalized)
state $\psi$ are described by the probability measure
\begin{equation}
   \Delta\mapsto\mu^{A}_\psi(\Delta) \equiv \langle \psi,
P^{A }(\Delta) \psi \rangle
\label{eq:spectrmeas}
\end{equation}
where $\Delta$ is a (Borel) set of real numbers and $P^A:
\Delta\mapsto P^{A }(\Delta)$ is the \emph{projection-valued-measure}
(PVM) uniquely associated with $A$ by the spectral theorem.  (We
recall \cite{RS80} that a PVM is a normalized, countably additive set
function whose values are, instead of nonnegative reals, orthogonal
projections on a Hilbert space \H{}.  Any PVM $P$ on \H\ determines,
for any given $\psi\in \H$, a probability measure
$\mu_\psi\equiv\mu_\psi^P : \Delta \mapsto \langle\psi ,
P(\Delta)\psi\rangle$ on $\R$.  Integration against \PV\ is analogous
to integration against ordinary measures, so that $B\equiv \int
f(\lambda) P(d\lambda) $ is well-defined, as an operator on $\H$.
Moreover, by the spectral theorem every \sa\ operator $A$ is of the
form $ A= \int \lambda\, P(d\lambda)$, for a unique \PV{} $ P =P^{A}$,
and $\int f(\lambda) P(d\lambda)= f(A)$. )

It is then rather clear how (\ref{eq:fretoe}) extends to general \sa{}
operators: \emph{a general experiment \E{} is a measurement of the
   \sa{} operator $A$ if the statistics of the results of \E{} are
   given by (\ref{eq:spectrmeas})}, i.e.,
\begin{equation}
\E\mapsto A \qquad
\mbox{if and only if}\qquad \rho^{ Z}_{\psi} =\mu^A_\psi \,.
\label{eq:prdeltan}
\end{equation}
In particular, if $\E\mapsto A $, then the moments of the result of
$\E$ are the moments of $A$:
$$
<Z^n>= \int \lambda^n \langle\psi ,P(d\lambda)\psi\rangle=
\langle\psi ,A^n\psi\rangle.  $$

   \section{The Quantum Formalism} \setcounter{equation}{0} The spirit of
   this section will be rather different {}from that of the previous
   one.  Here the focus will be on the formal structure of experiments
   measuring self-adjoint operators.  Our aim is to show that the
   standard quantum formalism emerges {}from a \emph{formal} analysis
   of the association $\E\mapsto A$ between operator and experiment
   provided by (\ref{eq:prdeltan}).  By ``formal analysis'' we mean not
   only that the detailed physical conditions under which might
   $\E\mapsto A$ hold (e.g., reproducibility) will play no role, but
   also that the practical requirement that \E{} be physically
   realizable will be of no relevance whatsoever.

   Note that such a formal approach is unavoidable in order to recover
   the quantum formalism.  In fact, within the quantum formalism one
   may consider measurements of arbitrary \sa{} operators, for example,
   the operator $A= \hat{X}^2\hat{P} + \hat{P}X^{2}$, where $\hat{X}$
   and $\hat{P}$ are respectively the position and the momentum
   operators.  However, it may very well be the case that no ``real
   world'' experiment measuring $A$ exists.  Thus, in order to allow
   for measurements of arbitrary self-adjoint operators we shall regard
   (\ref{eq:generalexperiment}) as characterizing an ``\emph{abstract
     experiment}''; in particular, we shall not regard the unitary map
   $U$ as arising necessarily {}from a (realizable) \Sc{} time
   evolution.  We may also speak of virtual experiments.

   In this regard one should observe that to resort to a formal
   analysis is indeed quite common in physics.  Consider, e.g., the
   Hamiltonian formulation of classical mechanics that arose {}from an
   abstraction of the physical description of the world provided by
   Newtonian mechanics.  Here we may freely speak of completely general
   Hamiltonians, e.g.  $H(p,q)= p^{6}$, without being concerned about
   whether they are physical or not.  Indeed, only very few
   Hamiltonians correspond to physically realizable motions!  \medskip

   A warning: As we have stressed in the introduction and in Section
   \ref{subsec.oao}, when we speak here of a measurement we don't
   usually mean a {\em genuine} measurement---an experiment revealing
   the pre-existing value of a quantity of interest, the measured
   quantity or property. (We speak in this unfortunate way because it
   is standard.)  Genuine measurement will be discussed much later, in
   Section \ref{secMO}.

\subsection{Weak Formal Measurements}
\label{sec:MO}
The first formal notion we shall consider is that of weak formal
measurement, formalizing the relevant data of an experiment measuring
a self-adjoint operator:
\begin{equation}
\mbox{%
\begin{minipage}{0.85\textwidth}\openup 1.4\jot
   \setlength{\baselineskip}{12pt}\emph{Any orthogonal decomposition
     ${\H}=\biga {\Ha}$, i.e., any complete collection $\{ \Ha\}$ of
     mutually orthogonal subspaces, paired with any set $\{\la \}$ of
     distinct real numbers, defines the weak formal measurement
     $\M\equiv\{(\Ha, \la )\}\equiv\{\Ha, \la \}$.}
   \end{minipage}}
\label{def:wfm}
\end{equation}
(Compare (\ref{def:wfm}) with (\ref{def:exptoa}) and note that now we
are not assuming that the spaces $\Ha$ are finite-dimensional.)  The
notion of weak formal measurement is aimed at expressing the minimal
structure that all experiments (some or all of which might be virtual)
measuring the same operator $A= \sum\la\Pa$ have in common ($\Pa$ is
the orthogonal projection onto the subspace $\Ha$).  Then, ``to
perform \M'' shall mean to perform (at least virtually) any one of
these experiments, i.e., any experiment such that
\begin{equation}
p_{\a}=\langle \psi, \Pa\psi \rangle
\label{eq:prdeltass}
\end{equation}
is the probability of obtaining the result $\la$ on a system initially
in the state $\psi$.  (This is of course equivalent to requiring that
the result $\la$ is definitely obtained if and only if the initial
wave function $\psi\in \Ha$.)

Given $\M\equiv\{\Ha, \la \}$ consider the set function
\begin{equation}
P:\Delta\mapsto P (\Delta)\equiv \sum_{\la\in \Delta} \Pa,
\label{eq:disfr}
\end{equation}
where $\Delta$ is a set of real numbers (technically, a Borel set).
Then
\begin{itemize}
\item[1)] $P$ is \emph{normalized}, i.e., $P(\R)= I$, where $I$ is the
   identity operator and $\R$ is the real line,

\item[2)] $P(\Delta)$ is an \emph{orthogonal projection}, i.e.,
   $P(\Delta)^{2}=P(\Delta)=P(\Delta)^{*}$,

\item[3)] $P$ is \emph{countably additive}, i.e., $ P(\bigcup_{n}
   \Delta_n) = \sum_{n} P(\Delta_n)$, for $\Delta_n$ disjoint sets.
\end{itemize}
Thus $P$ is a \PV{} and therefore the notion of weak formal
measurement is indeed equivalent to that of ``discrete'' PVM, that is,
a PVM supported by a countable set $\{\la\}$ of values.

More general PVMs, e.g. PVMs supported by a continuous set of values,
will arise if we extend (\ref{def:wfm}) and base the notion of weak
formal measurement upon the general association (\ref{eq:prdeltan})
between experiments and operators. If we stipulate that
\begin{equation}
\mbox{%
\begin{minipage}{0.90\textwidth}\openup 1.4\jot
   \setlength{\baselineskip}{12pt}\emph{any \PV{} $P$ on $\H$ defines a
     weak formal measurement $\M\equiv P$,}
   \end{minipage}}
\label{def:wfmg}
\end{equation}
then ``to perform $\M$'' shall mean to perform any experiment $\E$
associated with $A=\int \lambda P(d\lambda)$ in the sense of
(\ref{eq:prdeltan}).

Note that since by the spectral theorem there is a natural one-to-one
correspondence between PVMs and \sa{} operators, we may speak
equivalently of \emph{the} operator $A=A_{\mathcal{M}}$, for given
$\M$, or of \emph{the} weak formal $\M=\M_A$, for given $A$.  In
particular, the weak formal measurement $\M_{A}$ represents the
equivalence class of \emph{all} experiments $\E{}\to A$.

\subsection{Strong Formal Measurements}

We wish now to classify the different experiments \E{} associated with
the same \sa{} operator $A$ by taking into account the effect of \E{}
on the state of the system, i.e., the state transformations $\psi \to
\psia$ induced by the occurrence of the various results $\la$ of \E{}.
Accordingly, unless otherwise stated, {}from now on we shall assume
$\E$ to be a discrete experiment measuring $A=\sum\la\Pa$, for which
the state transformation $\psi \to \psia$ is defined by
\eq{eq:ormfin}.  This leads to the notion of strong formal
measurements. For the most important types of strong formal
measurements, ideal, normal and standard, there is a one-to-one
correspondence between $\a$'s and numerical results $\la$.

\subsubsection{Ideal Measurements} \label{sec:IM}
Given a weak formal measurement of $A$, the simplest possibility for
the transition $\psi\to\psia$ is that when the result $\la$ is
obtained, the initial state $\psi$ is projected onto the corresponding
space $\Ha$, i.e., that
\begin{equation}
\psi \to \psia = \Pa \psi.
\label{eq:ideal}
\end{equation}
This prescription defines uniquely the \emph{ideal measurement} of
$A$.  (The transformation $\psi\to\psia$ should be regarded as defined
only in the projective sense: $\psi \to \psi_\a$ and $\psi \to
c\psi_\a$ ($c\neq 0$) should be regarded as the same transition.)
``To perform an ideal measurement of $A$'' shall then mean to perform
a discrete experiment \E{} whose results are statistically distributed
according to (\ref{eq:prdeltass}) and whose state transformations
\eq{eq:ormfin} are given by (\ref{eq:ideal}).

Under an ideal measurement the \wf{} changes as little as possible: an
initial $\psi \in \Ha$ is unchanged by the measurement.  Ideal
measurements have always played a privileged role in quantum
mechanics. It is the ideal measurements that are most frequently
discussed in textbooks. It is for ideal measurements that the standard
collapse rule is obeyed.  When Dirac \cite{Dir30} wrote: ``a
measurement always causes the system to jump into an eigenstate of the
dynamical variable that is being measured" he was referring to an
ideal measurement.

\subsubsection{Normal Measurements}
\label{sec:NM}
The rigid structure of ideal measurements can be weakened by requiring
only that $\Ha$ as a whole, and not the individual vectors in $\Ha$,
is unchanged by the measurement and therefore that the state
transformations induced by the measurement are such that when the
result $\la$ is obtained the transition
\begin{equation}
\psi \to\psia = U_\a \Pa \psi
\label{eq:norm}
\end{equation}
occurs, where the $U_\a$ are operators on $\Ha$ ( $U_\a :\Ha\to\Ha$).
Then for any such discrete experiment \E{} measuring $A$, the $U_\a$
can be chosen so that \eq{eq:norm} agrees with \eq{eq:ormfin}, i.e.,
so that for $\psi \in \Ha$, $U(\psi\otimes\Phi_0) =
U_\a\psi\otimes\Phi_\a$, and hence so that $U_\a$ is unitary (or at
least a partial isometry). Such a measurement, with unitaries $U_\a
:\Ha\to\Ha$, will be called a \emph{normal measurement} of $A$.

In contrast with an ideal measurement, a normal measurement of an
operator is not uniquely determined by the operator itself: additional
information is needed to determine the transitions, and this is
provided by the family $\{U_{\a}\}$.  Different families define
different normal measurements of the same operator.  Note that ideal
measurements are, of course, normal (with $U_{\a}= I_{\a} \equiv$
identity on $\Ha$), and that normal measurements with one-dimensional
subspaces $\Ha$ are necessarily ideal.

Since the transformations (\ref{eq:norm}) leave invariant the
subspaces $\Ha$, the notion of normal measurement characterizes
completely the class of reproducible measurements of \sa{} operators.
Following the terminology introduced by Pauli \cite{Pau58}, normal
measurement are sometimes called {\it measurements of first kind\/}.
Normal measurements are also \emph{quantum non demolition (QND)
   measurements\/} \cite{Brag}, defined as measurements such that the
operators describing the induced state transformations, i.e, the
operators $\Aa\equiv U_{\a}\Pa$, commute with the measured operator
$A=\sum\la\Pa$. (This condition is regarded as expressing that the
measurement leaves the measured observable $A$ unperturbed).

\subsubsection{Standard Measurements}
\label{sec:SM}
We may now drop the condition that the $\Ha$ are left invariant by the
measurement and consider the very general state transformations
\begin{equation}
\psi \to \psia=T_\a \Pa \psi
\label{eq:stsm}
\end{equation}
with operators $T_\a : \Ha\to\H$. Then, exactly as for the case of
normal measurements, it follows that $T_\a$ can be chosen to be
unitary {}from $\Ha$ onto its range $\widetilde{\Ha}$.  The subspaces
$\widetilde{\Ha}$ need be neither orthogonal nor distinct. We shall
write $R_\a=T_\a \Pa$ for the general transition operators. With
$T_\a$ as chosen, $R_\a$ is characterized by the equation
$\Aa^{\ast}\Aa = \Pa$ (where $\Aa^{\ast}$ denotes the adjoint of
$\Aa$).

The state transformations (\ref{eq:stsm}), given by unitaries $T_\a:
\Ha\to\widetilde{\Ha}$, or equivalently by bounded operators $R_\a$ on
$\H$ satisfying $\Aa^{\ast}\Aa = \Pa$, define what we shall call a
\emph{standard measurement} of $A$.  Note that normal measurements are
standard measurements with $\widetilde{\Ha}=\Ha$ (or $
\widetilde{\Ha}\subset \Ha$).  Although standard measurements are in a
sense more realistic than normal measurements (real world measurements
are seldom reproducible in a strict sense), they are very rarely
discussed in textbooks. We emphasize that the crucial data in a
standard measurement is given by $R_\a$, which governs both the state
transformations ($\psi\to R_a\psi$) and the probabilities ($p_\a =
\langle\psi, \Pa\psi\rangle= \norm R_\a\psi\norm^2$).

We shall illustrate the main features of standard measurements by
considering a very simple example: Let $\{e_0, e_{1}, e_{2}, \ldots
\}$, be a fixed orthonormal basis of \H{} and consider the standard
measurement whose results are the numbers $0,1,2,\ldots $ and whose
state transformations are defined by the operators
\begin{displaymath}
\Aa\equiv |e_0\rangle\langle e_\a| \qquad \mbox{i.e.,}\qquad
R_{\a} \psi = \langle e_\a, \psi \rangle
e_{0},\qquad\a=0,1,2,\ldots
\end{displaymath}
With such $\Aa$'s are associated the projections
$P_{\a}=\Aa^{\ast}\Aa=|e_\a\rangle\langle e_\a|\,$, i.e., the
projections onto the one dimensional spaces $\Ha$ spanned respectively
by the vectors $e_{\a}$.  Thus, this is a measurement of the operator
$ A = \sum_{\a} \a |e_\a\rangle\langle e_\a| $.  Note that the spaces
$\widetilde{\Ha}$, i.e.  the ranges of the $\Aa$'s, are all the same
and equal to the space $\H_{0}$ generated by the vector $e_0$. The
measurement is then not normal since $\Ha\neq \widetilde{\Ha}$.
Finally, note that this measurement could be regarded as giving a
simple model for a photo detection experiment, where any state is
projected onto the ``vacuum state'' $e_0$ after the detection.

\subsubsection{Strong Formal Measurements}
\label{sec:StrM}

We shall now relax the condition that $\a\mapsto \la$ is one-to-one,
as we would have to do for an experiment having a general calibration
$\a\mapsto\la$, which need not be invertible. This leads to (what we
shall call) a \emph{strong formal measurement}.  Since this notion
provides the most general formalization of the notion of a
``measurement of a \sa{} operator'' that takes into account the effect
of the measurement on the state of the system, we shall spell it out
precisely as follows:
\begin{equation}
\mbox{%
   \begin{minipage}{0.85\textwidth}\openup 1.4\jot
     \setlength{\baselineskip}{12pt}\emph{Any complete (labelled)
       collection $\{ \Ha\}$ of mutually orthogonal subspaces, any
       (labelled) set $\{\la \}$ of not necessarily distinct real
       numbers, and any (labelled) collection $\{\Aa\}$ of bounded
       operators on $\H$, such that $\Aa^{\ast}\Aa\equiv\Pa$ (the
       projection onto $\Ha$), defines a strong formal measurement.}
   \end{minipage}}
\label{def:sfm}
\end{equation}

A strong formal measurement will be compactly denoted by $\M\equiv
\{(\Ha, \la, \Aa) \}\equiv\{\Ha, \la, \Aa \}$, or even more compactly
by $\M\equiv \{\la, \Aa \}$ (the spaces $\Ha$ can be extracted {}from
the projections $\Pa= \Aa^{\ast}\Aa$).  With \M{} is associated the
operator $A=\sum\la\Pa$.  Note that since the $\la$ are not
necessarily distinct numbers, $\Pa$ need not be the spectral
projection $P^A (\la)$ associated with $\la$; in general $$P^A
(\lambda) = \sum_{\a: \la =\lambda}\Pa,$$
i.e., it is the sum of all
the $\Pa$'s that are associated with the value $\lambda$.\footnote{ It
   is for this reason that it would be pointless and inappropriate to
   similarly generalize weak measurements. It is only when the state
   transformation is taken into account that the distinction between
   the outcome $\a$ (which determines the transformation) and the
   result $\la$ (whose probability the formal measurement is to supply)
   becomes relevant.}  ``\emph{To perform the measurement $\M$}'' on a
system initially in $\psi$ shall accordingly mean to perform a
discrete experiment \E{} such that: 1) the probability $p(\lambda)$ of
getting the result $\lambda$ is governed by $A$, i.e., $ p(\lambda) =
\langle \psi, P^A (\lambda) \psi \rangle$, and 2) the state
transformations of \E{} are those prescribed by \M{}, i.e., $ \psi \to
\psia= \Aa\psi$.

Observe that strong formal measurements do provide a more realistic
formalization of the notion of measurement of an operator than
standard measurements: the notion of discrete experiment does not
imply a one-to-one correspondence between outcomes, i.e, final
macroscopic configurations of the pointer, and the numerical results
of the experiment.

The relationship between (weak or strong) formal measurements, \sa{}
operators, and experiments can be summarized by the following sequence
of maps:
\begin{equation}
\E \mapsto \M \mapsto A
\label{eq:etomtoa}
\end{equation}
The first map expresses that $\M$ (weak or strong) is a formalization
of \E{}---it contains the ``relevant data'' about \E{}---and it will
be many-to-one if \M{} is a weak formal measurement\footnote{There is
   an obvious natural unitary equivalence between the preimages \E{} of
   a strong formal measurement \M{}.}; the second map expresses that
\M{} is a formal measurement of $A$ and it will be many-to-one if \M{}
is (required to be) strong and one-to-one if \M{} is weak.  \emph{Note
   that $\E\mapsto A$ is always many-to-one}.

\subsection{From Formal Measurements to Experiments}\label{subsec.exp}

Given a strong measurement $\M\equiv \{\Ha, \la, \Aa \}$ one may
easily construct a map \eq{eq:ormfin} defining a discrete experiment
$\E{}=\E_{\mathcal{M}}$ associated with $\M$:
\begin{equation}
U: \;\psi \ot \Phi_0  \mapsto
\sum_{\a} (\Aa\psi) \ot \Phia
\label{standu}
\end{equation}
The unitarity of $U$ ( {}from $ \H\ot\Phi_0 $ onto the range of $U$)
follows then immediately {}from the orthonormality of the $\{\Phia\}$
since
\begin{equation}
\sum_{\a} \|\Aa\psi\|^{2} = \sum_{\a} \langle \psi, \Aa^{\ast}\Aa
\psi \rangle
= \langle \psi, \sum_{\a} \Pa \psi \rangle = \langle
\psi,\psi\rangle = \|\psi\|^2
\label{eq:unide}
\end{equation}
This experiment is abstractly characterized by: 1) the finite or
countable set $I$ of outcomes $\a$, 2) the apparatus ready state
$\Phi_{0}$ and the set $\{\Phia\}$ of normalized apparatus states, 3)
the unitary map $U : \;{\H}\ot\Phi_0 \to \biga {\H} \ot\Phia$ given by
(\ref{standu}), 4) the calibration $\a \mapsto \la$ assigning
numerical values (or a vector of such values) to the various outcomes
$\a$.  Note that $U$ need not arise {}from a \Sc{} Hamiltonian
governing the interaction between system and apparatus. Thus \E{}
should properly be regarded as an ``abstract'' experiment as we have
already pointed out in the introduction to this section.

\subsection{Von Neumann Measurements}
\label{sec:vNM}

We shall now briefly comment on the relation between our approach,
based on formal measurements, and the widely used formulation of
quantum measurement in terms of von Neumann measurements \cite{vNe55}.

A {\it von Neumann measurement\/} of $A=\sum \la \Pa $ on a system
initially in the state $\psi$ can be described as follows (while the
nondegeneracy of the eigenvalues of $A$---i.e., that
$\mbox{dim}(\Ha)=1$---is usually assumed, we shall not do so): Assume
that the (relevant) configuration space of the apparatus, whose
generic configuration shall be denoted by $y$, is one-dimensional, so
that its Hilbert space $\H_{\mathcal{A}}\simeq L^{2}(\R)$, and that
the interaction between system and apparatus is governed by the
Hamiltonian
\begin{equation}
H= H_{\text{vN}}= \gamma A\otimes \hat{P_{y}}
\label{vontrans}
\end{equation}
where $\hat{P_{y}}\equiv i\hbar\partial/\partial y$ is (minus) the
momentum operator of the apparatus.  Let $\Phi_0 = \Phi_0 (y) $ be the
ready state of the apparatus.  Then for $\psi=\Pa\psi$ one easily sees
that the unitary operator $U\equiv e^{-i TH/\hbar}$ transforms the
initial state $\psi_\a \ot \Phi_0$ into $ \psi _\a \ot \Phia$ where
$\Phia = \Phi_{0}(y - \la\gamma T)$, so that the action of $U$ on
general $\psi =\sum \Pa \psi$ is
\begin{equation}
U: \;\psi \ot \Phi_0  \to
\sum_{\a} (\Pa\psi) \ot \Phia
\label{eq:vNm}
\end{equation}
If $\Phi_{0}$ has sufficiently narrow support, say around $y=0$, the
$\Phia$ will have disjoint support around the ``pointer positions''
$y_\a = \la\gamma T$, and thus will be orthogonal, so that, with
calibration $F(y)= y /\gamma T$ (more precisely, $F(y)= y_\a /\gamma
T$ for $y$ in the support of $\Phi_\a$), the resulting von Neumann
measurement becomes a discrete experiment measuring $A$; comparing
(\ref{eq:vNm}) and (\ref{eq:ideal}) we see that it is an ideal
measurement of $A$.\footnote{It is usually required that von Neumann
   measurements be impulsive ($\gamma$ large, $T$ small) so that only
   the interaction term \eq{vontrans} contributes significantly to the
   total Hamiltonian over the course of the measurement.}

Thus, the framework of von Neumann measurements is less general than
that of discrete experiments, or equivalently of strong formal
measurements; at the same time, since the Hamiltonian $H_{\text{vN}}$
is not of Schr\"odinger type, von Neumann measurements are just as
formal.  (We note that more general von Neumann measurements of $A$
can be obtained by replacing $H_{\text{vN}}$ with more general
Hamiltonians; for example, $H_{\text{vN}}'= H_{0} + H_{\text{vN}}$,
where $H_0$ is a self-adjoint operator on the system Hilbert space
which commutes with $A$, gives rise to a \emph{normal measurement} of
$A$, with $\Aa = e^{-iT H_0/\hbar}\Pa$.Thus by proper extension of the
von Neumann measurements one may arrive at a framework of measurements
completely equivalent to that of strong formal measurements.)

\subsection{Preparation Procedures}
\label{sec:PP}
Before discussing further extensions of the association between
experiments and operators, we shall comment on an implicit assumption
apparently required for the measurement analysis to be relevant: that
the system upon which measurements are to be performed can be prepared
in any prescribed state $\psi$.

Firstly, we observe that the system can be prepared in a prescribed
state $\psi$ by means of an appropriate standard measurement \M{}
performed on the system when it is initially in an unknown state
$\psi'$.  We have to choose $\M\equiv \{\Ha, \la, \Aa \}$ in such a
way that $R_{\a_{0}}\psi'=\psi$, for some $\a_{0}$ and all $\psi'$,
i.e., that Ran($R_{\a_{0}}$) = span($\psi$); then {}from reading the
result $\lambda_{\a_0}$ we may infer that the system has collapsed to
the state $\psi$.  The simplest possibility is for $\M$ to be an ideal
measurement with at least a one-dimensional subspace $\H_{\a_{0}}$
that is spanned by $\psi$.  Another possibility is to perform a
(nonideal) standard measurement like that of the example at the end of
Section \ref{sec:SM}, which can be regarded as defining a preparation
procedure for the state $e_{0}$.

Secondly, we wish to emphasize that the existence of preparation
procedures is not as crucial for relevance as it may seem.  If we had
only statistical knowledge about the initial state $\psi$, nothing
would change in our analysis of Bohmian experiments of Section 2, and
in our conclusions concerning the emergence of \sa{} operators, except
that the uncertainty about the final configuration of the pointer
would originate {}from both quantum equilibrium and randomness in
$\psi$. We shall elaborate upon this later when we discuss Bohmian
experiments for initial states described by a density matrix.

\subsection{Measurements of Commuting Families of Operators}
\label{secMCFO}

As hinted in Section \ref{sec:FDE}, the result of an experiment \E{}
might be more complex than we have suggested until now in Section 3:
it might be given by the vector $\lambda_{\a}\equiv(
\la^{(1)},\ldots,\la^{(m)})$ corresponding to the orientations of $m$
pointers. For example, the apparatus itself may be a composite of $m$
devices with the possible results $\la^{(i)}$ corresponding to the
final state of the $i$-th device.  Nothing much will change in our
discussion of measurements if we now replace the numbers $\la$ with
the vectors ${\lambda}_{\a}\equiv( \la^{(1)},\ldots,\la^{(m)})$, since
the dimension of the value space was not very relevant. However \E{}
will now be associated, not with a single self-adjoint operator, but
with a commuting family of such operators. In other words, we arrive
at the notion of an experiment \E{} that is a \emph{measurement of a
   commuting family} of \sa{} operators,\footnote{We recall some basic
   facts about commuting families of \sa{} operators
   \cite{vNe55,RN55,Pru71}.  The \sa{} operators $\avec{A}{m}$ form a
   commuting family if they are bounded and pairwise commute, or, more
   generally, if this is so for their spectral projections, i.e., if
   $[P^{A_{i}} (\Delta), P^{A_{j}} (\Gamma)] =0$ for all $i,j =
   1,\ldots,m$ and (Borel) sets $\Delta, \Gamma \subset \R $.  A
   commuting family $A\equiv(\avec{A}{m})$ of \sa{} operators is called
   \emph{complete} if every \sa{} operator $C$ that commutes with all
   members of the family can be expressed as $C= g(A_1,A_2,\dots )$ for
   some function $g$. The set of all such operators cannot be extended
   in any suitable sense (it is closed in all relevant operator
   topologies).  For any commuting family $(\avec{A}{m})$ of \sa{}
   operators there is a \sa{} operator $B$ and measurable functions
   $f_i$ such that $A_i= f_i(B)$.  If the family is complete, then this
   operator has simple (i.e., nondegenerate) spectrum.}  namely the
family
\begin{equation}
{A} \equiv \sum_{\a} {\lambda}_{\a} \Pa = \left(\sum_{\a}\la^{(1)}
\Pa ,\ldots, \sum_{\a}\la^{(m)}\Pa\right) \equiv (\avec{A}{m}).
\label{eq:ccff}
\end{equation}
Then the notions of the various kinds of formal measurements---weak,
ideal, normal, standard, strong---extend straightforwardly to formal
measurements of commuting families of operators. In particular, for
the general notion of weak formal measurement given by \ref{def:wfmg},
$P$ becomes a PVM on $\R^m$, with associated operators $ A_i=
\int_{\R^m} \lambda^{(i)} P(d\lambda)\quad [\lambda=(
\lambda^{(1)},\ldots,\lambda^{(m)})\in\R^m]$.  And just as for PVMs on
$\R$ and \sa{} operators, this association in fact yields, by the
spectral theorem, a one-to-one correspondence between PVMs on $\R^m$
and commuting families of $m$ \sa{} operators.The PVM corresponding to
the commuting family $(\avec{A}{m})$ is in fact simply the product PVM
$P= P^A= P^{A_1}\times \cdots\times P^{A_m}$ given on product sets by
   \begin{equation}
P^{{A}}(\Delta_1\times\cdots \times\Delta_m)= P^{A_{1}}
(\Delta_{1})\cdots  P^{A_{m}} (\Delta_{m}),
\label{eq:factpvm}
   \end{equation}
   where $P^{A_{1}} ,\ldots, P^{A_{m}}$ are the PVMs of $ \avec{A}{m}$,
   and $\Delta_i\subset \R$, with the associated probability
   distributions on $\R^m$ given by the spectral measures for $A$
\begin{equation}
\mu^{{A}}_{\psi}(\Delta)
=\langle\psi, P^{{A}}(\Delta)
\psi\rangle\
\label{plcf}
\end{equation}
for any (Borel) set $\Delta\subset\R^m$.

In particular, for a PVM on $\R^m$, corresponding to $A=
(\avec{A}{m})$, the $i$-marginal distribution, i.e., the distribution
of the $i$-th component $ \lambda^{(i)}$, is
$$
\mu^{A}_{\psi }(\R \times \cdots\R \times \Delta_i \times \R \times
\cdots \times \R) =\langle \psi, P^{A_i}( \Delta_i) \psi\rangle=
\mu^{A_i}_{\psi}(\Delta_i),
$$
the spectral measure for $A_i$. Thus, by focusing on the respective
pointer variables $\lambda^{(i)}$, we may regard an experiment
measuring (or a weak formal measurement of) $A= (\avec{A}{m})$ as
providing an experiment measuring (or a weak formal measurement of)
each $A_i$, just as would be the case for a genuine measurement of $m$
quantities $A_1,\ \ldots,A_m$. Note also the following: If $\{\Ha,
\la, \Aa \}$ is a strong formal measurement of $A= (\avec{A}{m})$,
then $\{\Ha, \lambda_\a^{(i)}, \Aa \}$ is a strong formal measurement
of $A_i$, but if $\{\Ha, \la, \Aa \}$ is an ideal, resp. normal, resp.
standard, measurement of $A$, $\{\Ha, \lambda_\a^{(i)}, \Aa \}$ need
not be ideal, resp. normal, resp. standard.

There is a crucial point to observe: the same operator may belong to
different commuting families.  Consider, for example, a measurement of
${A} = (\avec{A}{m})$ and one of ${B} = (\avec{B}{m})$, where
$A_{1}=B_{1}\equiv C$.  Then while both measurements provide a
measurement of $C$, they could be totally different: the operators
$A_{i}$ and $B_{i}$ for $i\neq1 $ need not commute and the PVMs of
${A}$ and ${B}$, as well as any corresponding experiments $\E_A$ and
$\E_B$, will be in general essentially different.

To emphasize this point we shall recall a famous example, the EPRB
experiment~\cite{EPR, Boh51}: A pair of spin one-half particles,
prepared in a spin-singlet state $$
\psi =
\frac{1}{\sqrt{2}}\left(\psi^{(+)}\otimes\psi^{(-)} +
   \psi^{(-)}\otimes\psi^{(+)}\right)\,,$$
are moving freely in
opposite directions.  Measurements are made, say by Stern-Gerlach
magnets, on selected components of the spins of the two particles.
Let $\bf {a} ,\, {b} ,\, {c}$ be three different unit vectors in
space, let $\mybold{\sigma}_{1} \equiv \mybold{\sigma}\ot \id$ and let
$\mybold{\sigma}_{2} \equiv \id \ot \mybold{\sigma}, $ where $
\mybold{\sigma} =(\sigma_x,\sigma_y,\sigma_z)$ are the Pauli matrices.
Then we could measure the operator $\mybold{\sigma}_{1}{\bf\cdot {a}}$
by measuring either of the commuting families $( \mybold{\sigma}_{1}
{\bf \cdot {a}}\,, \mybold{\sigma}_{2} {\bf \cdot {b}})$ and $(
\mybold{\sigma}_{1} {\bf \cdot {a}} \,, \mybold{\sigma}_{2} {\bf \cdot
   {c}}) $.  However these measurements are different, both as weak and
as strong measurements, and of course as experiments.  In \BM{} the
result obtained at one place at any given time will in fact depend
upon the choice of the measurement simultaneously performed at the
other place (i.e., on whether the spin of the other particle is
measured along $\bf {b}$ or along $\bf{c}$).  However, the statistics
of the results won't be affected by the choice of measurement at the
other place because both choices yield measurements of the same
operator and thus their results must have the same statistical
distribution.

\subsection{Functions of Measurements}

One of the most common experimental procedures is to recalibrate the
scale of an experiment \E{}: if $Z$ is the original result and $f$ an
appropriate function, recalibration by $f$ leads to $f({Z})$ as the
new result. Thus $f(\E)$ has an obvious meaning.  Moreover, if
$\E\mapsto A$ according to (\ref{eq:prdeltan}) then $ \mu^{
   f(Z)}_{\psi} =\mu^{ Z}_{\psi} \circ f^{-1} = \mu^A_\psi \circ f^{-1}
$, and
$$
\mu^A_\psi \circ f^{-1}(d\lam) =
\langle\psi,P^{A}(f^{-1}(d\lambda))\psi\rangle =
\langle\psi,P^{f(A)}(d\lambda)\psi\rangle
$$
where the last equality follows {}from the very definition of
$$f(A) = \int f(\lam) P^{A} (d\lam) = \int \lam
P^{A}(f^{-1}(d\lambda))$$
provided by the spectral theorem.  Thus,
\begin{equation}
\mbox{if }\quad \mu^{ Z}_{\psi} =\mu^A_\psi \qquad
\mbox{then }\qquad  \mu^{ f(Z)}_{\psi} =\mu^{f(A)}_\psi\,,
\label{eq:prfm}
\end{equation}
i.e.,
\begin{equation}
\text{if}\qquad \E \mapsto A \qquad\text{then}\qquad f(\E) \mapsto
f(A).
\end{equation}

The notion of \emph{function of a formal measurement} has then an
unequivocal meaning: if $\M$ is a weak formal measurement defined by
the PVM $P$ then $f(\M)$ is the weak formal measurement defined by the
PVM $P\circ f^{-1}$, so that if $\M$ is a measurement of $A$ then
$f(\M)$ is a measurement of $f(A)$; for a strong formal measurement
$\M=\{\Ha, \la, \Aa \}$ the self-evident requirement that the
recalibration not affect the \wf{} transitions induced by \M{} leads
to $ f(\M)= \{\Ha, f(\lambda_{\alpha}), \Aa \}$.  Note that if $\M$ is
a standard measurement, $f(\M)$ will in general not be standard (since
in general $f$ can be many--to--one).  \bigskip

To highlight some subtleties of the notion of function of measurement
we shall discuss two examples: Suppose that $\M$ and $\M'$ are
respectively measurements of the commuting families $A = (A_{1}, A_{
   2})$ and $ B = (B_{1}, B_{ 2})$, with $A_{1}A_{ 2}= B_{1}B_{ 2}=C$.
Let $f:\R^{2}\to\R$, $f (\lambda_{1}, \lambda_{2}) =
\lambda_{1}\lambda_{2}$.  Then both $f(\M)$ and $f(\M')$ are
measurement of the same \sa{} operator $C$.  Nevertheless, as strong
measurements or as experiments, they could be very different: if
$A_{2}$ and $B_{2}$ do not commute they will be associated with
different families of spectral projections. (Even more simply,
consider measurements $\M_x$ and $\M_y$ of $\sigma_x$ and $\sigma_y$
and let $f(\lambda)= \lambda^2$. Then $f(\M_x)$ and $f(\M_y)$ are
measurement of $I$---so that the result must be $1$)---but the two
strong measurements, as well as the corresponding experiments, are
completely different.)

The second example is provided by measurements designed to determine
whether the operator $A=\sum \la\Pa$ (the $\la$'s are distinct) has
values in some given set $\Delta$.  This determination can be
accomplished in at least two different ways: Suppose that $\M$ is an
ideal measurement of $A$ and let ${\sf 1}_\Delta(\lambda)$ be the
characteristic function of the set $\Delta$.  Then we could perform
${\sf 1}_\Delta(\M)$, that is, we measure $A$ and see whether
``$A\in\Delta$''.  But we could also perform an ``\emph{ideal
   determination} of $A\in\Delta$'', that is, an ideal measurement of
${\sf 1}_\Delta(A) = P^A(\Delta)$.  Now, both measurements provide a
``measurement of $A\in\Delta $'' (i.e., of the operator $ {\sf
   1}_\Delta(A)$), since in both cases the results 1 and 0 get assigned
the same probabilities.  However, as strong measurements, they are
different: when ${\sf 1}_\Delta(\M)$ is performed, and the result 1 is
obtained, $\psi$ undergoes the transition
$$
\psi \to \Pa \psi
$$
where $\a$ is the outcome with $\la\in \Delta$ that actually
occurs.  On the other hand, for an ideal measurement of ${\sf
   1}_\Delta(A)$, the occurrence of the result 1 will generate the
transition
$$
\psi \to P^{A}(\Delta)\psi = \sum_{\la\in \Delta} \Pa\psi.
$$
Note that in this case the state of the system is changed as little
as possible. For example, suppose that two eigenvalues, say
$\lambda_{\alpha_1}, \lambda_{\alpha_2}$, belong to $\Delta$ and $\psi
= \psi_{\alpha_1} + \psi_{\alpha_2}$; then determination by performing
${\sf 1}_\Delta(\M)$ will lead to either $\psi_{\alpha_1}$ or $
\psi_{\alpha_2}$, while the ideal determination of $A\in\Delta$ will
not change the state.

\subsection{Measurements of Operators with Continuous Spectrum}
\label{subsec.mocs}
We shall now reconsider the status of measurements of \sa{} operators
with continuous spectrum.  First of all, we remark that while on the
weak level such measurements arise very naturally---and, as already
stressed in Section \ref{sec:MO}, are indeed the first to appear in
\BM---there is no straightforward extension of the notion of strong
measurement to operators with continuous spectrum.

However, for given set of real numbers $\Delta$, one may consider any
determination of $A \in \Delta$, that is, any strong measurement of
the spectral projection $P^{A}(\Delta)$.  More generally, for any
choice of a \emph{simple function}
\begin{displaymath}
f (\lambda) = \sum_{i=1}^{N} c_i\, {\sf 1}_{\Delta_i}(\lambda) ,
\end{displaymath}
one may consider the strong measurements of $f(A)$.  In particular,
let $\{ f^{(n)} \}$ be a sequence of simple functions converging to
the identity, so that $f^{(n)}(A) \rightarrow A$, and let $\M_n $ be
measurements of $f^{(n)}(A)$.  Then $\M_n $ are \emph{ approximate
   measurements} of $A$.

Observe that the foregoing applies to operators with discrete
spectrum, as well as to operators with continuous spectrum.  But note
that while on the weak level we always have
\begin{displaymath}
\M_n \to \M \,,
\end{displaymath}
where $\M$ is a (general) weak measurement of $A$ (in the sense of
(\ref{def:wfmg})), if $A$ has continuous spectrum $\M$ will not exist
as a strong measurement (in any reasonable generalized sense, since
this would imply the existence of a bounded-operator-valued function
$R_\lambda$ on the spectrum of $A$ such that
$R^{\ast}_{\lambda}R_\lambda\, d\lambda = P^A (d\lambda)$, which is
clearly impossible).  In other words, in this case there can be no
actual (generalized) strong measurement that the approximate
measurements $\M_{n}$ approximate---which is perfectly reasonable.

\subsection{Sequential Measurements}
\label{sec:SeqM}

Suppose that $n$ measurements (with for each $i$, the
$\lambda^{(i)}_{\alpha_{i}}$ distinct)
\begin{displaymath}
\M_{1}\equiv \{ \H^{(1)}_{\alpha_{1}} , \lambda^{(1)}_{\alpha_{1}} ,
R^{(1)}_{\alpha_{1}}  \},\;
\dots\;,\; \M_{n}\equiv \{
\H^{(n)}_{\alpha_{n}},\lambda^{(n)}_{\alpha_{n}},
R^{(n)}_{\alpha_{n}}  \}
\end{displaymath}
of operators (which need not commute)
\begin{displaymath}
A_{1}= \sum_{\alpha_{1}}\lambda^{(1)}_{\alpha_{1}}
P^{(1)}_{\alpha_{1}},\;
\dots\;,\; A_{n}= \sum_{\alpha_{n}}\lambda^{(n)}_{\alpha_{n}}
P^{(n)}_{\alpha_{n}}
\end{displaymath}
are successively performed on our system at times $0 < t_1 < t_2<\dots
<t_N$.  Assume that the duration of any single measurement is small
with respect to the time differences $t_{i}-t_{i-1}$, so that the
measurements can be regarded as instantaneous.  If in between two
successive measurements the system's \wf{} changes unitarily with the
operators $U_{t}$ then, using obvious notation,
\begin{equation}
\prob (A_{1}=\lambda^{(1)}_{\alpha_{1}},\ldots , A_{n} =
\lambda^{(n)}_{\alpha_{n}} ) = \|
R^{(n)}_{\alpha_{n}}(t_{n})
\cdots\,R^{(1)}_{\alpha_{1}}(t_{1}) \, \psi\|^2 ,
\label{eq:conprop}
\end{equation}
where $R_{\a_{i}}^{(i)}(t) = U_{t}^{-1} R_{\a_{i}}^{(i)}U_{t}$ and
$\psi$ is the initial ($t=0$) \wf{}.

To understand how (\ref{eq:conprop}) comes about consider first the
case where $n=2$ and $t_2\approx t_1\approx 0$.  According to standard
probability rules, the probability of obtaining the results
$Z_{1}=\lambda^{(1)}_{\alpha_{1}}$ for the first measurement and
$Z_{2}=\lambda^{(2)}_{\alpha_{2}}$ for the second one is the
product\footnote{This is so because of the \textit{conditional
     independence} of the outcomes of two successive measurements
   \textit{given} the final \cwf{} for the first measurement.  More
   generally, the outcome of any measurement depends only on the \wf{}
   resulting {}from the preceding one.  For Bohmian experiments this
   independence is a direct consequence of (\ref{eq:fpfp}).  One may
   wonder about the status of this independence for \oqt{}.  We stress
   that while this issue might be problematical for \oqt{}, it is not a
   problem for \BM: the conditional independence of two successive
   measurements is a consequence of the theory. (For more on this
   point, see \cite{DGZ92a}).) We also would like to stress that this
   independence assumption is in fact crucial for \oqt{}.  Without it,
   it is hard to see how one could ever be justified in invoking the
   quantum formalism. Any measurement we may consider will follow many
   earlier measurements.}
\begin{displaymath}
\prob (Z_{2}= \lambda^{(2)}_{\alpha_{2}}| Z_{1} =
\lambda^{(1)}_{\alpha_{1}}) \cdot
\prob(Z_{1}=\lambda^{(1)}_{\alpha_{1}})
\end{displaymath}
where the first term is the probability of obtaining
$\lambda^{(2)}_{\alpha_{2}}$ given that the result of the first
measurement is $\lambda^{(1)}_{\alpha_{1}}$.  Since $\M_1$ then
transforms the \wf{} $\psi$ to $R^{(1)}_{\alpha_{1}}\psi$, the
(normalized) initial \wf{} for $\M_2$ is
${R^{(1)}_{\alpha_{1}}\psi}/{\|R^{(1)}_{\alpha_{1}}\psi\| }$, this
probability is equal to
\begin{displaymath}
\frac{\norm
R^{(2)}_{\alpha_{2}} R^{(1)}_{\alpha_{1}}\psi\norm^2}{\norm
R^{(1)}_{\alpha_{1}}\psi\norm^2}.
\end{displaymath}
The second term, the probability of obtaining
$\lambda^{(1)}_{\alpha_{1}}$, is of course $\norm R^{(1)}_{\alpha_{1}}
\psi\norm^2 $.  Thus
\begin{displaymath}
\prob(A^{(1)}=\lambda^{(1)}_{\alpha_{1}},A^{(2)}=
\lambda^{(2)}_{\alpha_{2}}) =\| R^{(2)}_{\alpha_{2}}
R^{(1)}_{\alpha_{1}}\psi\|^{2}
\end{displaymath}
in this case.  Note that, in agreement with the analysis of discrete
experiments (see Eq.~\eq{eq:pr}), the probability of obtaining the
results $\lambda^{(1)}_{\alpha_{1}}$ and $\lambda^{(2)}_{\alpha_{2}}$
turns out to be the square of the norm of the final system \wf{}
associated with these results.  Now, for general times $t_{1}$ and
$t_{2}-t_{1}$ between the preparation of $\psi$ at $t=0$ and the
performance of $\M_{1}$ and between $\M_{1}$ and $\M_{2}$,
respectively, the final system \wf{} is
\begin{math}
   R^{(2)}_{\alpha_{2}} U_{t_{2}-t_{1}}
   R^{(1)}_{\alpha_{1}}U_{t_{1}}\psi= R^{(2)}_{\alpha_{2}}U_{t_{2}}
   U^{-1}_{t_{1}} R^{(1)}_{\alpha_{1}} U_{t_{1}}\psi.
\end{math}
But
\begin{math}
   \|R^{(2)}_{\alpha_{2}} U_{t_{2}}U^{-1}_{t_{1}} R^{(1)}_{\alpha_{1}}
   U_{t_{1}}\psi\|= \|U^{-1}_{t_{2}}R^{(2)}_{\alpha_{2}}
   U_{t_{2}}U^{-1}_{t_{1}}R^{(1)}_{\alpha_{1}}U_{t_{1}}\psi\| ,
\end{math}
and it is easy to see, just as for the simple case just considered,
that the square of the latter is the probability for the corresponding
result, whence (\ref{eq:conprop}) for $n=2$.  Iterating, i.e., by
induction, we arrive at (\ref{eq:conprop}) for general $n$.

We note that when the measurements $\M_{1},\ldots \M_{n}$ are ideal,
the operators $R^{(i)}_{\alpha_{i}}$ are the orthogonal projections
$P^{(i)}_{\alpha_{i}}$, and equation (\ref{eq:conprop}) becomes the
standard formula for the joint probabilities of the results of a
sequence of measurements of quantum observables, usually known as
Wigner's formula \cite{Wig63}.

It is important to observe that, even for ideal measurements, the
joint probabilities given by (\ref{eq:conprop}) are not in general a
consistent family of joint distributions: summation in
(\ref{eq:conprop}) over the outcomes of the $i$-th measurement does
not yield the joint probabilities for the results of the measurements
of the operators
\begin{math}
   A_{1}, \ldots, A_{i-1},A_{i+1},\ldots A_{n}
\end{math}
performed at the times $t_{1}, \ldots, t_{i-1}, t_{i+1}, \ldots
t_{n}$.  (By rewriting the right hand side of (\ref{eq:conprop}) as
\begin{math}
   \langle \psi, R^{(1)}_{\alpha_{1}}(t_{n})^\ast \cdots
   R^{(n)}_{\alpha_{n}} (t_{n})^\ast R^{(n)}_{\alpha_{n}}(t_{n})
   R^{(1)}_{\alpha_{1}}(t_{1}) \psi\rangle
\end{math}
one easily sees that the ``sum rule'' will be satisfied when $i=n$ or
if the operators $ R^{(i)}_{\alpha_{i}}(t_{i})$ commute. More
generally, the consistency is guaranteed by the ``decoherence
conditions'' of Griffiths, Omn\`es, Gell-Mann and Hartle, and
Goldstein and Page~\cite{Gri84, GMH90, GoldPage}.

This failure of consistency means that the marginals of the joint
probabilities given by (\ref{eq:conprop}) are not themselves given by
the corresponding case of the formula. This should, however, come as
no surprise: Since performing the measurement $\M_{i}$ affects the
state of the system, the outcome of $\M_{i+1}$ should in general
depend on whether or not $\M_{i}$ has been performed.  Note that there
is nothing particularly quantum in the fact that measurements matter
in this way: They matter even for genuine measurements (unlike those
we have been considering, in which nothing need be genuinely
measured), and even in classical physics, if the measurements are such
that they affect the state of the system.

The sequences of results $
\lambda_{\alpha}\equiv(\lambda^{(1)}_{\alpha_{1}},
\ldots,\lambda^{(n)}_{\alpha_{n}}),$ the associated state
transformations $\Aa \equiv R^{(n)}_{\alpha_{n}} U_{t_n -t_{n-1}}
R^{(n-1)}_{\alpha_{n-1}} \cdots\,R^{(1)}_{\alpha_{1}} U_{t_1},$ and
the probabilities (\ref{eq:conprop}) (i.e., given by $p_\a = \norm
\Aa\norm^2$) define what we shall call a \emph{ sequential
   measurement} of $\M_1, \cdots\M_n$, which we shall denote by
$\M_{n}\otimes \ldots\otimes \M_{1}$.  A sequential measurement does
not in general define a formal measurement, neither weak nor strong,
since $R_{\alpha}^{\ast}R_{\alpha}$ need not be a projection. This
fact might seem disturbing (see, e.g., \cite{Dav76}); we shall take up
this issue in the next section.

\subsection{Some Summarizing Remarks}

The notion of formal measurement we have explored in this section is
at the heart of the quantum formalism.  It embodies the two essential
ingredients of a quantum measurement: the self-adjoint operator $A$
which represents the measured observable and the set of state
transformations $\Aa$ associated with the measured results.  The
operator always carries the information about the statistics of
possible results. The state transformations prescribe how the state of
the system changes when the measurement is performed. For ideal
measurement the latter information is also provided by the operator,
but in general additional structure (the $\Aa$'s) is required.

There are some important morals to draw.  \emph{The association
   between measurements and operators is many-to-one:} the same
operator $A$ can be measured by many different measurements, for
example ideal, or normal but not ideal. Among the possible
measurements of $A$, we must consider all possible measurements of
commuting families of operators that include $A$, each of which may
correspond to entirely different experimental setups.

A related fact: \emph{not all measurements are ideal
   measurements}.\footnote{In this regard we observe that the vague
   belief in a universal collapse rule is as old, almost, as quantum
   mechanics.  It is reflected in von Neumann's formulation of quantum
   mechanics \cite{vNe55}, based on two distinct dynamical laws: a
   unitary evolution {\it between measurements\/}, and a nonunitary
   evolution {\it when measurements are performed}.  However, von
   Neumann's original proposal \cite{vNe55} for the nonunitary
   evolution---that when a measurement of $A=\sum_{\a}\la \Pa$ is
   performed upon a system in the state given by the density matrix
   $W$, the state of the system after the measurement is represented by
   the density matrix $$
   W' = \sum_{\a} \sum_{\beta}\langle \phi_{\a
     \beta}, W\phi_{\a \beta} \rangle P_{[\phi_{\a \beta}]}$$
   where,
   for each $\a$, $\{\phi_{\a \beta}\}$ is a basis for ${\H}_{\a}
   $---does not treat the general measurement as ideal.  Moreover, this
   expression in general depends on the choice of the basis $\{\phi_{\a
     \beta}\}$, and was thus criticized by L\"uders \cite{Lud51}, who
   proposed the transformation $$
   W \to W' = \suma \Pa W \Pa\,,$$
   as it
   gives a {\it unique} prescription.  Note that for $W=P_{[\psi]}$,
   where $P_{[\psi]}$ is the projection onto the initial pure state
   $\psi$, $ W'= \suma p_{\alpha} P_{[ \psia]}$, where $ p_\a =
   |\langle \psi,\Pa \psi\rangle|^2 $ and $\psia = \Pa\psi$,
   corresponding to an ideal measurement.} No argument, physical or
mathematical, suggests that ideal measurements should be regarded as
``more correct'' than any other type.  In particular, the Wigner
formula for the statistics of a sequence of ideal measurements is no
more correct than the formula \eq{eq:conprop} for a sequence of more
general measurement.  Granting a privileged status to ideal
measurements amounts to a drastic and arbitrary restriction on the
quantum formalism {\it qua measurement formalism}, since many (in fact
most) real world measurements would be left out.

In this regard we note that the arbitrary restriction to ideal
measurements affects the research program of ``decoherent'' or
``consistent'' histories \cite{GMH90,Omn88,Gri84}, since Wigner's
formula for a sequence of ideal measurements is unquestionably at its
basis.  (It should be emphasized however that the special status
granted to ideal measurements is probably not the main difficulty with
this approach.  The no-hidden-variables theorems, which we shall
discuss in Section 7, show that the totality of different families of
weakly decohering histories, with their respective probability
formulas, is genuinely inconsistent. While such inconsistency is
perfectly acceptable for a measurement formalism, it is hard to see
how it can be tolerated as the basis of what is claimed to be a
fundamental theory. For more on this, see \cite {DGZ92a, ShellyPT}.

\section{The Extended Quantum Formalism} \setcounter{equation}{0}

As indicated in Section 2.9, the textbook \qf\ is merely an
idealization.  As just stressed, not all real world measurements are
ideal. In fact, in the real world the projection postulate---that when
the measurement of an observable yields a specific value, the \wf\ of
the system is replaced by its projection onto the corresponding
eigenspace---is rarely obeyed.  More importantly, a great many
significant real-world experiments are simply not at all associated
with operators in the usual way.  Consider for example an electron
with fairly general initial \wf, and surround the electron with a
``photographic'' plate, away {}from (the support of the \wf\ of) the
electron, but not too far away.  This setup measures the position of
``escape'' of the electron {}from the region surrounded by the plate.
Notice that since in general the time of escape is random, it is not
at all clear which operator should correspond to the escape
position---it should not be the Heisenberg position operator at a
specific time, and a Heisenberg position operator at a random time has
no meaning.  In fact, there is presumably no such operator, so that
for the experiment just described the probabilities for the possible
results cannot be expressed in the form \eq{eq:prdeltan}, and in fact
are not given by the spectral measure for any operator.

Time measurements, for example escape times or decay times, are
particularly embarrassing for the \qf.  This subject remains mired in
controversy, with various research groups proposing their own favorite
candidates for the ``time operator'' while paying little attention to
the proposals of the other groups.  For an analysis of time
measurements within the framework of \BM, see \cite{dau97}; in this
regard see also \cite{Lea90, leavens2, leavens3, grubl}.

Because of these and other difficulties, it has been proposed that we
should go beyond operators-as-observables, to ``{\it generalized
   observables\/},'' described by mathematical objects even more
abstract than operators (see, e.g., the books of Davies \cite{Dav76},
Holevo \cite{Hol82} and Kraus \cite{Kra83}).  The basis of this
generalization lies in the observation that, by the spectral theorem,
the concept of self-adjoint operator is completely equivalent to that
of (a normalized) projection-valued measure (PVM), an
orthogonal-projection-valued additive set function, on the value space
$\R$.  Orthogonal projections are among the simplest examples of
positive operators, and a natural generalization of a ``quantum
observable'' is provided by a positive-operator-valued measure (POVM):
a normalized, countably additive set function $O$ whose values are
positive operators on a Hilbert space \H{}.  When a POVM is sandwiched
by a \wf{} it generates a probability distribution
\begin{equation}
\mu^O_\psi: \Delta\mapsto \mu^O_\psi (\Delta) \equiv \langle\psi ,
O(\Delta)\psi\rangle
\label{mupsipov}
\end{equation}
in exactly the same manner as a PVM.

\subsection{POVMs and Bohmian Experiments}
\label{secpovbe}
{}From a fundamental perspective, it may seem that we would regard
this generalization, to positive-operator-valued measures, as a step
in the wrong direction, since it supplies us with a new, much larger
class of fundamentally unneeded abstract mathematical entities far
removed {}{}from the basic ingredients of \BM{}.  However {}{}from the
perspective of Bohmian phenomenology positive-operator-valued measures
form an extremely natural class of objects---\emph{indeed more natural
   than projection-valued measures}.

To see how this comes about observe that \eq{eq:ormfin} defines a
family of bounded linear operators $\Aa$ by
\begin{equation}
P_{[\Phia]}\left[ U({\psi }\ot\Phi_0) \right] = (\Aa
\psi)\ot\Phia,
\label{DISCRE}
\end{equation}
in terms of which we may rewrite the probability \eq{eq:pr} of
obtaining the result $\la$ (distinct) in a generic discrete experiment
as
\begin{equation}
p_{\a} = \|\psia \|^2 = \|\Aa \psi \|^2= \langle
\psi, \Aad \Aa \psi\rangle\, .
\label{paA}
\end{equation}
By the unitarity of the overall evolution of system and apparatus we
have that $ \sum_{\a} \|\psia \|^2 = \sum_{\a}\langle \psi, \Aad \Aa
\psi\rangle = 1 $ for all $\psi\in \H$, whence
\begin{equation}
\suma   \Aad \Aa = \id \, .
\label{eq:uni}
\end{equation}
The operators $ O_\a \equiv \Aad \Aa $ are obviously positive, i.e.,
\begin{equation}
\langle \psi, O_{\a}\psi \rangle\ge 0\qquad\mbox{for all}\quad
\psi\in \H
\label{eq:posoper}
\end{equation}
and by (\ref{eq:uni}) sum up to the identity,
\begin{equation}
\sum_{\a} O_{\a}= \id  \, .
\label{eq:sumone}
\end{equation}
Thus we may associate with a generic discrete experiment \E---with no
assumptions about reproducibility or anything else, but merely
\emph{unitarity}---a POVM
\begin{equation}
O (\Delta) = \sum_{\la \in \Delta} O_\a \equiv \sum_{\la \in
\Delta} \Aad \Aa,
\label{oeslaa}
\end{equation}
in terms of which the statistics of the results can be expressed in a
compact way: the probability that the result of the experiment lies in
a set $\Delta$ is given by
\begin{equation}
\sum_{\la \in \Delta} p_{\a} =
\sum_{\la \in \Delta} \langle
\psi,O_\a \psi\rangle   =\langle
\psi, O(\Delta) \psi\rangle \, .
\label{pro}
\end{equation}
Moreover, it follows {}{}from \eq{eq:ormfin} and \eq{DISCRE} that \E{}
generates state transformations
\begin{equation}
    \psi \to \psia=\Aa \psi\,.
\label{gentr}
\end{equation}

\subsection{Formal Experiments}\label{secFE}\label{subsec.comfefm}
The association between experiments and POVMs can be extended to a
general experiment (\ref{eq:generalexperiment}) in a straightforward
way.  In analogy with (\ref{eq:prdeltan}) we shall say that the POVM
$O$ is associated with the experiment \E{} whenever the probability
distribution (\ref{eq:indumas}) of the results of \E{} is equal to the
probability measure (\ref{mupsipov}) generated by $O$,
i.e.,\footnote{Whenever (\ref{etoo}) is satisfied we may say that the
   experiment \E{} is a measurement of the generalized observable $O$.
   We shall however avoid this terminology in connection with
   generalized observables; even when it is standard (so that we use
   it), i.e., when $ O$ is a PVM and thus equivalent to a \sa\
   operator, it is in fact improper.}
\begin{equation}
\E\mapsto O \qquad\mbox{if and only if}\qquad \rho^{ Z}_{\psi}
=\mu^O_\psi,
\label{etoo}
\end{equation}

We may now proceed as in Section 3 and analyze on a formal level the
association (\ref{etoo}) by introducing the notions of \emph{weak} and
\emph{strong} formal experiment as the obvious generalizations of
(\ref{def:wfmg}) and (\ref{def:sfm}):
\begin{equation}
\mbox{%
\begin{minipage}{0.85\textwidth}\openup 1.4\jot
   \setlength{\baselineskip}{12pt}\emph{Any positive-operator-valued
     measure $O$ defines the weak formal experiment $\Ex\equiv O$.  Any
     set $\{\la \}$ of not necessarily distinct real numbers (or
     vectors of real numbers) paired with any collection $\{\Aa\}$ of
     bounded operators on $\H$ such that $\sum\Aa^{\ast}\Aa=\id$
     defines the strong formal experiment $ \Ex\equiv\{\la, \Aa \}$
     with associated POVM \eq{oeslaa} and state transformations
     \eq{gentr}.  }
   \end{minipage}}
\label{def:wfe}
\end{equation}

The notion of formal experiment is a genuine extension of that of
formal measurement, the latter being the special case in which $O$ is
a PVM and $\Aad\Aa$ are the projections.

Formal experiments share with formal measurements many features.  This
is so because all measure-theoretic properties of projection-valued
measures extend to positive-operator-valued measures.  For example,
just as for PVMs, integration of real functions against
positive-operator-valued measure is a meaningful operation that
generates \sa\ operators: for given real (and measurable) function
$f$, the operator $B=\int f(\lambda) O(d\lambda)$ is a \sa{} operator
defined, say, by its matrix elements $\langle \phi,B \psi\rangle =\int
\lambda \mu_{\phi,\psi}(d\lambda)$ for all $\phi$ and $\psi$ in $\H$,
where $\mu_{\phi,\psi}$ is the complex measure
$\mu_{\phi,\psi}(d\lambda) = \langle \phi,O(d\lambda) \psi\rangle$.
(We ignore the difficulties that might arise if $f$ is not bounded.)
In particular, with $O$ is associated the \sa{} operator
\begin{equation}
\label{sawpov}
A_{O} \equiv \int \lam \, O (d\lam).
\end{equation}

It is however important to observe that this association (unlike the
case of PVMs, for which the spectral theorem provides the inverse) is
not invertible, since the \sa{} operator $A_{O}$ is always associated
with the PVM provided by the spectral theorem. Thus, unlike PVMs,
POVMs are not equivalent to \sa{} operators. In general, the operator
$A_{O}$ will carry information only about the mean value of the
statistics of the results,
$$
\int \lam\;\langle \psi, O(d\lam)\psi\rangle = \langle\psi,
A_{O}\psi\rangle \,,
$$
while for the higher moments we should expect that
$$
\int \lam^n\;\langle \psi, O(d\lam)\psi\rangle \neq \langle\psi,
A_{O}^n\psi\rangle \,
$$
unless $O$ is a PVM.

What we have just described is an important difference between general
formal experiments and formal measurements.  This and other
differences originate {}from the fact that a POVM is a much weaker
notion than a PVM. For example, a POVM $O$ on $\R^m$---like ordinary
measures and unlike PVMs---need not be a product measure: If $
O_1,\ldots, O_m$ are the \emph{marginals} of $ O$,
$$
O_1(\Delta_1) = O(\Delta_1 \times \R^{m-1})\,,\;\ldots\,,\;
O_m(\Delta_m) = O(\R^{m-1} \times \Delta_m ),
$$
the product POVM $ O_1\times\cdots\times O_m$ will be in general
different {}from $ O$. (This is trivial since any probability measure
on $\R^m$ times the identity is a POVM.)

Another important difference between the notion of POVM and that of
PVM is this: while the projections $P(\Delta)$ of a PVM, for different
$\Delta$'s, commute, the operators $O(\Delta)$ of a generic POVM need
not commute. An illustration of how this may naturally arise is
provided by sequential measurements.

A sequential measurement (see Section \ref{sec:SeqM}) $\M_{n}\otimes
\ldots\otimes \M_{1}$ is indeed a very simple example of a formal
experiment that in general is not a formal measurement (see also
Davies \cite{Dav76}). We have that
$$\M_{n}\otimes \ldots\otimes \M_{1}= \{\la, \Aa\}$$
where
$$
\lambda_{\alpha}\equiv(\lambda^{(1)}_{\alpha_{1}},
\ldots,\lambda^{(n)}_{\alpha_{n}})
$$
and
$$
\Aa \equiv R^{(n)}_{\alpha_{n}} U_{t_n -t_{n-1}}
R^{(n-1)}_{\alpha_{n-1}} \cdots\,R^{(1)}_{\alpha_{1}}\,.  U_{t_1 }.
$$
Note that since $p_\a = \norm\Aa\psi\norm^2$, we have that
$$
\sum_{\a} \Aad\Aa =\id$$
\, , which also follows directly using
$$
\sum_{\a_{j}}R^{(j)}_{\alpha_{j}}\,^\dagger R^{(j)}_{\alpha_{j}} =
\id\,,\qquad j= 1,\ldots,n
$$

Now, with $\M_{n}\otimes \ldots\otimes \M_{1}$ is associated the POVM
\begin{displaymath}
O (\Delta) = \sum_{\la \in \Delta} \Aad\Aa \,.
\end{displaymath}
Note that $O(\Delta)$ and $O(\Delta ')$ in general don't commute since
in general $R_{\a}$ and $R_{\beta}$ may fail to do so.  \bigskip

An interesting class of POVMs for which $O(\Delta)$ and $O(\Delta ')$
do commute arises in association with the notion of an
``\emph{approximate measurement}'' of a \sa{} operator: suppose that
the result $Z$ of a measurement $\M=P^A$ of a \sa{} operator $A$ is
distorted by the addition of an independent noise $N$ with symmetric
probability distribution $\eta (\lambda)$.  Then the result $Z+N$ of
the experiment, for initial system \wf{} $\psi$, is distributed
according to $$\Delta \mapsto \int_{\Delta}\int_{\R} \eta(\lambda -
\lambda ') \langle \psi, P_A (d\lambda ')\psi\rangle \, d\lambda \, ,
$$
which can be rewritten as
$$
\Delta \mapsto \langle\psi,\int_{\Delta} \eta(\lambda -A)
d\lambda\; \psi\rangle\, .$$
Thus the result $Z+N$ is governed by the
POVM
   \begin{equation}
O (\Delta)=\int_{\Delta} \eta(\lambda
-A)\, d\lambda \, .
\label{eq:appro}
\end{equation}
The formal experiment defined by this POVM can be regarded as
providing an approximate measurement of $A$. For example, let
\begin{equation}
\eta (\lambda) = \frac{1}{\sigma\sqrt{2\pi}}
e^{-\frac{\lambda^2}{2\,\sigma^2}}\, .
\label{gauss}
\end{equation}
Then for $\sigma\to 0$ the POVM (\ref{eq:appro}) becomes the PVM of
$A$ and the experiment becomes a measurement of $A$.

Concerning the POVM (\ref{eq:appro}) we wish to make two remarks.  The
first is that the $O(\Delta)$'s commute since they are all functions
of $A$.  The second is that this POVM has a continuous density, i.e.,
$$
O(d\lambda) = o(\lambda)\, d\lambda\qquad\mbox{where}\qquad
o(\lambda)= \eta(\lambda -A)\,. $$
This is another difference between
POVMs and PVMs: like ordinary measures and unlike PVMs, POVMs may have
a continuous density. The reason this is possible for POVMs is that,
for a POVM $O$, unlike for a PVM, given $\psi\in H$, the vectors $
O(\Delta)\psi$ and $ O(\Delta ')\psi$, for $\Delta$ and $\Delta '$
disjoint and arbitrarily small, need not be orthogonal.  Otherwise, no
density $o(d\lambda)$ could exist, because this would imply that there
is a continuous family $\{o(\lambda)\psi\}$ of orthogonal vectors in
$\H$.

Finally, we observe that unlike strong measurements, the notion of
strong formal experiment can be extended to POVM with continuous
spectrum (see Section \ref{subsec.mocs}).  One may in fact define a
strong experiment by $\Ex =\{\lambda, R_{\lambda}\}$, where $\lam
\mapsto\Al$ is a continuous \emph{bounded-operator-valued function}
such that $\int \Ald \Al \,d\,\lam \,=\,\id $.  Then the statistics
for the results of such an experiment is governed by the POVM
$O(d\lam) \equiv \Ald \Al\, d\lam$.  For example, let
$$
R_\lambda = \xi\, ( \lambda -A) \quad\mbox{where}\quad \xi\,
(\lambda) = \frac{1}{\sqrt{\sigma}\sqrt[4]{2\pi}}
e^{-\frac{\lambda^2}{4\,\sigma^2}}\,.
$$
Then $O(d\lambda) = \Ald \Al \,d\,\lam$ is the POVM
(\ref{eq:appro}) with $\eta$ given by (\ref{gauss}).  We observe that
the state transformations (cf. the definition \eq{eq:con} of the
\cwf{})
\begin{equation}
\psi \to  R_{\lam}\psi= \frac{1}{\sqrt{\sigma}\sqrt[4]{2\pi}}
e^{-\frac{(\lambda -A)^2}{4\,\sigma^2}} \psi
\label{eq:aha}
\end{equation}
can be regarded as arising {}from a von Neumann interaction with
Hamiltonian (\ref{vontrans}) (and $\gamma T=1$) and ready state of the
apparatus
$$
\Phi_{0}(y) = \frac{1}{\sqrt{\sigma}\sqrt[4]{2\pi}}
e^{-\frac{y^2}{4\,\sigma^2}}.
$$
Experiments with state transformations (\ref{eq:aha}), for large
$\sigma$, have been considered by Aharonov and coworkers (see, e.g.,
Aharonov, Anandan, and Vaidman \cite{AAV93}) as providing ``weak
measurements'' of operators. (The effect of the measurement on the
state of the system is ``small'' if $\sigma$ is sufficiently large).
This terminology notwithstanding, it is important to observe that such
experiments are not measurements of $A$ in the sense we have discussed
here.  They give information about the average value of $A$, since $
\int \lam\;\langle \psi,\Ald \Al\, \psi\rangle\,d\lam = \langle\psi,
A\psi\rangle $, but presumably none about its higher moments.

\subsection{From Formal Experiments to Experiments}\label{subsec.ffete}
\label{subsec.repexp}

Just as with a formal measurement (see Section \ref{subsec.exp}), with
a formal experiment $ \Ex\equiv\{\la, \Aa \}$, we may associate a
discrete experiment \E{}.  The unitary map (\ref{eq:ormfin}) of \E{}
will be given again by (\ref{standu}), i.e.,
\begin{equation}
U: \;\psi \ot \Phi_0  \mapsto
\sum_{\a} (\Aa\psi) \ot \Phia,
\label{standun}
\end{equation}
but now $\Aad\Aa$ of course need not be projection.  The unitarity of
$U$ follows immediately {}from the orthonormality of the $\Phia$ using
$\sum \Aad \Aa = \id $.  (Note that with a weak formal experiment
$\Ex\equiv O=\{O_{\a}\}$ we may associate many inequivalent discrete
experiments, defined by \eq{standun} with operators $\Aa\equiv
U_\alpha \sqrt{O_\alpha}$, for \emph{any} choice of unitary operators
$U_\a$.)

We shall now discuss a concrete example of a discrete experiment
defined by a formal experiment which will allow us to make some more
further comments on the issue of reproducibility discussed in Section
\ref{sec:RC}.

Let $\{ \dots, e_{-1},e_0,e_1,\dots \}$ be an orthonormal basis in the
system Hilbert space \H, let $P_{-}\,,P_0\,, P_{+}$ be the orthogonal
projections onto the subspaces $\widetilde{\mathcal{H}}_{-}$, $\H_0$,
$\widetilde{\mathcal{H}}_{+}$ spanned by $\{e\}_{\a < 0}$, $\{e_0\}$,
$\{e\}_{\a >0}$ respectively, and let $V_+$, $V_-$ be the right and
left shift operators,
$$V_+ e_{\a} = e_{\a+1}\,, \qquad V_-e_\a = e_{\a-1}\,.$$
Consider the
strong formal experiment \Ex\ with the two possible results
$\lam_{\pm}=\pm 1$ and associated state transformations
\begin{equation}
R_{\pm1} = V_{\pm}(P_{\pm} +\frac{1}{\sqrt{2}}P_0).
\label{eq:pove}
\end{equation}
Then the unitary $U$ of the corresponding discrete experiment \E{} is
given by
$$
U: \;\psi \ot \Phi_0 \to R_{-}\psi \ot \Phi_{-} + R_{+}\psi \ot
\Phi_{+},$$
where $\Phi_0$ is the ready state of the apparatus and
$\Phi_{\pm}$ are the apparatus states associated with the results $\pm
1$.  If we now consider the action of $U$ on the basis vectors
$e_{\a}$,
\begin{eqnarray}
U(e_\a\ot\Phi_0)&=&e_{\a + 1}\ot \Phi_{+}\qquad \mbox{for $\a>0$}
\nonumber\\
U(e_\a\ot\Phi_0)&=&e_{\a - 1}\ot \Phi_{-}\qquad \mbox{for $\a
<0$}
\nonumber\\
U(e_0\ot\Phi_0)&=&
\frac{1}{\sqrt{2}}(e_1\ot\Phi_{+}
+e_{-1}\ot\Phi_{-})\, ,\nonumber
\end{eqnarray}
we see immediately that $$U(\widetilde{\mathcal{H}}_{\pm}\ot\Phi_0)
\subset \widetilde{\mathcal{H}}_{\pm}\ot\Phi_{\pm1}.$$
Thus
(\ref{eq:repconold}) is satisfied and \E{} is a reproducible
experiment.  Note however that the POVM $ O = \{ O_{-1}, O_{+1}\}$
associated with (\ref{eq:pove}),
$$
O_{\pm1} =\As_{\pm1}^{\dagger}\As_{\pm1} = P_{\pm} +
\frac{1}{2}P_0\,,
$$
is not a PVM since the positive operators $ O_{\pm 1}$ are not
projections, i.e, $ O_{\pm 1}^2 \ne O_{\pm 1}$.  Thus \E{} is not a
measurement of any self-adjoint operator, which shows that without the
assumption of the finite dimensionality of the subspaces
$\widetilde{\mathcal{H}}_{\a}$ a reproducible discrete experiment need
not be a measurement of a self-adjoint operator.

\subsection{Measure-Valued Quadratic Maps}\label{sec:mvqf}
We conclude this section with a remark about POVMs. Via \eq{mupsipov}
every POVM $O$ defines a ``normalized quadratic map'' {}from \H{} to
measures on some space (the value-space for the POVM). Moreover, every
such map comes {}from a POVM in this way. Thus the two notions are
equivalent:
   \begin{equation}
\mbox{%
\begin{minipage}{0.85\textwidth}\openup 1.4\jot
   \setlength{\baselineskip}{12pt}\emph{ \eq{mupsipov} defines a
     canonical one-to-one correspondence between POVMs and normalized
     measure-valued quadratic maps on \H. }
   \end{minipage}}
\label{def:mvqm}
\end{equation}
To say that a measure-valued map on \H{}
\begin{equation}
\psi \mapsto \mu_{\psi}
\label{eq:qumap}
\end{equation}
is quadratic means that
\begin{equation}
\mu_{\psi}= B(\psi, \psi)
\label{eq:qumapb}
\end{equation}
is the diagonal part of a sesquilinear map $B$, {}from $\H\times\H$ to
the complex measures on some value space $\Lambda$. If $B(\psi, \psi)$
is a probability measure whenever $\norm\psi\norm =1$, we say that the
map is normalized.\footnote{A sesquilinear map $B(\phi,\psi)$ is one
   that is linear in the second slot and conjugate linear in the first:
\begin{eqnarray}
B(\phi, \a \psi_1 +\beta\psi_2) &=& \a B(\phi,\psi_1)+\beta
B(\phi,\psi_2)
\nonumber\\
B(\a\phi_1 +\beta\phi_2,\psi)&=&\bar {\a} B(\phi_1,\psi)+\bar
{\beta}B(\phi_2,\psi) \,. \nonumber \end{eqnarray}
Clearly any such normalized $B$ can be chosen to be conjugate
symmetric,
$  B(\psi, \phi)= \overline{B(\phi, \psi)}$,
without affecting its diagonal, and it follows
{}from  polarization that any such $B$ must in fact
   \emph{be} conjugate symmetric.}

Proposition (\ref{def:mvqm}) is a consequences of the following
considerations: For a given POVM $O$ the map $\psi \mapsto
\mu_{\psi}^O$, where $ \mu^O_\psi (\Delta) \equiv \langle\psi ,
O(\Delta)\psi\rangle$, is manifestly quadratic, with $B(\phi,\psi) =
\langle\phi , O(\cdot)\psi\rangle$, and it is obviously normalized.
Conversely, let $\psi \mapsto \mu_{\psi}$ be a normalized
measure-valued quadratic map, corresponding to some $B$, and write
$B_\Delta (\phi,\psi)= B (\phi,\psi)[\Delta]$ for the complex measure
$B$ at the Borel set $\Delta$.  By the Schwartz inequality, applied to
the positive form $ B_\Delta (\phi,\psi) $, we have that $ |B_\Delta
(\phi,\psi)|\le \norm\psi\norm \norm\phi\norm $. Thus, using Riesz's
lemma \cite{RS80}, there is a unique bounded operator $ O(\Delta)$ on
\H\ such that
$$
B_\Delta(\phi,\psi) = \langle\phi, O(\Delta)\psi\rangle .
$$
Moreover, $ O(\Delta) $, like $B_\Delta$, is countably additive in
$\Delta$, and since $B (\psi,\psi)$ is a (positive) measure, $O$ is a
positive-operator-valued measure, normalized because $B$ is.

A simple example of a normalized measure-valued quadratic map is
\begin{equation}
\label{qem}
\Psi\mapsto \rho^{\Psi} (dq) = |\Psi|^2 dq \, ,
\end{equation}
whose associated POVM is the PVM $P^{\hat{Q}}$ for the position
(configuration) operator
\begin{equation}
{\hat Q}\Psi(q) = q\Psi(q)\,.
\label{eq:posiope}
\end{equation}
Note also that if the quadratic map $\mu_\psi$ corresponds to the POVM
$O$, then, for any unitary $U$, the composite map
$\psi\mapsto\mu_{_{U\psi}}$ corresponds to the POVM $U^*OU$, since $
\langle U\psi, O(\Delta)U\psi\rangle = \langle\psi,
U^*O(\Delta)U\psi\rangle$.  In particular for the map \eq{qem} and
$U=U_T$, the composite map corresponds to the PVM $P^{\hat{Q}_T}$,
with $ \hat{Q}_T= U^*\hat{Q} U $, the Heisenberg position
(configuration) at time $T$, since $ U_T^* P^{\hat{Q}} U_T = P^{U_T^*
   \hat{Q} U_T } $.

\section{The General Emergence of Operators}\label{secGEO}\label{5}
\setcounter{equation}{0} \label{GEBM}

For \BM\ POVMs emerge naturally, not for discrete experiments, but for
a general experiment (\ref{eq:generalexperiment}).  To see how this
comes about consider the probability measure (\ref{eq:indumas}) giving
the probability distribution of the result ${Z}= F(Q_T)$ of the
experiment, where $Q_T$ is the final configuration of system and
apparatus and $F$ is the calibration function expressing the numerical
result, for example the orientation $\Theta$ of a pointer. Then the
map
\begin{equation}
\psi \mapsto \rho^{Z}_{\psi}  = \rho_{\Psi_{T}} \circ F^{-1},
\label{eq:basquamap}
\end{equation}
{}from the initial \wf{} of the system to the probability distribution
of the result, is quadratic since it arises {}from the sequence of
maps
\begin{equation}
\psi \mapsto
\Psi = \psi \otimes \Phi_0 \mapsto
   \Psi_T = U( \psi \otimes \Phi_0) \mapsto \rho_{\Psi_{T}}(dq) =
\Psi_T^{*} \Psi_T dq
\mapsto \rho^{Z}_{\psi} = \rho_{\Psi_{T}} \circ F^{-1},
\label{seqmap}
\end{equation}
where the middle map, to the \qe\ distribution, is obviously
quadratic, while all the other maps are linear, all but the second
trivially so.  Now, by (\ref{def:mvqm}), the notion of such a
quadratic map (\ref{eq:basquamap}) is completely equivalent to that of
a POVM on the system Hilbert space \H{}.  (The sesquilinear map $B$
associated with \eq{seqmap} is $B(\psi_1,\psi_2)= \Psi_{1\,T}^{*}
\Psi_{2\,T} dq \circ F^{-1}$, where $\Psi_{i\,T}= U (\psi_i \otimes
\Phi_0)$.)

Thus the emergence and role of POVMs as generalized observables in
\BM\ is merely an expression of the sesquilinearity of \qe\ together
with the linearity of the \Sc{} evolution.  Thus the fact that with
every experiment is associated a POVM, which forms a compact
expression of the statistics for the possible results, is a near
mathematical triviality.  It is therefore rather dubious that the
occurrence of POVMs---the simplest case of which is that of PVMs---as
observables can be regarded as suggesting any deep truths about
reality or about epistemology.  \bigskip

An explicit formula for the POVM defined by the quadratic map
(\ref{eq:basquamap}) follows immediately {}from (\ref{seqmap}):
$$
\rho^{Z}_{\psi}(d\lam) = \langle\psi\otimes\Phi_{0}, U^{*}
P^{\hat{Q}}(F^{-1}(d\lambda)) U\, \psi\otimes\Phi_{0}\rangle =
\langle\psi\otimes\Phi_{0}, P_0U^{*} P^{\hat{Q}}(F^{-1}(d\lambda))
UP_0\, \psi\otimes\Phi_{0}\rangle
$$
where $P^{\hat{Q}}$ is the PVM for the position (configuration)
operator (\ref{eq:posiope}) and $P_{0}$ is the projection onto
$\H\ot\Phi_{0}$, whence
\begin{equation}
O(d\lam) = 1_{\Phi_0}^{-1}
P_{0}\, U^{*} P^{\hat{Q}}(F^{-1}(d\lam)) U P_0 1_{\Phi_0}\,,
\label{eq:genpovm}
\end{equation}
where $ 1_{\Phi_0}\psi = \psi\ot\Phi_0 $ is the natural identification
of \H{} with $\H\ot\Phi_0$. This is the obvious POVM reflecting the
essential structure of the experiment.\footnote{This POVM can also be
   written as
\begin{equation}
   O(d\lam) = \tr_A\left[
P_{0}\, U^{*} P^{\hat{Q}}(F^{-1}(d\lam)) U \right],
\label{eq:pto}
\end{equation}
where $\tr_A$ is the partial trace over the apparatus variables.  The
partial trace is a map $\tr_{A}\,:\, W \mapsto \tr_{A}(W)$, {}from
trace class operators on the Hilbert space $\H_{S}\otimes\H_{A}$ to
trace class operators on $\H_{S}$, uniquely defined by $ \tr_{S} (
\tr_{A}(W) B)= \tr_{S+A} (W B\otimes I)$, where $\tr_{S+A}$ and
$\tr_{S}$ are the usual (scalar-valued) traces of operators on
$\H_{S}\otimes\H_{A}$ and $\H_{S}$, respectively.  For a trace class
operator $B$ on $L^2(dx)\otimes L^2(dy)$ with kernel $B(x,y, x',y')$
we have $\tr_{A}\left(B\right) (x,x') = \int\, B(x,y, x',y) dy .$ In
\eq{eq:pto} $\tr_A$ is applied to operators that need not be trace
class---nor need the operator on the left be trace class---since,
e.g., $O(\Lambda)= I$. The formula nonetheless makes sense.  }

Note that the POVM \eq{eq:genpovm} is unitarily equivalent to
\begin{equation}
P_0 P^{F(\hat{Q}_T)}(d\lam) P_0
\label{eq:genpovmue}
\end{equation}
where $\hat{Q}_T$ is the Heisenberg configuration of system and
apparatus at time $T$. This POVM, acting on the subspace
$\H\ot\Phi_{0}$, is the projection to that subspace of a PVM, the
spectral projections for $F(\hat{Q}_T)$. Naimark has shown (see, e.g.,
\cite{Dav76}) that every POVM is equivalent to one that arises in this
way, as the orthogonal projection of a PVM to a subspace.\footnote{If
   $O(d\lam)$ is a POVM on $\Sigma$ acting on \H{}, then the Hilbert
   space on which the corresponding PVM acts is the natural Hilbert
   space associated with the data at hand, namely $L^2(\Sigma, \H,
   O(d\lam))$, the space of \H-valued functions $\psi(\lam)$ on
   $\Sigma$, with inner product given by $\int \langle\psi(\lam),
   O(d\lam)\phi(\lam)\rangle$. (If this is not, in fact, positive
   definite, then the quotient with its kernel should be
   taken---$\psi(\lam)$ should, in other words, be understood as the
   appropriate equivalence class.) Then $ O(d\lam)$ is equivalent to
   $PE(d\lam)P$, where $E(\Delta) =\hat{{\sf 1}}_{\Delta}(\lam)$,
   multiplication by ${\sf 1}_{\Delta}(\lam)$, and $P$ is the
   orthogonal projection onto the subspace of constant \H-valued
   functions $\psi(\lam)=\psi$.}

We shall now illustrate the association of POVMs with experiments by
considering some special cases of (\ref{seqmap}).

\subsection{``No Interaction'' Experiments }
\label{sec:nie}
Let $U=U_{S} \ot U_{A}$ in (\ref{seqmap}) (hereafter the indices
``$S$'' and ``$A$'' shall refer, respectively, to system and
apparatus).  Then for $F(x,y)=y$ the measure-valued quadratic map
defined by (\ref{seqmap}) is
$$
\psi\mapsto c(y) \norm\psi\norm^2 dy
$$
where $c(y) = |U_{A}\Phi_{0}|^{2}(y)$, with POVM $O_1(dy)= c(y) dy
\;I_S$, while for $F(q)=q=(x,y)$ the map is
$$
\psi \mapsto c(y)\, |U_{S}\psi|^{2}(x)\, dq$$
with corresponding
POVM $O_2(dq) = c(y)\, U_{S}^\dagger P^{\hat X}(dx)U_{S}\, dy$.
Neither $O_1$ nor $O_2$ is a PVM. However, if $F$ is independent of
$y$, $F(x,y) =F(x)$, then the apparatus can be ignored in
(\ref{seqmap}) or (\ref{eq:genpovm}) and $O= U_{S}^{*}
P^{\hat{X}}U_{S}\circ F^{-1}$, i.e.,
$$
O(d\lam) = U_{S}^{*} P^{\hat{X}} (F^{-1}(d\lam))U_{S} \, ,
$$
which is manifestly a PVM---in fact corresponding to $
F(\hat{X}_T)$, where $\hat{X}_T$ is the Heisenberg configuration of
the system at the end of the experiment.

This case is somewhat degenerate: with no interaction between system
and apparatus it hardly seems anything like a measurement.  However,
it does illustrate that it is ``true'' POVMs (i.e., those that aren't
PVMs) that typically get associated with experiments---i.e., unless
some special conditions hold (here that $F=F(x)$).

\subsection{``No $X$'' Experiments}\label{noXexp}

The map (\ref{seqmap}) is well defined even when the system (the
$x$-system) has no translational degrees of freedom, so that there is
no $x$ (or $X$). This will be the case, for example, when the system
Hilbert space $\H_S$ corresponds to the spin degrees of freedom. Then
$\H_S=\CC^n$ is finite dimensional.

In such cases, the calibration $F$ of course is a function of $y$
alone, since there is no $x$.  For $F=y$ the measure-valued quadratic
map defined by (\ref{seqmap}) is
\begin{equation}
   \psi \mapsto |
[U(\psi\otimes\Phi_{0})](y)|^{2} dy\,,
\label{eq:nox}
\end{equation}
where $|\cdots |$ denotes the norm in $\CC^n$.

This case is physically more interesting than the previous one, though
it might appear rather puzzling since until now our measured systems
have always involved configurations.  After all, without
configurations there is no Bohmian mechanics!  However, what is
relevant {}from a Bohmian perspective is that the composite of system
and apparatus be governed by \BM{}, and this may well be the case if
the apparatus has configurational degrees of freedom, even if what is
called the system doesn't. Moreover, this case provides the prototype
of many real-world experiments, e.g., spin measurements.

For the measurement of a spin component of a spin--$1/2$
particle---recall the description of the Stern-Gerlach experiment
given in Section \ref{secSGE}---we let $\H_{S}= \CC^2$, the spin
space, with ``apparatus'' configuration $y= {\bf x}$, the position of
the particle, and with suitable calibration $F({\bf x}) $.  (For a
real world experiment there would also have to be a genuine
apparatus---a detector---that measures where the particle
\emph{actually is} at the end of the experiment, but this would not in
any way affect our analysis.  We shall elaborate upon this below.)
The unitary $U$ of the experiment is the evolution operator up to time
$T$ generated by the Pauli Hamiltonian (\ref{sgh}), which under the
assumption (\ref{consg}) becomes
\begin{equation}
H = -\frac{\hbar^{2}}{2m} \boldsymbol{\nabla}^{2}  - ( b+ az) \sigma_{z}
\label{eq:pahamagain}
\end{equation}

Moreover, as in Section \ref{secSGE}, we shall assume that the initial
particle \wf{} has the form $\Phi_{0}({\bf x)}= \Phi_{0}(z)
\phi(x,y)$.\footnote{We abuse notation here in using the notation $ y
   = {\bf x} = (x,y,z)$. The $y$ on the right should of course not be
   confused with the one on the left.}  Then for $F({\bf x}) = z$ the
quadratic map (\ref{seqmap}) is
\begin{eqnarray*}
\psi &\mapsto& \left(
|\langle\psi^+, \psi\rangle|^2 |\Phi^{(+)}_{T}(z)|^{2}   +
|\langle\psi^-, \psi\rangle|^2 |\Phi^{(-)}_{T}(z)|^{2} \right)dz\\
&=& \left\langle\psi\,,\;
|\psi^+\rangle\langle\psi^+||\Phi^{(+)}_{T}(z)|^{2}
+
|\psi^-\rangle\langle\psi^-||\Phi^{(-)}_{T}(z)|^{2}\;
   \psi \right\rangle\, dz
\end{eqnarray*}
with POVM
\begin{equation}
O(dz)\; = \;
\left( \begin{array}{cc} |\Phi^{(+)}_{T}(z)|^{2} & 0 \\ 0 &
   |\Phi^{(-)}_{T}(z)|^{2} \end{array} \right) \, dz \, ,
\label{eq:povmspin}
\end{equation}
where $\psi^{\pm}$ are the eigenvectors (\ref{eq:spinbasis}) of
$\sigma_{z}$ and $\Phi^{(\pm)}_{T}$ are the solutions of
(\ref{eq:SGequ}) computed at $t=T$, for initial conditions
${\Phi_0}^{(\pm)}=\Phi_0(z)$.

Consider now the appropriate calibration for the Stern-Gerlach
experiment, namely the function
\begin{equation}
F({\bf x}) =\begin{cases} +1 & \text{if $z>0$},\\
-1& \text{if $z<0$}
\end{cases}
\label{eq:rigcal}
\end{equation}
which assigns to the outcomes of the experiment the desired numerical
results: if the particle goes up in the $z$- direction the spin is +1,
while if the particle goes down the spin is -1.  The corresponding
POVM $O_{T}$ is defined by
$$
O_{T}(+1) \,=\, \left( \begin{array}{cc} p_{T}^{+} & 0 \\ 0 &
     p_{T}^- \end{array} \right) \qquad O_{T}(-1) \,=\, \left(
    \begin{array}{cc} 1-p_{T}^{+} & 0 \\ 0 & 1-p_{T}^-
   \end{array} \right)
$$
where $$
p_{T}^+ = \int_0^\infty|{\Phi_T}^{(+)}|^2(z)dz,\qquad
p_{T}^- = \int_0^\infty |{\Phi_T}^{(-)}|^2(z)dz\, .$$

It should be noted that $O_{T}$ is not a PVM. However, as indicated in
Section \ref{secSGE}, as $T\to\infty$, $p_{T}^+\to 1$ and $p_{T}^-\to
0$, and the POVM $O_{T}$ becomes the PVM of the operator $\sigma_{z}$,
i.e., $O_{T}\to P^{\sigma_{z}}$, defined by
\begin{equation}
\label{opm}
P(+1) \,=\,
\left( \begin{array}{cc} 1 & 0 \\ 0 & 0 \end{array} \right) \qquad
P{(-1)}
\,=\, \left( \begin{array}{cc} 0 & 0 \\ 0 & 1 \end{array} \right)
\end{equation}
and the experiment becomes a measurement of the operator $\sigma_{z}$.

\subsection{``No $Y$'' Experiments}
Suppose now that the ``apparatus''involves no translational degrees of
freedom, i.e., that there is no $y$ (or $Y$). For example, suppose the
apparatus Hilbert space $\H_A$ corresponds to certain spin degrees of
freedom, with $\H_{A}= \CC^n$ finite dimensional. Then, of course,
$F=F(x)$.

This case illustrates what measurements are not.  If the apparatus has
no configurational degrees of freedom, then neither in \BM{} nor in
orthodox quantum mechanics is it a \emph{bona fide} apparatus:
Whatever virtues such an apparatus might otherwise have, it certainly
can't generate any directly observable results (at least not when the
system itself is microscopic).  According to Bohr (\cite{Boh58}, pages
73 and 90): ``Every atomic phenomenon is closed in the sense that its
observation is based on registrations obtained by means of suitable
amplification devices with irreversible functioning such as, for
example, permanent marks on the photographic plate'' and ``the
quantum-mechanical formalism permits well-defined applications only to
such closed phenomena.''  To stress this point, discussing particle
detection Bell has said~\cite{Bel80}: ``Let us suppose that a
discharged counter pops up a flag sayings `Yes' just to emphasize that
it is a macroscopically different thing {}from an undischarged
counter, in a very different region of configuration space.''

Experiments based on certain micro-apparatuses, e.g., ``one-bit
detectors'' \cite{SEW91}, provide a nice example of ``No Y''
experiments.  We may think of a one-bit detector as a spin-$1/2$-like
system (e.g., a two-level atom), with ``down'' state $\Phi_{0}$ (the
ready state) and ``up'' state $\Phi_{1}$ and which is such that its
configurational degrees of freedom can be ignored.  Suppose that this
``spin-system,'' in its ``down'' state, is placed in a small spatial
region $\Delta_1$ and consider a particle whose \wf{} has been
prepared in such a way that at $t=0$ it has the form $\psi = \psi_1 +
\psi_2$, where $\psi_{1}$ is supported by $\Delta_1$ and $\psi_{2}$ by
$\Delta_2$ disjoint {}from $\Delta_1$.  Assume that the particle
interacts locally with the spin-system, in the sense that were
$\psi=\psi_1$ the ``spin'' would flip to the ``up'' state, while were
$\psi=\psi_2$ it would remain in its ``down'' state, and that the
interaction time is negligibly small, so that other contributions to
the Hamiltonian can be ignored.  Then the initial state $\psi \ot
\Phi_0$ undergoes the unitary transformation
\begin{equation}
\label{unos}
U\,:\,\psi  \ot \Phi_0  {\to}
\Psi \,=\, \psi_{1} \ot \Phi_1  + \psi_{2} \ot\Phi_0 \,.
\end{equation}

We may now ask whether $U$ defines an experiment genuinely measuring
whether the particle is in $\Delta_{1}$ or $\Delta_{2}$.  The answer
of course is no (since in this experiment there is no apparatus
property at all with which the position of the particle could be
correlated) \emph{unless} the experiment is (quickly) completed by a
measurement of the ``spin'' by means of another (macroscopic)
apparatus.  In other words, we may conclude that the particle is in
$\Delta_{1}$ only if the spin-system in effect pops up a flag saying
``up''.

\subsection{``No $Y$ no $\Phi$'' Experiments}\label{secnoy}
Suppose there is no apparatus at all: no apparatus configuration $y$
nor Hilbert space $\H_A$, or, what amounts to the same thing,
$\H_{A}=\CC$. For calibration $F=x$ the measure-valued quadratic map
defined by (\ref{seqmap}) is $$
\psi \mapsto | U\psi(x)|^{2}\,,$$
with
POVM $U^* P^{\hat{X}}U$, while the POVM for general calibration $F(x)$
is
\begin{equation}
O(d\lam) = U^{*} P^{\hat{X}}(F^{-1}(d\lam)) U\,.
\label{eq:noYnophi}
\end{equation}
$O$ is a PVM, as mentioned in Section \ref{sec:nie}, corresponding to
the operator $U^* F(\hat{X})U= F(\hat{X}_T)$, where $\hat{X}_T$ is the
Heisenberg position (configuration) operator at time $T$.

It is important to observe that even though these experiments suffer
{}from the defect that no correlation is established between the
system and an apparatus, this can easily be remedied---by adding a
final {\it detection measurement} that measures the final actual
configuration ${X}_T$---without in any way affecting the essential
formal structure of the experiment. For these experiments the
apparatus thus does not introduce any additional randomness, but
merely reflects what was already present in ${X}_T$.  All randomness
in the final result
\begin{equation}
Z=F(X_{T})
\label{eq:finresnoy}
\end{equation}
arises {}from randomness in the initial configuration of the
system.\footnote{Though passive, the apparatus here plays an important
   role in recording the final configuration of the system.  However,
   for experiments involving detections at different times, the
   apparatus plays an active role: Consider such an experiment, with
   detections at times $t_1,\ldots,t_n$, and final result $ Z =
   F(X_{t_1}, \ldots, X_{t_n})$.  Though the apparatus introduces no
   extra randomness, it plays an essential role by changing the wave
   function of the system at the times $t_1,\ldots,t_n$ and thus
   changing the evolution of its configuration. These changes are
   reflected in the POVM structure that governs the statistical
   distribution of $Z$ for such experiments (see Section
   \ref{sec:SeqM}).}

For $F=x$ and $U=\id$ the quadratic map is $\psi \mapsto
|\psi(x)|^{2}$ with PVM $P^{\hat{X}}$, so that this (trivial)
experiment corresponds to the simplest and most basic operator of
quantum mechanics: the position operator.  How other basic operators
arise {}from experiments is what we are going to discuss next.

\subsection{The Basic Operators of Quantum  Mechanics}
\label{subsec.basop}

According to \BM{}, a particle whose \wf{} is real (up to a global
phase), for example an electron in the ground state of an atom, has
vanishing velocity, even though the \qf{} assigns a nontrivial
probability distribution to its momentum.  It might thus seem that we
are faced with a conflict between the predictions of \BM{} and those
of the quantum formalism. This, however, is not so. The quantum
predictions about momentum concern the results of an experiment that
happens to be called a momentum measurement and a conflict with \BM{}
with regard to momentum must reflect disagreement about the results of
such an experiment.

One may base such an experiment on free motion followed by a final
measurement of position.\footnote{The emergence of the momentum
   operator in such so-called time-of-flight measurements was discussed
   by Bohm in his 1952 article \cite{Boh52}.  A similar derivation of
   the momentum operator can be found in Feynman and Hibbs
   \cite{FH65}.}  Consider a particle of mass $m$ whose \wf{} at $t=0$
is $\psi=\psi({\bf x})$. Suppose no forces are present, that is, that
all the potentials acting on the particle are turned off, and let the
particle evolve freely.  Then we measure the position ${\bf X}_T$ that
it has reached at the time $t=T$.  It is natural to regard ${\bf V}_T
={\bf X}_T/T $ and ${\bf P}_T =m {\bf X}_T/T $ as providing, for large
$T$, approximations to the asymptotic velocity and momentum of the
particle. It turns out that the probability distribution of ${\bf P}_T
$, in the limit $T\to\infty$, is exactly what quantum mechanics
prescribes for the momentum, namely $|\tilde{\psi}({\bf p})|^2$, where
$$
\tilde{\psi}({\bf p}) = (\mathcal{F}\psi)({\bf
   p})=\frac{1}{\sqrt{(2\pi \hbar)^3}} \int e^{ -\frac{i}{\hbar} {\bf p
     \cdot x}} \psi( {\bf x})\, d{\bf x}
$$
is the Fourier transform of $\psi$.

This result can be easily understood: Observe that $|\psi_{T}({\bf
   x})|^{2}\, d{\bf x}$, the probability distribution of ${\bf X}_{T}$,
is the spectral measure $\mu_{\psi}^{ \hat{\bf X}_{T}} (d\,{\bf x})
=\langle\psi, P^{\hat{\bf X}_{T} }(d\,{\bf x})\,\psi\rangle\, $ of
$\,\hat{\bf X}_{T}= U_{T}^{*} \hat{\bf X} U_{T}$, the (Heisenberg)
position operator at time $t=T$; here $U_{t}$ is the free evolution
operator and $\hat{\bf X}$ is, as usual, the position operator at time
$t=0$.  By elementary quantum mechanics (specifically, the Heisenberg
equations of motion), $\hat{\bf X}_T = \frac{1}{m}\hat{\bf P}\,T\, +\,
\hat{\bf X}$, where $\hat{\bf P}\equiv -i\hbar\boldsymbol{\nabla}$ is
the momentum operator.  Thus as $T\to\infty$ the operator $ m{\hat{\bf
     X}_T}/{T}$ converges to the momentum operator $ \hat{\bf P} $,
since $ \hat{\bf X}/T $ is $ O(1/T) $, and the distribution of the
random variable ${\bf P}_{T}$ accordingly converges to the spectral
measure of $\hat {\bf P}$, given by $|\tilde{\psi}({\bf
   p})|^2$.\footnote{\label{foot:conv} This formal argument can be
   turned into a rigorous proof by considering the limit of the
   characteristic function of ${\bf P}_T $, namely of the function
   $f_{T}(\boldsymbol{\lambda})= \int e^{i \boldsymbol{\lambda} \cdot
     {\bf p} }\,\rho_T(d{\bf p})$, where $\rho_{T}$ is the distribution
   of $m{{\bf X}_T}/{ T} $: $f_{T}(\boldsymbol{\lambda})=\left\langle
     \psi,\, \exp \left(i\boldsymbol{\lambda} \cdot m{\hat{{\bf
             X}}_T}/{ T} \right) \, \psi \right\rangle $, and using the
   dominated convergence theorem \cite{RS80} this converges as
   $T\to\infty$ to $ f(\boldsymbol{\lambda})=\left\langle \psi,
     \exp\left(i \boldsymbol{\lambda \cdot} \hat{\bf
         P}\right)\psi\right\rangle$, implying the desired result.  The
   same result can also be obtained using the well known asymptotic
   formula (see, e.g., \cite{RS75}) for the solution of the free \Sc{}
   equation with initial condition $\psi =\psi({\bf x})$,
   $$
   \psi_T({\bf x}) \sim \left(\frac{m}{iT}\right)^{\frac{3}{2}} e^{i
     \frac{m{\bf x}^2}{2\hbar T}}\; \tilde{\psi}(\frac{m{\bf x}}{T})
   \quad\mbox{for}\quad T\to\infty.  $$}

The momentum operator arises {}from a ($T\to\infty$) limit of ``no $Y$
no $\Phi$'' single-particle experiments, each experiment being defined
by the unitary operator $U_{T}$ (the free particle evolution operator
up to time $T$) and calibration $ F_{T}({\bf x}) = m{\bf x}/{ T}$.
Other standard quantum-mechanical operators emerge in a similar
manner, i.e., {}from a $T\to\infty$ limit of appropriate
single-particle experiments.

This is the case, for example, for the spin operator $\sigma_{z}$. As
in Section \ref{noXexp}, consider the evolution operator $U_{T}$
generated by Hamiltonian (\ref{eq:pahamagain}), but instead of
(\ref{eq:rigcal}), consider the calibration $ F_{T}({\bf x}) = 2 m
\,z/\,a\, T^2 $. This calibration is suggested by (\ref{eq:mmsd}), as
well as by the explicit form of the $z$-component of the position
operator at time $t=T$,
\begin{equation}
\hat{Z}_T= U_{T}^{*} \hat{Z} U_{T} = \hat{Z} +
\frac{\hat{P}_z}{m}\,T + \frac{a}{2m}
\sigma_z\,T^{2}\,,
\label{eq:heispin}
\end{equation}
which follows {}from the Heisenberg equations
$$
m \frac{d^{2} \hat{Z}_{t}}{d\,t^2} = a\,\sigma_{z} \,, \qquad
\left.\frac{d\, \hat{Z}_{t}}{d\,t}\right|_{t=0}\!  = \hat{P}_{z}
\equiv -i\hbar\frac{\partial}{\partial z}\,,\qquad
\hat{Z}_{0}=\hat{Z}\,.
$$
Then, for initial state $\Psi = \psi\otimes\Phi_0$ with suitable $
\Phi_0 $, where $\psi= \alpha \psi^{(+)}\,+\, \beta\psi^{(-)}$, the
distribution of the random variable
$$
{\Sigma_{z}}_{T} = F_T({\bf X}_T) = \frac{2\,m \, Z_T }{a\, T^2}$$
converges as $T\to\infty$ to the spectral measure of $\sigma_z$, with
values $+1$ and $-1$ occurring with probabilities $|\a|^{2}$ and
$|\beta|^{2}$, respectively.\footnote{For the Hamiltonian
   \eq{eq:pahamagain} no assumption on the initial state $\Psi$ is
   required here; however \eq{eq:pahamagain} will be a reasonably good
   approximation only when $\Psi$ has a suitable form, expressing in
   particular that the particle is appropriately moving towards the
   magnet.}  This is so, just as with the momentum, because as
$T\to\infty$ the operator $ \frac{2\,m \, \hat{Z}_T }{a\, T^2}$
converges to $\sigma_{z}$.  \bigskip

We remark that we've made use above of the fact that simple algebraic
manipulations on the level of random variables correspond
automatically to the same manipulations for the associated operators.
More precisely, suppose that
\begin{equation}
Z \mapsto A
\label{eq:assrvop}
\end{equation}
in the sense (of \eq{eq:prdeltan}) that the distribution of the random
variable $Z$ is given by the spectral measure for the self-adjoint
operator $A$. Then it follows {}from (\ref{eq:prfm}) that
\begin{equation}
f(Z) \to f(A)
\label{eq:funcom}
\end{equation}
for any (Borel) function $f$.  For example, since ${\bf X}_T\mapsto
\hat{\bf X}_T$, $m{\bf X}_T/T\mapsto m\hat{\bf X}_T/T$, and since $Z_T
\to \hat{Z}_T$, $ \frac{2\,m \, {Z}_T }{a\, T^2} \to \frac{2\,m \,
   \hat{Z}_T }{a\, T^2}$. Similarly, if a random variable $P\mapsto
\hat{P}$, then $ P^2/(2m)\mapsto H_0= \hat{P}^2/(2m)$. This is rather
trivial, but it is not as trivial as the failure even to distinguish
$Z$ and $\hat{Z}$ would make it seem.

\subsection{From Positive-Operator-Valued Measures to
   Experiments}\label{subsec.fpovtoex} We wish here to point out that
to a very considerable extent the association $\E\mapsto O(d\lam)$ of
experiments with POVMs is onto. It is more or less the case that every
POVM arises {}from an experiment.

We have in mind two distinct remarks. First of all, it was pointed out
in the first paragraph of Section 4.3 that every discrete POVM $O_\a$
(weak formal experiment) arises {}from some discrete experiment \E{}.
Thus, for every POVM $O(d\lam)$ there is a sequence $\E^{(n)}$ of
discrete experiments for which the corresponding POVMs $O^{(n)}$
converge to $O$.

The second point we wish to make is that to the extent that every PVM
arises {}from an experiment $\E=\{\Phi_0,U, F\}$, so too does every
POVM.  This is based on the fact, mentioned at the end of the
introduction to Section 5, that every POVM $O(d\lam)$ can be regarded
as arising {}from the projection of a PVM $E(d\lam)$, acting on
$\H^{(1)}$, onto the subspace $\H\subset\H^{(1)}$. We may assume
without loss of generality that both $\H$ and $\H^{(1)}\ominus\H$ are
infinite dimensional (by some otherwise irrelevant enlargements if
necessary). Thus we can identify $\H^{(1)}$ with
$\H\ot\H_{\text{apparatus}^{(1)}}$ and the subspace with
$\H\ot\Phi_0^{(1)}$, for any choice of $\Phi_0^{(1)}$.  Suppose now
that there is an experiment $\E^{(1)}=\{\Phi_0^{(2)},U, F\}$ that
measures the PVM $E$ (i.e., that measures the observable $A=\int
\lam{} E(d\lam{})$) where $\Phi_0^{(2)}\in
\H_{\text{apparatus}^{(2)}}$, $U$ acts on $\H\ot\H_{\text{apparatus}}$
where $\H_{\text{apparatus}}= \H_{\text{apparatus}^{(1)}}\ot
\H_{\text{apparatus}^{(2)}}$ and $F$ is a function of the
configuration of the composite of the 3 systems: system,
apparatus$^{(1)}$ and apparatus$^{(2)}$.  Then, with $\Phi_0=
\Phi_0^{(1)}\ot \Phi_0^{(2)}$, $\E=\{\Phi_0,U, F\}$ is associated with
the POVM $O$.

\subsection{Invariance Under Trivial Extension}\label{subsec:iute}

Suppose we change an experiment $\E$ to $\E'$ by regarding its
$x$-system as containing more of the universe that the $x$-system for
$\E$, without in any way altering what is physically done in the
experiment and how the result is specified. One would imagine that
$\E'$ would be equivalent to $\E$. This would, in fact, be trivially
the case classically, as it would if $\E$ were a genuine measurement,
in which case $\E'$ would obviously measure the same thing as $\E$.
This remains true for the more formal notion of measurement under
consideration here.  The only source of nontriviality in arriving at
this conclusion is the fact that with $\E'$ we have to deal with a
different, larger class of initial \wf s.

We will say that $\E'$ is a trivial extension of $\E$ if the only
relevant difference between $\E$ and $\E'$ is that the $x$-system for
$\E'$ has generic configuration $x'=(x,\hat{x})$, whereas the
$x$-system for $\E$ has generic configuration $x$. In particular, the
unitary operator $U'$ associated with $\E'$ has the form $U'=
U\ot\hat{U}$, where $U$ is the unitary associated with \E{},
implementing the interaction of the $x$-system and the apparatus,
while $\hat{U}$ is a unitary operator describing the independent
evolution of the $\hat{x}$-system, and the calibration $F$ for $\E'$
is the same as for $\E$. (Thus $F$ does not depend upon $\hat{x}$.)

The association of experiments with (generalized) observables (POVMs)
is \emph{invariant under trivial extension}: if $\E\mapsto O$ in the
sense of \eq{etoo} and $\E'$ is a trivial extension of $\E$, then
$\E'\mapsto O\ot I$, where $I$ is the identity on the Hilbert space of
the $\hat{x}$-system.

To see this note that if $\E\mapsto O$ then the sesquilinear map $B$
arising {}from \eq{seqmap} for $\E'$ is of the form
$$
B(\psi_1\ot \hat{\psi}_1, \psi_2\ot \hat{\psi}_2) = \langle\psi_1,
O\psi_2\rangle \langle\hat{\psi}_1, \hat{\psi}_2\rangle$$
on product
\wf{}s $\psi'= \psi\ot \hat{\psi}$, which easily follows {}from the
form of $U'$ and the fact that $F$ doesn't depend upon $\hat{x}$, so
that the $\hat{x}$-degrees of freedom can be integrated out. Thus the
POVM $O'$ for $\E'$ agrees with $ O\ot I$ on product \wf{}s, and since
such wave functions span the Hilbert space for the
$(x,\hat{x})$-system, we have that $O'= O\ot I$. Thus $\E'\mapsto O\ot
I$.

In other words, if \E{}{} is a measurement of $O$, then $\E'$ is a
measurement of $O\ot I$. In particular, if \E{} is a measurement the
self-adjoint operator $A$, then $\E'$ is a measurement of $A\ot I$.
This result is not quite so trivial as it would be were it concerned
with genuine measurements, rather than with the more formal notion
under consideration here.

Now suppose that $\E'$ is a trivial extension of a discrete experiment
$\E$, with state transformations given by $\Aa$. Then the state
transformations for $\E{}'$ are given by $\Aa' = \Aa \ot \hat{U}$.
This is so because $\Aa'$ must agree with $\Aa \ot \hat{U}$ on product
\wf{}s $\psi'= \psi\ot \hat{\psi}$, and these span the Hilbert space
of the $(x,\hat{x}$)-system.

\subsection{POVMs and the Positions of Photons and Dirac Electrons}
We have indicated how POVMs emerge naturally in association with
Bohmian experiments. We wish here to indicate a somewhat different
role for a POVM: to describe the probability distribution of the
actual (as opposed to measured\footnote{The accurate measurement of
   the position of a Dirac electron is presumably impossible.})
position. The probability distribution of the position of a Dirac
electron in the state $\psi$ is $\psi^+\psi$. This is given by a PVM
$E(d{\bf x})$ on the one-particle Hilbert space $\H$ spanned by
positive and negative energy electron wave functions. However the
physical one-particle Hilbert-space $\H_+$ consists solely of positive
energy states, and this is not invariant under the projections $E$.
Nonetheless the probability distribution of the position of the
electron is given by the POVM $P_+ E(d{\bf x}) P_+$ acting on $\H_+$,
where $P_+$ is the orthogonal projection onto $\H_+$.  Similarly,
constraints on the photon wave function require the use of POVMs for
the localization of photons~\cite{Kraus, Emch}.\footnote{For example,
   on the one-photon level, both the proposal
   $\boldsymbol{\Psi}=\mathbf{E}+i \mathbf{B}$ (where $\mathbf{E}$ and
   $\mathbf{B}$ are the electric and the magnetic free fields)
   \cite{Birula}, and the proposal $\boldsymbol{\Psi}=\mathbf{A}$
   (where $\mathbf{A}$ is the vector potential in the Coulomb gauge)
   \cite{Emch}, require the constraint $\boldsymbol{\nabla}\cdot
   \boldsymbol{\Psi}=0$.}

\section{Density Matrices} \setcounter{equation}{0}

The notion of a density matrix, a positive (trace class) operator with
unit trace on the Hilbert space of a system, is often regarded as
providing the most general characterization of a quantum state of that
system.  According to the quantum formalism, when a system is
described by the density matrix $W$, the expected value of an
observable $A$ is given by $ \tr(WA)$.  If $A$ has PVM $O$, and more
generally for any POVM $O$, the probability that the (generalized)
observable $O$ has value in $\Delta$ is given by
   \begin{equation}
\pro(O\in\Delta) = \tr (W O(\Delta)).
\label{eq:den1}
   \end{equation}
   A density matrix that is a one-dimensional projection, i.e., of the
   form $|\psi\rangle\langle\psi|$ where $\psi$ is a unit vector in the
   Hilbert space of the system, describes a \emph{pure state} (namely,
   $\psi$), and a general density matrix can be decomposed into a
   \emph{mixture} of pure states $\psi_{k}$,
\begin{equation}
W =\sum_k p_k |\psi_k\rangle\langle\psi_k| \qquad\mbox{where}\qquad
\sum_{k} p_{k} =1.
\label{eq:dmsd}
\end{equation}

Naively, one might regard $p_{k}$ as the probability that the system
\emph{is} in the state $\psi_{k}$.  This interpretation is, however,
untenable, for a variety of reasons. First of all, the decomposition
\eq{eq:dmsd} is not unique. A density matrix $W$ that does not
describe a pure state can be decomposed into pure states in a variety
of different ways.

It is always possible to decompose a density matrix $W$ in such a way
that its components $\psi_k$ are orthonormal. Such a decomposition
will be unique except when $W$ is degenerate, i.e., when some $p_k$'s
coincide. For example, if $p_1=p_2$ we may replace $\psi_{1}$ and
$\psi_{2}$ by any other orthonormal pair of vectors in the subspace
spanned by $\psi_{1}$ and $\psi_{2}$. And even if $W$ were
nondegenerate, it need not be the case that the system is in one of
the states $\psi_k$ with probability $p_k$, because for any
decomposition \eq{eq:dmsd}, regardless of whether the $\psi_k$ are
orthogonal, if the \wf{} of the system were $\psi_k$ with probability
$p_k$, this situation would be described by the density matrix $W$.

Thus a general density matrix carries no information---not even
statistical information---about the actual \wf{} of the system.
Moreover, a density matrix can describe a system that has no wave
function at all!  This happens when the system is a subsystem of a
larger system whose \wf{} is entangled, i.e., does not properly
factorize (in this case one usually speaks of the reduced density
matrix of the subsystem).

This impossibility of interpreting density matrices as real mixtures
of pure states has been regarded by many authors (e.g., von Neumann
\cite{vNe55} and Landau \cite{LL}) as a further indication that
quantum randomness is inexplicable within the realm of classical logic
and probability.  However, {}from the point of view of Bohmian
mechanics, there is nothing mysterious about density matrices.
Indeed, their role and status within the quantum formalism can be
understood very easily in terms of the general framework of
experiments of Section \ref{GEBM}. (It can, we believe, be reasonably
argued that even {}from the perspective of orthodox quantum theory,
density matrices can be understood in a straightforward way.)

\subsection{Density Matrices and Bohmian Experiments}
\label{secRWF}

Consider a general experiment $\E\mapsto O$ (see equation \eq{etoo})
and suppose that the initial \wf{} of the system is random with
probability distribution $p (d\psi)$ (on the set of unit vectors in
\H).  Then nothing will change in the general argument of Section
\ref{GEBM} except that now $\rho^Z_\psi$ in \eq{etoo} and \eq{seqmap}
should be interpreted as the conditional probability {\it given}
$\psi$.  It follows then {}from (\ref{eq:den1}), using the fact that
$\langle \psi , O (\Delta) \psi \rangle = \tr (|\psi \rangle \langle
\psi| \, O(\Delta) ) $, that the probability that the result of \E{}
lies in $\Delta$ is given by
\begin{equation}
\label{stcon} \int p(d \psi )\,\langle \psi , O
(\Delta) \psi \rangle = \tr\left( \int p(d\psi )\,| \psi \rangle
\langle \psi| \, O(\Delta)\right)= \tr\left( W
O(\Delta)\right)
\end{equation}
where\footnote{Note that since $p$ is a probability measure on the
   unit sphere in $\H$, $W$ is a positive trace class operator with
   unit trace.}
   \begin{equation}
W\equiv \int p(d\psi )\,| \psi \rangle \langle \psi |
\label{eq:ensdm}
   \end{equation}
   is the \emph{ensemble density matrix} arising {}from a random wave
   function with (ensemble) distribution~$p$.

   Now suppose that instead of having a random \wf{}, our system has no
   \wf{} at all because it is entangled with another system.  Then
   there is still an object that can naturally be regarded as the state
   of our system, an object associated with the system itself in terms
   of which the results of experiments performed on our system can be
   simply expressed. This object is a density matrix $W$ and the
   results are governed by \eq{eq:den1}. $W$ is the \emph{reduced
     density matrix} arising {}from the state of the larger system.
   This is more or less an immediate consequence of invariance under
   trivial extension, described in Section \ref{subsec:iute}:

   Consider a trivial extension $\E'$ of an experiment $\E\mapsto O$ on
   our system---precisely what we must consider if the larger system
   has a \wf{} $\psi'$ while our (smaller) system does not. The
   probability that the result of $\E'$ lies in $\Delta$ is given by
\begin{equation}
\label{stcon2}
\langle \psi' , O(\Delta)\otimes I \psi' \rangle = \tr ' \left( |
\psi' \rangle \langle \psi'| \, O(\Delta)\otimes I\right)=
\tr\left( W O(\Delta)\right) \, ,
\end{equation}
where $\tr'$ is the trace for the $x'$-system (the big system) and
$\tr$ is the trace for the $x$-system.  In agreement with standard
quantum mechanics, the last equality of (\ref{stcon2}) defines $W$ as
the reduced density matrix of the $x$-system, i.e,
\begin{equation}
W\equiv \widehat{\tr}\left( | \psi' \rangle \langle \psi' |\right)
\label{eq:reddm}
\end{equation}
where $\widehat{\tr}$ denotes the partial trace over the coordinates
of the $\hat{x}$-system.

\subsection{Strong Experiments and Density Matrices}\label{secSEI}

A strong formal experiment $\mathcal{E}\equiv\{\la, \Aa\}$ generates
state transformations $\psi\to\Aa\psi$. This suggests the following
action on an initial state described by a density matrix $W$: When the
outcome is $\a$, we have the transformation
\begin{equation}
W \to \frac{\mathcal{R}_\a W}{\tr\left(\mathcal{R}_\a W\right) }
    \equiv \frac{\Aa W \Aad}{\tr\left( \Aa W \Aad \right)}
\label{eq:axdens}
\end{equation}
where
\begin{equation}
\mathcal{R}_\a W = \Aa W \Aad\,.
\label{eq:axdens2}
\end{equation}
After all, (\ref{eq:axdens}) is a density matrix naturally associated
with $\Aa$ and $W$, and it agrees with $\psi\to\Aa\psi$ for a pure
state, $W=| \psi \rangle \langle\psi|$. In order to show that
(\ref{eq:axdens}) is indeed correct, we must verify it for the two
different ways in which our system might be assigned a density matrix
$W$, i.e., for $W$ an ensemble density matrix and for $W$ a reduced
density matrix.

Suppose the initial wave function is random, with distribution
$p(d\psi)$. Then the initial state of our system is given by the
density matrix \eq{eq:ensdm}. When the outcome $\a$ is obtained, two
changes must be made in \eq{eq:ensdm} to reflect this information: $|
\psi \rangle \langle\psi|$ must be replaced by $ (\Aa| \psi \rangle
\langle\psi|\Aad)/ \norm\Aa\psi\norm^2 $, and $p(d\psi)$ must be
replaced by $p(d\psi|\a)$, the conditional distribution of the initial
\wf{} given that the outcome is $\a$. For the latter we have
$$
p(d\psi|\a)= \frac{\norm\Aa\psi\norm^2}{\tr( \Aa W\Aad)} p(d\psi)
$$
($ \norm\Aa\psi\norm^2p(d\psi)$ is the joint distribution of $\psi$
and $\a$ and the denominator is the probability of obtaining the
outcome $\a$.) Therefore $W$ undergoes the transformation
$$
W= \int p(d\psi )\,| \psi \rangle \langle \psi | \quad\to\quad \int
p(d\psi|\a) \,\frac{\Aa| \psi \rangle \langle
   \psi|\Aad}{\norm\Aa\psi\norm^2} = \int p (d\psi )\, \frac{\Aa | \psi
   \rangle \langle \psi| \Aad}{\tr( \Aa W\Aad)} = \frac{\Aa W
   \Aad}{\tr( \Aa W\Aad)} .
$$

We wish to emphasize that this demonstrates in particular the
nontrivial fact that the density matrix $\mathcal{R}_\a
W/\tr(\mathcal{R}_\a W)$ produced by the experiment depends only upon
the initial density matrix $W$. Though $W$ can arise in many different
ways, corresponding to the multiplicity of different probability
distributions $p(d\psi)$ yielding $W$ via \eq{eq:ensdm}, insofar as
the final state is concerned, these differences don't matter.

This does not, however, establish \eq{eq:axdens} when $W$ arises not
{}from a random wave function but as a reduced density matrix. To deal
with this case we consider a trivial extension $\E'$ of a discrete
experiment \E{} with state transformations $\Aa$. Then $\E'$ has state
transformations $\Aa\ot \hat{U}$ (see Section \ref{subsec:iute}).
Thus, when the initial state of the $x'$-system is $\psi'$, the final
state of the $x$-system is given by the partial trace
$$
\frac{\widehat{\tr} \left(\Aa\otimes \hat{U}| \psi '\rangle \langle
     \psi' |\Aad\otimes \hat{U}^{*}\right)}{\tr'\left(\Aa\otimes
     \hat{U}| \psi '\rangle \langle \psi' |\Aad\otimes
     \hat{U}^{*}\right)} = \frac{\widehat{\tr} \left(\Aa\otimes I| \psi
     '\rangle \langle \psi' |\Aad\otimes I\right)}{\tr'
   \left(\Aa\otimes I| \psi '\rangle \langle \psi' |\Aad\otimes
     I\right)} =\frac{\Aa \, \widehat{\tr} (| \psi' \rangle \langle
   \psi '|) \Aad}{\tr\!\left(\Aa \widehat{\tr} (| \psi' \rangle \langle
     \psi '|) \Aad\right)}$$
$$= \frac{\Aa W \Aad}{\tr\!\left(\Aa W \Aad\right)} \, ,
$$
where the cyclicity of the trace has been used.

To sum up, when a strong experiment $\mathcal{E}\equiv\{\la, \Aa\}$ is
performed on a system described by the initial density matrix $W$ and
the outcome $\a$ is obtained, the final density matrix is given by
(\ref{eq:axdens}); moreover, {}from the results of the previous
section it follows that the outcome $\a$ will occur with probability
\begin{equation}
p_{\a}= \tr( W O_{\a})= \tr\left( W \Aad\Aa\right) = \tr\left(
\mathcal{R}_\a W \right),
\label{eq:pexdm}
\end{equation}
where the last equality follows {}from the cyclicity of the trace.

\subsection{The Notion of Instrument}

We shall briefly comment on the relationship between the notion of
strong formal experiment and that of \emph{instrument} (or
\emph{effect}) discussed by Davies \cite{Dav76}.

Consider an experiment $\mathcal{E}\equiv\{\la, \Aa\}$ on a system
with initial density matrix $W$.  Then a natural object associated
with $\Ex$ is the set function
\begin{equation}
\mathcal{R}(\Delta) W \equiv\sum_{\lambda_\alpha \in
\Delta}\mathcal{R}_\a W =\sum_{\la\in \Delta}\Aa W\Aad \, .
\label{eq:ins}
\end{equation}
The set function $\mathcal{R}: \Delta \mapsto \mathcal{R} (\Delta)$
compactly expresses both the statistics of \Ex\ for a general initial
system density matrix $W$ and the effect of \Ex\ on $W$
\emph{conditioned} on the occurrence of the event ``the result of
\Ex{} is in $\Delta$''.

To see this, note first that it follows {}from \eq{eq:pexdm} that the
probability that the result of the experiment lies in the set $\Delta$
is given by
$$
p(\Delta)= \tr\left(\mathcal{R}(\Delta) W \right)\,. $$
The
conditional distribution $p(\a|\Delta)$ that the outcome is $\a$ given
that the result $\la\in\Delta$ is then $\tr(\mathcal{R}_\a W)/ \tr(
\mathcal{R}(\Delta) W )$. The density matrix that reflects the
knowledge that the result is in $\a$, obtained by averaging
\eq{eq:axdens} over $\Delta$ using $p(\a|\Delta)$, is thus
$\mathcal{R}(\Delta) W / \tr (\mathcal{R}(\Delta) W )$.

It follows {}from (\ref{eq:ins}) that $\mathcal{R}$ is a countably
additive set function whose values are positive preserving linear
transformations in the space of trace-class operators in \H. Any map
with these properties, not necessarily of the special form
(\ref{eq:ins}), is called an \emph{instrument}.

\subsection {On the State Description Provided by Density Matrices} So
far we have followed the standard terminology and have spoken of a
density matrix as describing the {\it state} of a physical system.  It
is important to appreciate, however, that this is merely a frequently
convenient way of speaking, for \BM{} as well as for \oqt{}.  Insofar
as \BM{} is concerned, the significance of density matrices is neither
more nor less than what is implied by their role in the quantum
formalism as described in Sections \ref{secRWF} and \ref{secSEI}.
While many aspects of the notion of (effective) \wf\ extend to density
matrices, in particular with respect to weak and strong experiments,
density matrices lack the dynamical implications of \wf{}s for the
evolution of the configuration, a point that has been emphasized by
Bell \cite{Bel80}:
\begin{quotation}\setlength{\baselineskip}{12pt}\noindent
   In the de Broglie-Bohm theory a fundamental significance is given to
   the wave function, and it cannot be transferred to the density
   matrix.  \ldots Of course the density matrix retains all its usual
   practical utility in connection with quantum statistics.
\end{quotation}
That this is so should be reasonably clear, since it is the \wf{} that
determines, in \BM{}, the evolution of the configuration, and the
density matrix of a system does not determine its \wf{}, even
statistically. To underline the point we shall recall the analysis of
Bell \cite{Bel80}: Consider a particle described by a density matrix
$W_t$ evolving autonomously, so that $W_{t} =U_{t}W_{0}U_{t}^{-1}$,
where $U_{t}$ is the unitary group generated by a \Sc{} Hamiltonian.
Then $ \rho^{W_{t}}(x) \equiv W_{t}(x,x)\equiv \langle x| W_{t}|
x\rangle $ gives the probability distribution of the position of the
particle.  Note that $\rho^{W}$ satisfies the continuity equation
$$
\frac{\partial \rho^W}{\partial t} + \hbox{\rm div}\, J^W
\,=\,0\qquad\mbox{where}\qquad J^{W} (x) = \frac{\hbar}{m}{\rm Im}\,
\left[ \nabla_x W(x,x')\right]_{x'=x}.
$$
This might suggest that the velocity of the particle should be
given by $ v =J^W /\rho^W $, which indeed agrees with the usual
formula when $W$ is a pure state ($W(x,x') = \psi (x) \psi^*(x')$).
However, this extension of the usual formula to arbitrary density
matrices, though mathematically ``natural,'' is not consistent with
what \BM\ prescribes for the evolution of the configuration. Consider,
for example, the situation in which the \wf\ of a particle is random,
either $\psi_1$ {\it or } $\psi_2$, with equal probability.  Then the
density matrix is $ W(x,x') = \frac12\left( \psi_1(x) \psi_1^* (x')+
   \psi_2(x)\psi_2^*(x')\right) $. But the velocity of the particle
will be always {\it either} $v_1$ or $v_2$ (according to whether the
actual \wf{} is $\psi_1$ or $\psi_2$), and---unless $\psi_1$ and
$\psi_2$ have disjoint supports---this does not agree with $J^W /
\rho^W $, an average of $v_1$ and $v_2$.

What we have just said is correct, however, only when spin is ignored.
For particles with spin a novel kind of density matrix emerges, a {\em
   conditional density matrix}, analogous to the conditional wave
function \eq{eq:con} and with an analogous dynamical role: Even though
no conditional wave function need exist for a system entangled with
its environment when spin is taken into account, a conditional density
matrix $W$ always exists, and is such that the velocity of the system
is indeed given by $ J^W /\rho^W $. See \cite{Rodiden} for details.

A final remark: the statistical role of density matrices is basically
different {}from that provided by statistical ensembles, e.g, by Gibbs
states in classical statistical mechanics. This is because, as
mentioned earlier, even when it describes a random \wf{} via
\eq{eq:ensdm}, a density matrix $W$ does not determine the ensemble
$p(d\psi)$ {}from which it emerges. The map defined by
(\ref{eq:ensdm}) {}from probability measures $p$ on the unit sphere in
\H{} to density matrices $W$ is many-to-one.\footnote{This is relevant
   to the foundations of quantum statistical mechanics, for which the
   state of an isolated thermodynamic system is usually described by
   the microcanonical density matrix $\mathcal{Z}^{-1} \delta ( H-E)$,
   where $\mathcal{Z}=\tr \delta ( H-E)$ is the partition function.
   Which ensemble of \wf s should be regarded as forming the
   thermodynamic ensemble?  A natural choice is the uniform measure on
   the subspace $H=E$, which should be thought of as fattened in the
   usual way. Note that this choice is quite distinct {}from another
   one that people often have in mind: a uniform distribution over a
   basis of energy eigenstates of the appropriate energy.  Depending
   upon the choice made, we obtain different notions of typical
   equilibrium \wf{}.} Consider, for example, the density matrix
$\frac{1}{n} \id $ where $\id$ is the identity operator on an
$n$-dimensional Hilbert space \H{}.  Then a uniform distribution over
the vectors of any given orthonormal basis of \H{} leads to this
density matrix, as well as does the continuous uniform measure on the
sphere $\|\psi\|=1$.  However, since the statistical distribution of
the results of any experiment depends on $p$ only through $W$,
different $p$'s associated with the same $W$ are {\it empirically
   equivalent} in the sense that they can't be distinguished by
experiments performed on a system prepared somehow in the state $W$.

\section{Genuine Measurements}
\label{secMO}
\setcounter{equation}{0}

We have so far discussed various interactions between a system and an
apparatus relevant to the quantum measurement formalism, {}from the
very special ones formalized by ``ideal measurements'' to the general
situation described in section 5. It is important to recognize that
nowhere in this discussion was there any implication that anything was
actually being measured. The fact that an interaction with an
apparatus leads to a pointer orientation that we call the result of
the experiment or ``measurement'' in no way implies that this result
reflects anything of significance concerning the system under
investigation, let alone that it reveals some preexisting property of
the system---and this is what is supposed to be meant by the word
measurement. After all \cite{Sch35}, ``any old playing around with an
indicating instrument in the vicinity of another body, whereby at any
old time one then takes a reading, can hardly be called a measurement
of this body,'' and the fact the experiment happens to be associated,
say, with a self-adjoint operator in the manner we have described, so
that the experiment is spoken of, in the quantum formalism, as a
measurement of the corresponding observable, certainly offers little
support for using language in this way.

We shall elaborate on this point later on. For now we wish to observe
that the very generality of our analysis, particularly that of section
5, covering as it does all possible interactions between system and
apparatus, covers as well those particular situations that in fact are
genuine measurements. This allows us to make some definite statements
about what can be measured in Bohmian mechanics.

For a physical quantity, describing an objective property of a system,
to be measurable means that it is possible to perform an experiment on
the system that measures the quantity, i.e., an experiment whose
result conveys its value.  In \BM\ a physical quantity $\xi$ is
expressed by a function
\begin{equation} {\xi}= f (X, \psi) \label{pq}
\end{equation}
of the complete state $(X, \psi)$ of the system.  An experiment \E\
measuring $\xi$ is thus one whose result ${Z}=F(X_T,Y_T)\equiv
{Z}(X,Y,\Psi)$ equals $\xi=f(X,\psi)\equiv {\xi}(X,\psi)$,
\begin{equation}
{Z}(X,Y,\Psi)={\xi}(X,\psi),\label{xz}
\end{equation}
where $X$, $Y$, $\psi$ and $\Psi$ refer, as in Section 5, to the
initial state of system and apparatus, immediately prior to the
measurement, and where the equality should be regarded as approximate,
holding to any desired degree of accuracy.

The most basic quantities are, of course, the state components
themselves, namely $X$ and $\psi$, as well as the velocities
\begin{equation}\label{velox}{{\bf v}_k} = \frac{\hbar}{m_k}{\rm
Im}\frac{{\boldsymbol{\nabla}_k}\psi(X)}{\psi(X)}
\end{equation}
of the particles. One might also consider quantities describing the
future behavior of the system, such as the configuration of an
isolated system at a later time, or the time of escape of a particle
{}from a specified region, or the asymptotic velocity discussed in
Section \ref{subsec.basop}. (Because the dynamics is deterministic,
all of these quantities are functions of the initial state of the
system and are thus of the form (\ref{pq}).)

We wish to make a few remarks about the measurability of these
quantities.  In particular, we wish to mention, as an immediate
consequence of the analysis at the beginning of Section 5, a condition
that must be satisfied by any quantity if it is to be measurable.

\subsection{A Necessary Condition for Measurability}

Consider any experiment \E\ measuring a physical quantity $\xi$. We
showed in Section 5 that the statistics of the result $Z$ of \E\ must
be governed by a POVM, i.e., that the probability distribution of $Z$
must be given by a measure-valued quadratic map on the system Hilbert
space \H.  Thus, by (\ref{xz}),
\begin{equation}
\mbox{%
\begin{minipage}{0.85\textwidth}\openup 1.4\jot
   \setlength{\baselineskip}{12pt}\emph{$\xi$ is measurable only if its
     probability distribution $\mu_{\xi}^{\psi}$ is a measure-valued
     quadratic map on \H.  }
\end{minipage}}
\label{MC1}
\end{equation}

As indicated earlier, the position ${\bf X}$ and the asymptotic
velocity or momentum ${\bf P}$ have distributions quadratic in $\psi$,
namely $\mu_{\bf X}^{\psi}(d{\bf x})=|\psi({\bf x})|^2$ and $\mu_{\bf
   P}^{\psi}(d{\bf p})=|\tilde{\psi}({\bf p})|^2$, respectively.
Moreover, they are both measurable, basically because suitable local
interactions exist to establish appropriate correlations with the
relevant macroscopic variables.  For example, in a bubble chamber a
particle following a definite path triggers a chain of reactions that
leads to the formation of (macroscopic) bubbles along the path.

The point we wish to make now, however, is simply this: the
measurability of these quantities is not a consequence of the fact
that these quantities obey this measurability condition. We emphasize
that this condition is merely a necessary condition for measurability,
and not a sufficient one. While it does follow that if $\xi$ satisfies
this condition there exists a discrete experiment that is an
approximate formal measurement of $\xi$ (in the sense that the
distribution of the result of the experiment is approximately $
\mu_{\xi}^{\psi}$), this experiment need not provide a genuine
measurement of $\xi$ because the interactions required for its
implementation need not exist and because, even if they did, the
result $Z$ of the experiment might not be related to the quantity
$\xi$ in the right way, i.e, via (\ref{xz}).

We now wish to illustrate the use of this condition, first
transforming it into a weaker but more convenient form. Note that any
quadratic map $\mu^\psi$ must satisfy
\[
\mu^{\psi_1 + \psi_2} + \mu^{\psi_1 - \psi_2} = 2(\mu^{\psi_1} +
\mu^{\psi_2})
\]
and thus if $\mu^\psi$ is also positive we have the inequality
\begin{equation}\label{ineq} \mu^{\psi_1+\psi_2} \le 2(\mu^{\psi_1} +
\mu^{\psi_2}).
\end{equation}
Thus it follows {}from \eq{MC1} that a quantity\footnote{This
   conclusion is also a more or less direct consequence of the
   linearity of the Schr\"odinger evolution: If
   $\psi_i\otimes\Phi_0\mapsto\Psi_i$ for all $i$, then
   $\sum\psi_i\otimes\Phi_0\mapsto\sum\Psi_i$. But, again, our purpose
   here has been mainly to illustrate the use of the measurability
   condition itself.}
\begin{equation}
\mbox{%
\begin{minipage}{0.85\textwidth}\openup 1.4\jot
   \setlength{\baselineskip}{12pt}\emph{$\xi$ must fail to be
     measurable if it has a possible value (one with nonvanishing
     probability or probability density) when the wave function of the
     system is $\psi_1+ \psi_2$ that is neither a possible value when
     the wave function is $\psi_1$ nor a possible value when the wave
     function is $\psi_2$. }
\end{minipage}}
\label{MC2}
\end{equation}
(Here neither $\psi_1$ nor $\psi_2$ need be normalized.)

\subsection{The Nonmeasurability of Velocity, Wave Function and
   Deterministic Quantities}\label{vwf}

It is an immediate consequence of \eq{MC2} that neither the velocity
nor the wave function is measurable, the latter because the value ``$
\psi_1+ \psi_2$'' is neither ``$\psi_1$'' nor ``$\psi_2$,'' and the
former because every wave function $\psi$ may be written as
$\psi=\psi_1+ \psi_2$ where $\psi_1$ is the real part of $\psi$ and
$\psi_2$ is $i$ times the imaginary part of $\psi$, for both of which
the velocity (of whatever particle) is 0.

Note that this is a very strong and, in a sense, surprising
conclusion, in that it establishes the {\it impossibility} of
measuring what is, after all, a most basic dynamical variable for a
{\it deterministic} mechanical theory of particles in motion. It
should probably be regarded as even more surprising that the proof
that the velocity---or wave function---is not measurable seems to rely
almost on nothing, in effect just on the linearity of the evolution of
the \wf. However, one should not overlook the crucial role of quantum
equilibrium.

We observe that the nonmeasurability of the \wf\ is related to the
{\it impossibility of copying} the \wf. (This question arises
sometimes in the form, ``Can one clone the wave function?"
\cite{ghiraun, WoZu, ghira}.) Copying would be accomplished, for
example, by an interaction leading, for all $\psi$, {}from
$\psi\otimes\phi_0\otimes\Phi_0$ to $\psi\otimes\psi\otimes\Phi$, but
this is clearly incompatible with unitarity. We wish here merely to
remark that the impossibility of cloning can also be regarded as a
consequence of the nonmeasurability of the \wf. In fact, were cloning
possible one could---by making many copies---measure the \wf\ by
performing suitable measurements on the various copies. After all, any
wave function $\psi$ is determined by $\langle \psi, A \psi\rangle$
for sufficiently many observables $A$ and these expectation values can
of course be computed using a sufficiently large ensemble.

By a deterministic quantity we mean any function ${\xi}=f(\psi)$ of
the wave function alone (which thus does not inherit any irreducible
randomness associated with the random configuration $X$).  It follows
easily {}from \eq{MC2} that no (nontrivial) deterministic quantity is
measurable.\footnote{Note also that
   $\mu_{\xi}^\psi(d\lambda)=\delta(\lambda-f(\psi)) d\lambda$ seems
   manifestly nonquadratic in $\psi$ (unless $f$ is constant).} In
particular, the mean value $\langle \psi, A \psi\rangle$ of an
observable $A$ (not a multiple of the identity) is not
measurable---though it would be were it possible to copy the wave
function, and it can of course be measured by a nonlinear experiment,
see Section \ref{secnl}.

\subsection{Initial Values and Final Values}\label{secIVFV}

Measurement is a tricky business. In particular, one may wonder how,
if it is not measurable, we are ever able to know the \wf{} of a
system---which in orthodox quantum theory often seems to be the only
thing that we do know about it.

In this regard, it is important to appreciate that we were concerned
in the previous section only with initial values, with the wave
function and the velocity {\it prior\/} to the measurement. We shall
now briefly comment upon the measurability of final values, produced
by the experiment.

The nonmeasurability argument of Section \ref{vwf} does not cover
final values.  This may be appreciated by noting that the crucial
ingredient in the analysis involves a fundamental time-asymmetry: The
probability distribution $\mu^\psi$ of the result of an experiment is
a quadratic functional of the {\it initial\/} wave function $\psi$,
not the final one---of which it is not a functional at all. Moreover,
the final velocity can indeed be measured, by a momentum measurement
as described in Section \ref{subsec.basop}.  (That such a measurement
yields also the final velocity follows {}from the formula in footnote
\ref{foot:conv} for the asymptotic wave function.) And the final wave
function can be measured by an ideal measurement of any nondegenerate
observable, and more generally by any strong formal measurement whose
subspaces $\Ha$ are one-dimensional, see Section \ref{sec:PP}: If the
outcome is $\a$, the final \wf{} is $\Aa \psi= \Aa \Pa \psi$, which is
independent of the initial \wf{} $\psi$ (up to a scalar multiple).

We also wish to remark that this distinction between measurements of
initial values and measurements of final values has no genuine
significance for passive measurements, that merely reveal preexisting
properties without in any way affecting the measured system. However,
quantum measurements are usually active; for example, an ideal
measurement transforms the wave function of the system into an
eigenstate of the measured observable.  But passive or active, a
measurement, by its very meaning, is concerned strictly speaking with
properties of a system just before its performance, i.e., with initial
values. At the same time, to the extent that any property of a system
is conveyed by a typical quantum ``measurement,'' it is a property
defined by a final value.

For example, according to orthodox quantum theory a position
measurement on a particle with a spread-out wave function, to the
extent that it measures anything at all, measures the final position
of the particle, created by the measurement, rather than the initial
position, which is generally regarded as not existing prior to the
measurement. And even in Bohmian mechanics, in which such a
measurement may indeed reveal the initial position, which---if the
measurement is suitably performed---will agree with the final
position, this measurement will still be active since the wave
function of the system must be transformed by the measurement into one
that is compatible with the sharper knowledge of the position that it
provides, see Section 2.1.

\subsection{Nonlinear Measurements and the Role of Prior Information}
\label{secnl}

The basic idea of measurement is predicated on initial ignorance. We
think of a measurement of a property of a system as conveying that
property by a procedure that does not seriously depend upon the state
of the system,\footnote{This statement must be taken with a grain of
   salt. Some things must be known about the system prior to
   measurement, for example, that it is in the vicinity the measurement
   apparatus, or that an atom whose angular momentum we wish to measure
   is moving towards the relevant Stern Gerlach magnets, as well as a
   host of similar, often unnoticed, pieces of information. This sort
   of thing does not much matter for our purposes in this paper and can
   be safely ignored. Taking them into account would introduce
   pointless complications without affecting the analysis in an
   essential way.} any details of which must after all be unknown prior
to at least some engagement with the system. Be that as it may, the
notion of measurement as codified by the quantum formalism is indeed
rooted in a standpoint of ignorance: the experimental procedures
involved in the measurement do not depend upon the state of the
measured system. And our entire discussion of measurement up to now
has been based upon that very assumption, that \E\ itself does not
depend on $\psi$ (and certainly not on $X$).

If, however, some prior information on the initial system wave
function $\psi$ were available, we could exploit this information to
measure quantities that would otherwise fail to be measurable. For
example, for a single-particle system, if we somehow knew its initial
\wf{} $\psi$ then a measurement of the initial position of the
particle would convey its initial velocity as well, via
(\ref{velox})---even though, as we have shown, this quantity isn't
measurable without such prior information.

By a nonlinear measurement or experiment $\E=\E^\psi$ we mean one in
which, unlike those considered so far, one or more of the defining
characteristics of the experiment depends upon $\psi$. For example, in
the measurement of the initial velocity described in the previous
paragraph, the calibration function $F=F^\psi$ depends upon
$\psi$.\footnote{Suppose that ${Z}_1=F_1(Q_T)=X$ is the result of the
   measurement of the initial position. Then $F^\psi=G^\psi\circ F_1$
   where $G^\psi(\cdot)= \frac{\hbar}{m}{\rm
     Im}\frac{\boldsymbol{\nabla}\psi}{\psi}(\cdot)$.}  More generally
we might have that $U=U^\psi$ or $\Phi_0=\Phi_0^\psi$.

The wave function can of course be measured by a nonlinear
measurement---just let $F^\psi\equiv\psi$. Somewhat less trivially,
the initial wave function can be measured, at least formally, if it is
known to be a member of a given orthonormal basis, by measuring any
nondegenerate observable whose eigenvectors form that basis. The
proposals of Aharonov, Anandan and Vaidman \cite{AAV93} for measuring
the wave function, though very interesting, are of this
character---they involve nonlinear measurements that depend upon a
choice of basis containing $\psi$---and thus remain
controversial.\footnote{In one of their proposals the wave function is
   ``protected'' by a procedure that depends upon the basis; in
   another, involving adiabatic interactions, $\psi$ must be a
   nondegenerate eigenstate of the Hamiltonian $H$ of the system, but
   it is not necessary that the latter be known.}

\subsection{A Position Measurement that Does not Measure  Position}
\label{secapm}
We began this section by observing that what is spoken of as a
measurement in quantum theory need not really measure anything. We
mentioned, however, that in Bohmian mechanics the position can be
measured, and the experiment that accomplishes this would of course be
a measurement of the position operator. We wish here to point out, by
means of a very simple example, that the converse is not true, i.e.,
that a measurement of the position operator need not be a measurement
of the position.

Consider the harmonic oscillator in 2 dimensions with Hamiltonian
$$H = -\frac{\hbar^2}{2m}\big( \frac{\partial^2}{\partial x^2} +
\frac{\partial^2}{\partial y^2}\big) \ + \frac{\omega^2 m}{2} (x^2
+y^2)\,.
$$
Except for an irrelevant time-dependent phase factor, the evolution
$\psi_t$ is periodic, with period $\tau =2\pi/\omega$.  The Bohm
motion of the particle, however, need not have period $\tau$. For
example, the $(n=1, m=1)$-state, which in polar coordinates is of the
form
\begin{equation}
\psi_t (r, \phi)
=\frac{m\omega}{\hbar\sqrt \pi} r e^{-\frac{m\omega}{2\hbar}r^2}
e^{i\phi}e^{-i\frac 32 \omega t},
\label{nm}
\end{equation}
generates a circular motion of the particle around the origin with
angular velocity $\hbar/(mr^2)$, and hence with periodicity depending
upon the initial position of the particle---the closer to the origin,
the faster the rotation. Thus, in general, $${\bf X}_\tau \neq {\bf
   X}_0.$$
Nonetheless, ${\bf X}_\tau$ and ${\bf X}_0 $ are identically
distributed random variables, since
$|\psi_\tau|^2=|\psi_0|^2\equiv|\psi|^2$.

We may now focus on two different experiments: Let \E\ be a
measurement of the actual position ${\bf X}_0$, the {\it initial\/}
position, and hence of the position operator, and let $\E'$ be an
experiment beginning at the same time as \E\ but in which it is the
position ${\bf X}_\tau$ at time $\tau$ that is actually measured.
Since for all $\psi$ the result of $\E'$ has the same distribution as
the result of \E, $\E'$ is also a measurement of the position
operator.  But $\E'$ is not a measurement of the initial position
since the position at time $\tau$ does not in general agree with the
initial position: A measurement of the position at time $\tau$ is not
a measurement of the position at time $0$.  Thus, while a measurement
of position is always a measurement of the position operator,
\begin{quotation}\setlength{\baselineskip}{12pt}\noindent{\it
     A measurement of the position operator is not necessarily a
     genuine measurement of position!}
\end{quotation}

\subsection{Theory Dependence of Measurement}

The harmonic oscillator example provides a simple illustration of an
elementary point that is often ignored: in discussions of measurement
it is well to keep in mind the theory under consideration. The theory
we have been considering here has been Bohmian mechanics. If, instead,
we were to analyze the harmonic oscillator experiments described above
using different theories our conclusions about results of measurements
would in general be rather different, even if the different theories
were empirically equivalent. So we shall analyze the above experiment
$\E'$ in terms of various other formulations or interpretations of
quantum theory.

In strict orthodox quantum theory there is no such thing as a genuine
particle, and thus there is no such thing as the genuine position of a
particle. There is, however, a kind of operational definition of
position, in the sense of an experimental setup, where a measurement
device yields results the statistics of which are given by the
position operator.

In naive orthodox quantum theory one does speak loosely about a
particle and its position, which is thought of---in a somewhat
uncritical way---as being revealed by measuring the position operator.
Any experiment that yields statistics given by the position operator
is considered a genuine measurement of the particle's
position.\footnote{This, and the failure to appreciate the theory
   dependence of measurements, has been a source of unfounded
   criticisms of \BM (see \cite{ESSW92, DFGZ93f, DHS93}).}  Thus $\E'$
would be considered as a measurement of the position of the particle
at time zero.

The decoherent (or consistent) histories formulation of quantum
mechanics \cite{GMH90, Omn88, Gri84} is concerned with the
probabilities of certain coarse-grained histories, given by the
specification of finite sequences of events, associated with
projection operators, together with their times of occurrence.  These
probabilities are regarded as governing the occurrence of the
histories, regardless of whether any of the events are measured or
observed, but when they are observed, the probabilities of the
observed histories are the same as those of the unobserved histories.
The experiments \E{} and $\E'$ are measurements of single-event
histories corresponding to the position of the particle at time $0$
and at time $\tau$, respectively.  Since the Heisenberg position
operators $\hat{\bf X}_\tau =\hat{\bf X}_0$ for the harmonic
oscillator, it happens to be the case, according to the decoherent
histories formulation of quantum mechanics, that for this system the
position of the particle at time $\tau$ is the same as its position at
time $0$ when the positions are unobserved, and that $\E'$ in fact
measures the position of the particle at time $0$ (as well as the
position at time $\tau$).

The spontaneous localization or dynamical reduction models
\cite{GRW,GRP90} are versions of quantum theory in which there are no
genuine particles; in these theories reality is represented by the
wave function alone (or, more accurately, by entities entirely
determined by the \wf{}).  In these models \Sc{}'s equation is
modified by the addition of a stochastic term that causes the \wf{} to
collapse during measurement in a manner more or less consistent with
the quantum formalism. In particular, the performance of \E{} or $\E'$
would lead to a random collapse of the oscillator wave function onto a
narrow spatial region, which might be spoken of as the position of the
particle at the relevant time. But $\E'$ could not be regarded in any
sense as measuring the position at time $0$, because the localization
does not occur for $\E'$ until time $\tau$.

Finally we mention stochastic mechanics \cite{Nel85}, a theory
ontologically very similar to \BM\, in that the basic entities with
which it is concerned are particles described by their positions.
Unlike \BM{}, however, the positions evolve randomly, according to a
diffusion process. Just as with \BM{}, for stochastic mechanics the
experiment $\E'$ is not a measurement of the position at time zero,
but in contrast to the situation in \BM{}, where the result of the
position measurement at time $\tau$ determines, given the \wf{}, the
position at time zero (via the Bohmian equation of motion), this is
not so in stochastic mechanics because of the randomness of the
motion.

\section{Hidden Variables}\label{secHV}

The issue of hidden variables concerns the question of whether quantum
randomness arises in a completely ordinary manner, merely {}from the
fact that in orthodox quantum theory we deal with an incomplete
description of a quantum system. According to the hidden-variables
hypothesis, if we had at our disposal a sufficiently complete
description of the system, provided by supplementary parameters
traditionally called hidden variables, the totality of which is
usually denoted by $\lambda$, the behavior of the system would thereby
be determined, as a function of $\lambda$ (and the wave function).  In
such a hidden-variables theory, the randomness in results of
measurements would arise solely {}from randomness in the unknown
variables $\lambda$. On the basis of a variety of ``impossibility
theorems,'' the hidden-variables hypothesis has been widely regarded
as having been discredited.

Note that Bohmian mechanics is just such a hidden-variables theory,
with the hidden variables $\lambda$ given by the configuration $Q$ of
the total system. We have seen in particular that in a Bohmian
experiment, the result $Z$ is determined by the initial configuration
$Q=(X,Y)$ of the system and apparatus.  Nonetheless, there remains
much confusion about the relationship between Bohmian mechanics and
the various theorems supposedly establishing the impossibility of
hidden variables.  In this section we wish to make several comments on
this matter.

\subsection{Experiments and Random Variables}\label{seerv}

In \BM\ we understand very naturally how random variables arise in
association with experiments: the initial complete state $(Q, \Psi)$
of system and apparatus evolves deterministically and uniquely
determines the outcome of the experiment; however, as the initial
configuration $Q$ is in quantum equilibrium, the outcome of the
experiment is random.

A general experiment \E\ is then {\it always} associated a random
variable (RV) $Z$ describing its result. In other words, according to
\BM, there is a natural association
\begin{equation}\label{extrv}  \EE \mapsto  {Z}, \end{equation}
between experiments and RVs.  Moreover, whenever the statistics of the
result of \E\ is governed by a \sa\ operator $A$ on the Hilbert space
of the system, with the spectral measure of $A$ determining the
distribution of $Z$, for which we shall write $Z\mapsto A$ (see
\eq{eq:prdeltan}), \BM{} establishes thereby a natural association
between \E\ and $A$
\begin{equation}\label{extop}  \E \mapsto A . \end{equation}

While for Bohmian mechanics the result $Z$ depends in general on both
$X$ and $Y$, the initial configurations of the system and of the
apparatus, for many real-world experiments $Z$ depends only on $X$ and
the randomness in the result of the experiment is thus due solely to
randomness in the initial configuration of the system alone.  This is
most obvious in the case of genuine position measurements (for which
$Z(X,Y)= X$).  That in fact the apparatus need not introduce any extra
randomness for many other real-world experiments as well follows then
{}from the observation that the role of the apparatus in many
real-world experiments is to provide suitable background fields, which
introduce no randomness, as well as a final detection, a measurement
of the actual positions of the particles of the system.  In
particular, this is the case for those experiments most relevant to
the issue of hidden variables, such as Stern-Gerlach measurements of
spin, as well as for momentum measurements and more generally
scattering experiments, which are completed by a final detection of
position.

The result of these experiments is then given by a random variable $$
{Z}= F(X_T)= G(X)\, ,$$
where $T$ is the final time of the
experiment,\footnote{Concerning the most common of all real-world
   quantum experiments, scattering experiments, although they are
   completed by a final detection of position, this detection usually
   occurs, not at a definite time $T$, but at a random time, for
   example when a particle enters a localized detector. Nonetheless,
   for computational purposes the final detection can be regarded as
   taking place at a definite time $T$. This is a consequence of the
   flux-across-surfaces theorem \cite{dau96,det3,det2}, which
   establishes an asymptotic equivalence between flux across surfaces
   (detection at a random time) and scattering into cones (detection at
   a definite time).} on the probability space $\{ \Omega, {\P} \}$,
where $\Omega=\{ X\}$ is the set of initial configurations of the
system and ${\P}(dx)= |\psi|^2dx$ is the quantum equilibrium
distribution associated with the initial \wf\ $\psi$ of the system.
For these experiments (see Section \ref{secnoy}) the distribution of
${Z}$ is always governed by a PVM, corresponding to some \sa{}
operator $A$, $Z\mapsto A$, and thus \BM\ provides in these cases a
natural map $\E \mapsto A$.

\subsection{Random Variables, Operators, and the Impossibility
   Theorems}
\label{sec:RVOIT}
We would like to briefly review the status of the so-called
impossibility theorems for hidden variables, the most famous of which
are due to von Neumann~\cite{vNe55}, Gleason~\cite{Glea57}, Kochen and
Specker~\cite{KoSp67}, and Bell~\cite{Bel64}.  Since Bohmian mechanics
exists, these theorems can't possibly establish the impossibility of
hidden variables, the widespread belief to the contrary
notwithstanding.  What these theorems do establish, in great
generality, is that there is no ``{\it good''} map {}from \sa\
operators on a Hilbert space \H{} to random variables on a common
probability space,
\begin{equation}\label{dax} A\mapsto {\rv}\equiv {\rv}_A\, ,
\end{equation}
where ${\rv}_A={\rv}_A(\lam)$ should be thought of as the result of
``measuring $A$'' when the hidden variables, that complete the quantum
description and restore determinism, have value $\lam$.  Different
senses of ``good'' correspond to different impossibility theorems.

For any particular choice of $\lam$, say $\lam_0$, the map \eq{dax} is
transformed to a \emph{value} map
\begin{equation}\label{dax2} A\mapsto v(A)
\end{equation}
{}from \sa{} operators to real numbers (with $v(A)= {\rv}_A(\lam_0)$).
The stronger impossibility theorems establish the impossibility of a
good value map, again with different senses of ``good'' corresponding
to different theorems.

Note that such theorems are not very surprising. One would not expect
there to be a ``good'' map {}from a noncommutative algebra to a
commutative one.

One of von Neumann's assumptions was, in effect, that the map \eq{dax}
be linear. While mathematically natural, this assumption is physically
rather unreasonable and in any case is entirely unnecessary. In order
to establish that there is no good map \eq{dax}, it is sufficient to
require that the map be good in the minimal sense that the following
{\it agreement condition} is satisfied:
\begin{quotation}\setlength{\baselineskip}{12pt}{\it
     Whenever the quantum mechanical joint distribution of a set of
     \sa{} operators $(A_1,\ldots, A_m)$ exists, i.e., when they form a
     commuting family, the joint distribution of the corresponding set
     of random variables, i.e., of $(\rv_{A_1}, \ldots, \rv_{A_m})$,
     agrees with the quantum mechanical joint
     distribution.}\end{quotation}

The agreement condition implies that all deterministic relationships
among commuting observables must be obeyed by the corresponding random
variables. For example, if $A$, $B$ and $C$ form a commuting family
and $C=AB$, then we must have that $\rv_C =\rv_A\rv_B$ since the joint
distribution of $\rv_A$, $\rv_B$ and $\rv_C$ must assign probability
$0$ to the set $\{(a,b,c)\in \R^3 | c\neq ab\}$.  This leads to a
minimal condition for a good value map $A\mapsto v(A)$, namely that it
preserve functional relationships among commuting observables: For any
commuting family $A_1,\ldots, A_m$, whenever $f(A_1, \dots, A_m)=0$
(where $f:\R^m\to\R$ represents a linear, multiplicative, or any other
relationship among the $A_i$'s), the corresponding values must satisfy
the same relationship, $f(v(A_1),\ldots, v(A_m))=0 $.

The various impossibility theorems correctly demonstrate that there
are no maps, {}from self-adjoint operators to random variables or to
values, that are good, merely in the minimal senses described
above.\footnote{Another natural sense of good map $A\mapsto v(A)$ is
   given by the requirement that $v({\bf A})\in \mbox{sp}\,({\bf A})$,
   where ${\bf A}=(A_1,\ldots, A_m)$ is a commuting family, $v({\bf
     A})= (v(A_1),\ldots, v(A_m))\in \R^m$ and $\mbox{sp}\,({\bf A})$
   is the joint spectrum of the family.  That a map good in this sense
   is impossible follows {}from the fact that if ${\bf
     \a}=(\a_1,\ldots\a_m)\in \mbox{sp}\,({\bf A})$, then
   $\a_1,\ldots\a_m $ must obey all functional relationships for
   $A_1,\ldots, A_m $.}  \bigskip

We note that while the original proofs of the impossibility of a good
value map, in particular that of the Kochen-Specker theorem, were
quite involved, in more recent years drastically simpler proofs have
been found (for example, by Peres \cite{Per91}, by Greenberg, Horne,
and Zeilinger \cite{GHSZ89}, and by Mermin \cite{merm93}).

In essence, one establishes the impossibility of a good map
$A\mapsto\rv_A$ or $A\mapsto v(A)$ by showing that the $v(A)$'s, or
$\rv_A$'s, would have to satisfy impossible relationships.  These
impossible relationships are very much like the following:
$\rv_A=\rv_B=\rv_C \neq\rv_A$. However no impossible relationship can
arise for only three quantum observables, since they would have to
form a commuting family, for which quantum mechanics would supply a
joint probability distribution. Thus the quantum relationships can't
possibly lead to an inconsistency for the values of the random
variables in this case.

With four observables $A,B,C$, and $D$ it may easily happen that
$[A,B]=0$, $[B,C]=0$, $[C,D]=0$, and $[D,A]=0$ even though they don't
form a commuting family (because, say, $[A,C]\neq 0$).  It turns out,
in fact, that four observables suffice for the derivation of
impossible quantum relationships.  Perhaps the simplest example of
this sort is due to Hardy~\cite{hardy}, who showed that for almost
every quantum state for two spin 1/2 particles there are four
observables $A,B,C$, and $D$ (two of which happen to be spin
components for one of the particles while the other two are spin
components for the other particle) whose quantum mechanical pair-wise
distributions for commuting pairs are such that a good map to random
variables must yield random variables $\rv_A,\rv_B,\rv_C$, and $\rv_D$
obeying the following relationships:
\begin{itemize}
\item[(1)] The event $\{ \rv_A=1\;\mbox{and}\; \rv_B =1\}$ has
   positive probability (with an optimal choice of the quantum state,
   about $.09$).
\item[(2)] If $\{ \rv_A=1\}$ then $\{ \rv_D=1\}$.
\item[(3)] If $\{ \rv_B=1\}$ then $\{ \rv_C=1\}$.
\item[(4)] The event $\{ \rv_D=1\;\mbox{and}\; \rv_C =1\}$ has
   probability $0$.
\end{itemize}
Clearly, there exist no such random variables.

The point we wish to emphasize here, however, is that although they
are correct and although their hypotheses may seem minimal, these
theorems are nonetheless far less relevant to the possibility of a
deterministic completion of quantum theory than one might imagine. In
the next subsection we will elaborate on how that can be so. We shall
explain why we believe such theorems have little physical significance
for the issues of determinism and hidden variables.  We will
separately comment later in this section on Bell's related nonlocality
analysis \cite{Bel64}, which does have profound physical implications.

\subsection{Contextuality}\label{sec:context}

It is a simple fact there can be no map $A\mapsto {\rv}_A$, {}from
\sa{} operators on \H{} (with $\mbox{dim}\,(\H)\ge{}3$) to random
variables on a common probability space, that is good in the minimal
sense that the joint probability distributions for the random
variables agree with the corresponding quantum mechanical
distributions, whenever the latter ones are defined. But does not
\BM{} yield precisely such a map? After all, have we not emphasized
how \BM{} naturally associates with any experiment a random variable
$Z$ giving its result, in a manner that is in complete agreement with
the quantum mechanical predictions for the result of the experiment?
Given a quantum observable $A$, let $Z_A$ be then the result of a
measurement of $A$.  What gives?

Before presenting what we believe to be the correct response, we
mention some possible responses that are off-target.  It might be
objected that measurements of different observables will involve
different apparatuses and hence different probability spaces. However,
one can simultaneously embed all the relevant probability spaces into
a huge common probability space.  It might also be objected that not
all \sa{} operators can be realistically be measured. But to arrive at
inconsistency one need consider, as mentioned in the last subsection,
only $4$ observables, each of which are spin components and are thus
certainly measurable, via Stern-Gerlach experiments. Thus, in fact, no
enlargement of probability spaces need be considered to arrive at a
contradiction, since as we emphasized at the end of Section
\ref{seerv}, the random variables giving the results of Stern-Gerlach
experiments are functions of initial particle positions, so that for
joint measurements of pairs of spin components for 2-particles the
corresponding results are random variables on the common probability
space of initial configurations of the 2 particles, equipped with the
quantum equilibrium distribution determined by the initial \wf{}.

There must be a mistake. But where could it be? The mistake occurs, in
fact, so early that it is difficult to notice it. It occurs at square
one.  The difficulty lies not so much in any conditions on the map
$A\mapsto\rv_A$, but in the conclusion that \BM{} supplies such a map
at all.

What \BM{} naturally supplies is a map $\E\mapsto{}Z_{\E}$, {}from
experiments to random variables. When $Z_{\E}\mapsto{}A$, so that we
speak of \E{} as a measurement of $A$ ($\E\mapsto{}A$), this very
language suggests that insofar as the random variable is concerned all
that matters is that \E{} measures $A$, and the map
$\E\mapsto{}Z_{\E}$ becomes a map $A\mapsto{}Z_A$.  After all, if \E{}
were a genuine measurement of $A$, revealing, that is, the preexisting
(i.e., prior to the experiment) value of the observable $A$, then $Z$
would have to agree with that value and hence would be an unambiguous
random variable depending only on $A$.

But this sort of argument makes sense only if we take the quantum talk
of operators as observables too seriously.  We have emphasized in this
paper that operators do naturally arise in association with quantum
experiments. But there is little if anything in this association,
beyond the unfortunate language that is usually used to describe it,
that supports the notion that the operator $A$ associated with an
experiment \E{} is in any meaningful way genuinely measured by the
experiment.  From the nature of the association itself, it is
difficult to imagine what this could possibly mean.  And for those who
think they imagine some meaning in this talk, the impossibility
theorems show they are mistaken.

The bottom line is this: in \BM{} the random variables $Z_{\E}$ giving
the results of experiments \E{} depend, of course, on the experiment,
and there is no reason that this should not be the case when the
experiments under consideration happen to be associated with the same
operator. Thus with any \sa{} operator $A$, \BM{} naturally may
associate many different random variables $Z_{\E}$, one for each
different experiment $\E\mapsto{}A$ associated with $A$. A crucial
point here is that the map $\E\mapsto{}A$ is many-to-one.\footnote{We
   wish to remark that, quite aside {}from this many-to-oneness, the
   random variables $Z_{\E}$ cannot generally be regarded as
   corresponding to any sort of natural property of the ``measured''
   system. $Z_{\E}$, in general a function of the initial configuration
   of the system-apparatus composite, may fail to be a function of the
   configuration of the system alone. And even when, as is often the
   case, $Z_{\E}$ does depend only on the initial configuration of the
   system, owing to chaotic dynamics this dependence could have an
   extremely complex character.}  \bigskip

Suppose we define a map $A\mapsto{}Z_A$ by selecting, for each $A$,
one of the experiments, call it $\E_A$, with which $A$ is associated,
and define $Z_A$ to be $Z_{\E_A}$. Then the map so defined can't be
good, because of the impossibility theorems; moreover there is no
reason to have expected the map to be good. Suppose, for example, that
$[A,B]=0$. Should we expect that the joint distribution of $Z_A$ and
$Z_B$ will agree with the joint quantum mechanical distribution of $A$
and $B$? Only if the experiments $\E_A$ and $\E_B$ used to define
$Z_A$ and $Z_B$ both involved a common experiment that
``simultaneously measures $A$ and $B$,'' i.e., an experiment that is
associated with the commuting family $(A,B)$. If we consider now a
third operator $C$ such that $[A,C]=0$, but $[B,C]\neq 0$, then there
is no choice of experiment \E{} that would permit the definition of a
random variable $Z_A$ relevant both to a ``simultaneous measurement of
$A$ and $B$'' and a ``simultaneous measurement of $A$ and $C$'' since
no experiment is a ``simultaneous measurement of $A$, $B$, and $C$.''
In the situation just described we must consider at least two random
variables associated with $A$, $Z_{A,B}$ and $Z_{A,C}$, depending upon
whether we are considering an experiment ``measuring $A$ and $B$'' or
an experiment ``measuring $A$ and $C$.''  It should be clear that when
the random variables are assigned to experiments in this way, the
possibility of conflict with the predictions of \oqt{} is eliminated.
It should also be clear, in view of what we have repeatedly stressed,
that quite aside {}from the impossibility theorems, this way of
associating random variables with experiments is precisely what
emerges in \BM.

The dependence of the result of a ``measurement of the observable
$A$'' upon the other observables, if any, that are ``measured
simultaneously together with $A$''---e.g., that $Z_{A,B}$ and
$Z_{A,C}$ may be different---is called \emph{contextuality}: the
result of an experiment depends not just on ``what observable the
experiment measures'' but on more detailed information that conveys
the ``context'' of the experiment. The essential idea, however, if we
avoid misleading language, is rather trivial: that the result of an
experiment depends on the experiment.

To underline this triviality we remark that for two experiments, $\E$
and $\E'$, that ``measure $A$ and only $A$'' and involve no
simultaneous ``measurement of another observable,'' the results
$Z_{\E}$ and $Z_{\E'}$ may disagree. For example in Section
\ref{secapm} we described experiments $\E$ and $\E'$ both of which
``measured the position operator'' but only one of which measured the
actual initial position of the relevant particle, so that for these
experiments in general $Z_{\E}\neq Z_{\E'}$.

One might feel, however, that in the example just described the
experiment that does not measure the actual position is somewhat
disreputable---even though it is in fact a ``measurement of the
position operator.''  We shall therefore give another example, due to
D. Albert~\cite{albert}, in which the experiments are as simple and
canonical as possible and are entirely on the same footing.  Let
$\E_{\uparrow}$ and $\E_{\downarrow}$ be Stern-Gerlach measurements of
$A=\sigma_z$, with $\E_{\downarrow}$ differing {}from $\E_{\uparrow}$
only in that the polarity of the Stern-Gerlach magnet for
$\E_{\downarrow}$ is the reverse of that for $\E_{\uparrow}$. (In
particular, the geometry of the magnets for $\E_{\uparrow}$ and
$\E_{\downarrow}$ is the same.)  If the initial \wf{}
$\psi_{\text{symm}}$ and the magnetic field $\pm B$ have sufficient
reflection symmetry with respect to a plane between the poles of the
Stern-Gerlach magnets, the particle whose spin component is being
``measured'' cannot cross this plane of symmetry, so that if the
particle is initially above, respectively below, the symmetry plane,
it will remain above, respectively below, that plane. But because
their magnets have opposite polarity, $\E_{\uparrow}$ and
$\E_{\downarrow}$ involve opposite calibrations: $F_{\uparrow}=
-F_{\downarrow}$. It follows that
$$
Z^{\psi_{\text{symm}}}_{\E_{\uparrow}}= -
Z^{\psi_{\text{symm}}}_{\E_{\downarrow}}
$$
and the two experiments completely disagree about the ``value of
$\sigma_z$'' in this case.

The essential point illustrated by the previous example is that
instead of having in \BM{} a natural association
$\sigma_z\mapsto{}Z_{\sigma_z}$, we have a rather different pattern of
relationships, given in the example by
$$
\genfrac{}{}{0pt}{}{\E_{\uparrow}
   \to{Z_{\E_{\uparrow}}}}{\E_{\downarrow} \to
   {Z_{\E_{\downarrow}}}}\,^{\searrow}_\nearrow\, \sigma_z,$$

\subsection{Against ``Contextuality''}\label{sec:agcontext}

The impossibility theorems require the assumption of noncontextuality,
that the random variable $Z$ giving the result of a ``measurement of
quantum observable $A$'' should depend on $A$ alone, further
experimental details being irrelevant. How big a deal is
contextuality, the violation of this assumption? Here are two ways of
describing the situation:
\begin{enumerate}
\item In quantum mechanics (or quantum mechanics supplemented with
   hidden variables), observables and properties have a novel, highly
   nonclassical aspect: they (or the result of measuring them) depend
   upon which other compatible properties, if any, are measured
   together with them.

   In this spirit, Bohm and Hiley~\cite{bohi} write that (page 109)
\begin{quotation}\setlength{\baselineskip}{12pt}\noindent
   the quantum properties imply \dots that measured properties are not
   intrinsic but are inseparably related to the apparatus. It follows
   that the customary language that attributes the results of
   measurements \dots to the observed system alone can cause confusion,
   unless it is understood that these properties are actually dependent
   on the total relevant context.
\end{quotation}
They later add that (page 122)
\begin{quotation}\setlength{\baselineskip}{12pt}\noindent
   The context dependence of results of measurements is a further
   indication of how our interpretation does not imply a simple return
   to the basic principles of classical physics. It also embodies, in a
   certain sense, Bohr's notion of the indivisibility of the combined
   system of observing apparatus and observed object.
\end{quotation}
\smallskip

\item The result of an experiment depends upon the experiment.  Or, as
   expressed by Bell \cite{Bel87} (pg.166),
\begin{quotation}\setlength{\baselineskip}{12pt}\noindent
   A final moral concerns terminology. Why did such serious people take
   so seriously axioms which now seem so arbitrary? I suspect that they
   were misled by the pernicious misuse of the word `measurement' in
   contemporary theory. This word very strongly suggests the
   ascertaining of some preexisting property of some thing, any
   instrument involved playing a purely passive role. Quantum
   experiments are just not like that, as we learned especially {}from
   Bohr. The results have to be regarded as the joint product of
   `system' and `apparatus,' the complete experimental set-up. But the
   misuse of the word `measurement' makes it easy to forget this and
   then to expect that the `results of measurements' should obey some
   simple logic in which the apparatus is not mentioned. The resulting
   difficulties soon show that any such logic is not ordinary logic. It
   is my impression that the whole vast subject of `Quantum Logic' has
   arisen in this way {}from the misuse of a word. I am convinced that
   the word `measurement' has now been so abused that the field would
   be significantly advanced by banning its use altogether, in favour
   for example of the word `experiment.'
\end{quotation}
\end{enumerate}

With one caveat, we entirely agree with Bell's observation. The caveat
is this: We do not believe that the difference between quantum
mechanics and classical mechanics is quite as crucial for Bell's moral
as his language suggests it is.  For any experiment, quantum or
classical, it would be a mistake to regard any instrument involved as
playing a purely passive role, unless the experiment is a genuine
measurement of a property of a system, in which case the result is
determined by the initial conditions of the system alone. However, a
relevant difference between classical and quantum theory remains:
Classically it is usually taken for granted that it is in principle
possible to measure any observable without seriously affecting the
observed system, which is clearly false in quantum mechanics (or
\BM{}).\footnote{The assumption could (and probably should) also be
   questioned classically.}  \bigskip

Mermin has raised a similar question~\cite{merm93} (pg. 811):
\begin{quotation}\setlength{\baselineskip}{12pt}\noindent
   Is noncontextuality, as Bell seemed to suggest, as silly a condition
   as von Neumann's~\dots?
\end{quotation}
To this he answers:
\begin{quotation}\setlength{\baselineskip}{12pt}\noindent
   I would not characterize the assumption of noncontextuality as a
   silly constraint on a hidden-variables theory. It is surely an
   important fact that the impossibility of embedding quantum mechanics
   in a noncontextual hidden-variables theory rests not only on Bohr's
   doctrine of the inseparability of the objects and the measuring
   instruments, but also on a straightforward contradiction,
   independent of one's philosophic point of view, between some
   quantitative consequences of noncontextuality and the quantitative
   predictions of quantum mechanics.
\end{quotation}
This is a somewhat strange answer. First of all, it applies to von
Neumann's assumption (linearity), which Mermin seems to agree is
silly, as well as to the assumption of noncontextuality. And the
statement has a rather question-begging flavor, since the importance
of the fact to which Mermin refers would seem to depend on the
nonsilliness of the assumption which the fact concerns.

Be that as it may, Mermin immediately supplies his real argument for
the nonsilliness of noncontextuality. Concerning two experiments for
``measuring observable $A$,'' he writes that
\begin{quotation}\setlength{\baselineskip}{12pt}\noindent
   it is \dots\ an elementary theorem of quantum mechanics that the
   joint distribution \dots\ for the first experiment yields precisely
   the same marginal distribution (for $A$) as does the joint
   distribution \dots\ for the second, in spite of the different
   experimental arrangements. \dots\ The obvious way to account for
   this, particularly when entertaining the possibility of a
   hidden-variables theory, is to propose that both experiments reveal
   a set of values for $A$ in the individual systems that is the same,
   regardless of which experiment we choose to extract them {}from.
   \dots\ A {\it contextual} hidden-variables account of this fact
   would be as mysteriously silent as the quantum theory on the
   question of why nature should conspire to arrange for the marginal
   distributions to be the same for the two different experimental
   arrangements.
\end{quotation}
A bit later, Mermin refers to the ``striking insensitivity of the
distribution to changes in the experimental arrangement.''

For Mermin there is a mystery, something that demands an explanation.
It seems to us, however, that the mystery here is very much in the eye
of the beholder. It is first of all somewhat odd that Mermin speaks of
the mysterious silence of quantum theory concerning a question whose
answer, in fact, emerges as an ``elementary theorem of quantum
mechanics.'' What better way is there to answer questions about nature
than to appeal to our best physical theories?

More importantly, the ``two different experimental arrangements,'' say
$\E_1$ and $\E_2$, considered by Mermin are not merely any two
randomly chosen experimental arrangements. They obviously must have
something in common. This is that they are both associated with the
same \sa{} operator $A$ in the manner we have described: $\E_1\mapsto
A $ and $\E_2\mapsto A$. It is quite standard to say in this situation
that both $\E_1$ and $\E_2$ measure the observable $A$, but both for
\BM{} and for \oqt{} the very meaning of the association with the
operator $A$ is merely that the distribution of the result of the
experiment is given by the spectral measures for $A$.  Thus there is
no mystery in the fact that $\E_1$ and $\E_2$ have results governed by
the same distribution, since, when all is said and done, it is on this
basis, and this basis alone, that we are comparing them.

(One might wonder how it could be possible that there are two
different experiments that are related in this way. This is a somewhat
technical question, rather different {}from Mermin's, and it is one
that \BM{} and quantum mechanics readily answer, as we have explained
in this paper. In this regard it would probably be good to reflect
further on the simplest example of such experiments, the Stern-Gerlach
experiments $\E_{\uparrow}$ and $\E_{\downarrow}$ discussed in the
previous subsection.)

It is also difficult to see how Mermin's proposed resolution of the
mystery, ``that both experiments reveal a set of values for $A$ \dots\
that is the same, regardless of which experiment we choose to extract
them {}from,'' could do much good. He is faced with a certain pattern
of results in two experiments that would be explained if the
experiments did in fact genuinely measure the same thing.  The
experiments, however, as far as any detailed quantum mechanical
analysis of them is concerned, don't appear to be genuine measurements
of anything at all.  He then suggests that the mystery would be
resolved if, indeed, the experiments did measure the same thing, the
analysis to the contrary notwithstanding. But this proposal merely
replaces the original mystery with a bigger one, namely, of how the
experiments could in fact be understood as measuring the same thing,
or anything at all for that matter. It is like explaining the mystery
of a talking cat by saying that the cat is in fact a human being,
appearances to the contrary notwithstanding.  \bigskip

A final complaint about contextuality: the terminology is misleading.
It fails to convey with sufficient force the rather definitive
character of what it entails: {\it ``Properties'' that are merely
   contextual are not properties at all; they do not exist, and their
   failure to do so is in the strongest sense possible!}

\subsection{Nonlocality, Contextuality and Hidden Variables}

There is, however, a situation where contextuality is physically
relevant. Consider the EPRB experiment, outlined at the end of Section
\ref{secMCFO}.  In this case the dependence of the result of a
measurement of the spin component $\boldsymbol{\sigma}_{1}\cdot
\mathbf{a}$ of a particle upon which spin component of a distant
particle is measured together with it---the difference between
$Z_{\boldsymbol{\sigma}_{1}\cdot \mathbf{a},\;
   \boldsymbol{\sigma}_{2}\cdot \mathbf{b}}$ and
$Z_{\boldsymbol{\sigma}_{1}\cdot \mathbf{a},\;
   \boldsymbol{\sigma}_{2}\cdot \mathbf{c}}$ (using the notation
described in the seventh paragraph of Section \ref{sec:context})---is
an expression of {\em nonlocality}, of, in Einstein words, a ``spooky
action at distance.'' More generally, whenever the relevant context is
distant, contextuality implies nonlocality.

Nonlocality is an essential feature of Bohmian mechanics: the
velocity, as expressed in the guiding equation (\ref{eq:velo}), of any
one of the particles of a many-particle system will typically depend
upon the positions of the other, possibly distant, particles whenever
the wave function of the system is entangled, i.e., not a product of
single-particle wave functions.  In particular, this is true for the
EPRB experiment under examination. Consider the extension of the
single particle Hamiltonian (\ref{sgh}) to the two-particle case,
namely $$
H = -\frac{\hbar^{2}}{2m_{1}} \boldsymbol{\nabla}_{1}^{2}
-\frac{\hbar^{2}}{2m_{2}} \boldsymbol{\nabla}_{2}^{2}-
\mu_{1}\boldsymbol{\sigma}_1 {\bf \cdot B(\mathbf{ x_{1}) }}
-\mu_{2}\boldsymbol{\sigma}_2 {\bf \cdot B(\mathbf{x_{2}) }} .  $$
Then for initial singlet state, and spin measurements as described in
Sections \ref{secSGE} and \ref{noXexp}, it easily follows {}from the
laws of motion of Bohmian mechanics that
$$Z_{\boldsymbol{\sigma}_{1}\cdot \mathbf{a},\;
   \boldsymbol{\sigma}_{2}\cdot \mathbf{b}} \neq
Z_{\boldsymbol{\sigma}_{1}\cdot \mathbf{a},\;
   \boldsymbol{\sigma}_{2}\cdot \mathbf{c}}\;.$$

This was observed long ago by Bell \cite{Bel66}.  In fact, Bell's
examination of Bohmian mechanics led him to his celebrated nonlocality
analysis. In the course of his investigation of Bohmian mechanics he
observed that (\cite{Bel87}, p.  11)
\begin{quotation}\setlength{\baselineskip}{12pt}\noindent
   in this theory an explicit causal mechanism exists whereby the
   disposition of one piece of apparatus affects the results obtained
   with a distant piece.

   Bohm of course was well aware of these features of his scheme, and
   has given them much attention.  However, it must be stressed that,
   to the present writer's knowledge, there is no {\em proof} that {\em
     any} hidden variable account of quantum mechanics {\em must} have
   this extraordinary character. It would therefore be interesting,
   perhaps, to pursue some further ``impossibility proofs," replacing
   the arbitrary axioms objected to above by some condition of
   locality, or of separability of distant systems.
\end{quotation}
\noindent In a footnote, Bell added that ``Since the completion of
this paper such a proof has been found." This proof was published in
his 1964 paper \cite{Bel64}, "On the Einstein-Podolsky-Rosen Paradox,"
in which he derives Bell's inequality, the basis of his conclusion of
quantum nonlocality.

We find it worthwhile to reproduce here the analysis of Bell, deriving
a simple inequality equivalent to Bell's, in order to highlight the
conceptual significance of Bell's analysis and, at the same time, its
mathematical triviality.  The analysis involves two parts. The first
part, the Einstein-Podolsky-Rosen argument applied to the EPRB
experiment, amounts to the observation that for the singlet state the
assumption of locality implies the existence of noncontextual hidden
variables.  More precisely, it implies, for the singlet state, the
existence of random variables $ Z^{i}_{\boldsymbol{\alpha}}=
Z_{\boldsymbol{\alpha}\cdot \boldsymbol{\sigma}_i}$, $i=1, 2$,
corresponding to all possible spin components of the two particles,
that obey the agreement condition described in Section
\ref{sec:RVOIT}.  In particular, focusing on components in only 3
directions $\mathbf{a}$, $\mathbf{b}$ and $\mathbf{c}$ for each
particle, locality implies the existence of 6 random variables
$$
Z^{i}_{\boldsymbol{\alpha}}\qquad i=1,2\quad {\boldsymbol{\alpha}}=
\mathbf{a},\; \mathbf{b},\; \mathbf{c}
$$
such that
\begin{eqnarray}
Z^{i}_{\boldsymbol{\alpha}} &=& \pm 1  \label{eq:pc1}\\
Z^{1}_{{\boldsymbol{\alpha}}}& =&
-Z^{2}_{{\boldsymbol{\alpha}}}\label{eq:pc2}
\end{eqnarray}
and, more generally,
\begin{equation}
\text{Prob}(Z^{1}_{\boldsymbol{\alpha}}\neq
Z^{2}_{\boldsymbol{\alpha}}) =
q_{ {\boldsymbol{\alpha}}
   {\boldsymbol{\beta}}   }, \label{eq:pc3}
\end{equation}
the corresponding quantum mechanical probabilities. This conclusion
amounts to the idea that measurements of the spin components reveal
preexisting values (the $Z^{i}_{\boldsymbol{\alpha}}$), which,
assuming locality, is implied by the perfect quantum mechanical
anticorrelations \cite{Bel64}:
\begin{quotation}\setlength{\baselineskip}{12pt}\noindent
   Now we make the hypothesis, and it seems one at least worth
   considering, that if the two measurements are made at places remote
   {}from one another the orientation of one magnet does not influence
   the result obtained with the other. Since we can predict in advance
   the result of measuring any chosen component of
   ${\boldsymbol{\sigma}}_2$, by previously measuring the same
   component of ${\boldsymbol{\sigma}}_1$, it follows that the result
   of any such measurement must actually be predetermined.
\end{quotation}
People very often fail to appreciate that the existence of such
variables, given locality, is not an assumption but a consequence of
Bell's analysis.  Bell repeatedly stressed this point (by determinism
Bell here means the existence of hidden variables):
   \begin{quotation}\setlength{\baselineskip}{12pt}
     It is important to note that to the limited degree to which {\em
       determinism} plays a role in the EPR argument, it is not assumed
     but {\em inferred}. What is held sacred is the principle of `local
     causality' -- or `no action at a distance'.  \ldots

     It is remarkably difficult to get this point across, that
     determinism is not a {\em presupposition} of the analysis.
     (\cite{Bel87}, p. 143)

     Despite my insistence that the determinism was inferred rather
     than assumed, you might still suspect somehow that it is a
     preoccupation with determinism that creates the problem. Note well
     then that the following argument makes no mention whatever of
     determinism.  \ldots\ Finally you might suspect that the very
     notion of particle, and particle orbit \ldots\ has somehow led us
     astray. \ldots\ So the following argument will not mention
     particles, nor indeed fields, nor any other particular picture of
     what goes on at the microscopic level.  Nor will it involve any
     use of the words `quantum mechanical system', which can have an
     unfortunate effect on the discussion.  The difficulty is not
     created by any such picture or any such terminology.  It is
     created by the predictions about the correlations in the visible
     outputs of certain conceivable experimental set-ups.
     (\cite{Bel87}, p. 150)
\end{quotation}

The second part of the analysis, which unfolds the ``difficulty
\ldots\ created by the \ldots\ correlations,'' involves only very
elementary mathematics. Clearly,
$$
\text{Prob}\left( \{Z^{1}_{\mathbf{a}} = Z^{1}_{\mathbf{b}}\} \cup
   \{Z^{1}_{\mathbf{b}} = Z^{1}_{\mathbf{c}}\} \cup
   \{Z^{1}_{\mathbf{c}} = Z^{1}_{\mathbf{a}}\} \right) =1\;.$$
since at
least two of the three (2-valued) variables
$Z^{1}_{\boldsymbol{\alpha}}$ must have the same value. Hence, by
elementary probability theory,
$$
\text{Prob} \left( Z^{1}_{\mathbf{a}} = Z^{1}_{\mathbf{b}}\right) +
\text{Prob} \left( Z^{1}_{\mathbf{b}} = Z^{1}_{\mathbf{c}}\right) +
\text{Prob} \left( Z^{1}_{\mathbf{c}} = Z^{1}_{\mathbf{a}} \right) \ge
1, $$
and using the perfect anticorrelations (\ref{eq:pc2}) we have
that
   \begin{equation}
\text{Prob}
\left( Z^{1}_{\mathbf{a}} =  -Z^{2}_{\mathbf{b}}\right)
+ \text{Prob}
\left( Z^{1}_{\mathbf{b}} = -Z^{2}_{\mathbf{c}}\right)
+ \text{Prob}
\left(  Z^{1}_{\mathbf{c}} =
-Z^{2}_{\mathbf{a}}
     \right) \ge 1, \label{eq:bellineq}
\end{equation}
which is equivalent to Bell's inequality and in conflict with
(\ref{eq:pc3}).  For example, when the angles between $\mathbf{a}$,
$\mathbf{b}$ and $\mathbf{c}$ are 120$^{0}$ the 3 relevant quantum
correlations $q_{ {\boldsymbol{\alpha}} {\boldsymbol{\beta}} }$ are
all $1/4$.

To summarize the argument, let H be the hypothesis of the existence of
the noncontextual hidden variables we have described above.  Then the
logic of the argument is:
\begin{eqnarray}
\text{Part 1:}&\qquad \mbox{quantum mechanics} + \mbox{locality}
&\Rightarrow\quad
\mbox{H} \label{qmlic}\\
\text{Part 2:}&\qquad \mbox{quantum mechanics} &\Rightarrow\quad
\mbox{not H} \label{qmic}\\
\text{Conclusion:}&\qquad \mbox{quantum mechanics} &\Rightarrow\quad
\mbox{not locality} \label{qmiccon}
\end{eqnarray}
To fully grasp the argument it is important to appreciate that the
identity of H---the existence of the noncontextual hidden
variables---is of little substantive importance. What is important is
not so much the identity of H as the fact that H is incompatible with
the predictions of quantum theory.  The identity of H is, however, of
great historical significance: It is responsible for the misconception
that Bell proved that hidden variables are impossible, a belief until
recently almost universally shared by physicists.

Such a misconception has not been the only reaction to Bell's
analysis.  Roughly speaking, we may group the different reactions into
three main categories, summarized by the following statements:
\begin{enumerate}
\item Hidden variables are impossible.
\item Hidden variables are possible, but they must be contextual.
\item Hidden variables are possible, but they must be nonlocal.
\end{enumerate}
Statement 1 is plainly wrong.  Statement 2 is correct but not terribly
significant.  Statement 3 is correct, significant, but nonetheless
rather misleading. It follow {}from (\ref{qmlic}) and (\ref{qmic})
that {\em any} account of quantum phenomena must be nonlocal, not just
any hidden variables account. Bell's argument shows that nonlocality
is implied by the predictions of standard quantum theory itself. Thus
if nature is governed by these predictions, then {\em nature is
   nonlocal}. (That nature is so governed, even in the crucial
EPR-correlation experiments, has by now been established by a great
many experiments, the most conclusive of which is perhaps that of
Aspect \cite{Aspect1982}.)

\section{Against Naive Realism About Operators}

Traditional naive realism is the view that the world {\it is\/} pretty
much the way it {\it seems,\/} populated by objects which force
themselves upon our attention as, and which in fact are, the locus of
sensual qualities.  A naive realist regards these ``secondary
qualities,'' for example color, as objective, as out there in the
world, much as perceived.  A decisive difficulty with this view is
that once we understand, say, how our perception of what we call color
arises, in terms of the interaction of light with matter, and the
processing of the light by the eye, and so on, we realize that the
presence out there of color per se would play no role whatsoever in
these processes, that is, in our understanding what is relevant to our
perception of ``color.'' At the same time, we may also come to realize
that there is, in the description of an object provided by the
scientific world-view, as represented say by classical physics,
nothing which is genuinely ``color-like.''

A basic problem with quantum theory, more fundamental than the
measurement problem and all the rest, is a naive realism about
operators, a fallacy which we believe is far more serious than
traditional naive realism: With the latter we are deluded partly by
language but in the main by our senses, in a manner which can scarcely
be avoided without a good deal of scientific or philosophical
sophistication; with the former we are seduced by language alone, to
accept a view which can scarcely be taken seriously without a large
measure of (what often passes for) sophistication.

Not many physicists---or for that matter philosophers---have focused
on the issue of naive realism about operators, but Schr\"odinger and
Bell have expressed similar or related concerns:

\begin{quotation}\setlength{\baselineskip}{12pt}\noindent \dots the
   new theory [quantum theory] \dots considers the [classical] model
   suitable for guiding us as to just which measurements can in
   principle be made on the relevant natural object.  \dots Would it
   not be pre-established harmony of a peculiar sort if the
   classical-epoch researchers, those who, as we hear today, had no
   idea of what {\it measuring\/} truly is, had unwittingly gone on to
   give us as legacy a guidance scheme revealing just what is
   fundamentally measurable for instance about a hydrogen
   atom!?~\cite{Sch35}
\end{quotation}

\begin{quotation}\setlength{\baselineskip}{12pt}\noindent
   Here are some words which, however legitimate and necessary in
   application, have no place in a {\it formulation\/} with any
   pretension to physical precision: {\it system; apparatus;
     environment; microscopic, macroscopic; reversible, irreversible;
     observable; information; measurement.\/}

   \dots The notions of ``microscopic'' and ``macroscopic'' defy
   precise definition. \dots Einstein said that it is theory which
   decides what is ``observable''.  I think he was right. \dots
   ``observation'' is a complicated and theory-laden business.  Then
   that notion should not appear in the {\it formulation\/} of
   fundamental theory. \dots

   On this list of bad words {}from good books, the worst of all is
   ``measurement''.  It must have a section to itself.~\cite{Bel90}
\end{quotation}

We agree almost entirely with Bell here.  We insist, however, that
``observable'' is just as bad as ``measurement,'' maybe even a little
worse.  Be that as it may, after listing Dirac's measurement
postulates Bell continues:

\begin{quotation}\setlength{\baselineskip}{12pt}\noindent
   It would seem that the theory is exclusively concerned about
   ``results of measurement'', and has nothing to say about anything
   else.  What exactly qualifies some physical systems to play the role
   of ``measurer''?  Was the wave function of the world waiting to jump
   for thousands of millions of years until a single-celled living
   creature appeared?  Or did it have to wait a little longer, for some
   better qualified system \dots with a Ph.D.?  If the theory is to
   apply to anything but highly idealized laboratory operations, are we
   not obliged to admit that more or less ``measurement-like''
   processes are going on more or less all the time, more or less
   everywhere.  Do we not have jumping then all the time?

   The first charge against ``measurement'', in the fundamental axioms
   of quantum mechanics, is that it anchors the shifty split of the
   world into ``system'' and ``apparatus''.  A second charge is that
   the word comes loaded with meaning {}from everyday life, meaning
   which is entirely inappropriate in the quantum context.  When it is
   said that something is ``measured'' it is difficult not to think of
   the result as referring to some {\it preexisting property\/} of the
   object in question.  This is to disregard Bohr's insistence that in
   quantum phenomena the apparatus as well as the system is essentially
   involved.  If it were not so, how could we understand, for example,
   that ``measurement'' of a component of ``angular momentum'' \dots
   {\it in an arbitrarily chosen direction\/} \dots yields one of a
   discrete set of values?  When one forgets the role of the apparatus,
   as the word ``measurement'' makes all too likely, one despairs of
   ordinary logic \dots hence ``quantum logic''.  When one remembers
   the role of the apparatus, ordinary logic is just fine.

   In other contexts, physicists have been able to take words {}from
   ordinary language and use them as technical terms with no great harm
   done.  Take for example the ``strangeness'', ``charm'', and
   ``beauty'' of elementary particle physics.  No one is taken in by
   this ``baby talk''. \dots Would that it were so with
   ``measurement''.  But in fact the word has had such a damaging
   effect on the discussion, that I think it should now be banned
   altogether in quantum mechanics.  ({\sl Ibid.\/})
\end{quotation}

While Bell focuses directly here on the misuse of the word
``measurement'' rather than on that of ``observable,'' it is worth
noting that the abuse of ``measurement'' is in a sense inseparable
{}from that of ``observable,'' i.e., {}from naive realism about
operators.  After all, one would not be very likely to speak of
measurement unless one thought that something, some ``observable''
that is, was somehow there to be measured.

Operationalism, so often used without a full appreciation of its
consequences, may lead many physicists to beliefs which are the
opposite of what one might expect.  Namely, by believing somehow that
a physical property {\it is} and {\it must be} defined by an
operational definition, many physicists come to regard properties such
as spin and polarization, which can easily be operationally defined,
as intrinsic properties of the system itself, the electron or photon,
despite all the difficulties that this entails.  If operational
definitions were banished, and ``real definitions'' were required,
there would be far less reason to regard these ``properties'' as
intrinsic, since they are not defined in any sort of intrinsic way; in
short, we have no idea what they really mean, and there is no reason
to think they mean anything beyond the behavior exhibited by the
system in interaction with an apparatus.  \medskip

There are two primary sources of confusion, mystery and incoherence in
the foundations of quantum mechanics: the insistence on the
completeness of the description provided by the \wf{}, despite the
dramatic difficulties entailed by this dogma, as illustrated most
famously by the measurement problem; and \nrao{}. While the second
seems to point in the opposite direction {}from the first, the dogma
of completeness is in fact nourished by \nrao{}. This is because
\nrao{} tends to produce the belief that a more complete description
is impossible because such a description should involve preexisting
values of the quantum observables, values that are revealed by
measurement. And this is impossible. But without \nrao{}---without
being misled by all the quantum talk of the measurement of
observables---most of what is shown to be impossible by the
impossibility theorems would never have been expected to begin with.

\addcontentsline{toc}{section}{Acknowledgments}
\section*{Acknowledgments}
An early version of this paper had a fourth author: Martin Daumer.
Martin left our group a long time ago and has not participated since
in the very substantial changes in both form and content that the
paper has undergone. His early contributions are very much
appreciated. We thank Roderich Tumulka for a careful reading of this
manuscript and helpful suggestions. This work was supported in part by
NSF Grant No. DMS--9504556, by the DFG, and by the INFN.  We are
grateful for the hospitality that we have enjoyed, on more than one
occasion, at the Mathematisches Institut of
Ludwig-Maximilians-Universit\"at M\"unchen, at the Dipartimento di
Fisica of Universit\`a degli Studi di Genova, and at the Mathematics
Department of Rutgers University.

\end{document}